\documentclass[sigconf,screen,authorversion,nonacm]{acmart}
\pdfoutput=1

\newcommand\vldbpagestyle{plain} 

\usepackage{multirow}

\usepackage{pifont}
\usepackage{booktabs}
\usepackage{graphicx}
\usepackage{enumitem}
\usepackage{bm}
\usepackage{xcolor}
\usepackage{algorithm}
\usepackage{algpseudocode}
\usepackage{array}

\usepackage{amsmath}
\usepackage{xcolor}
\usepackage{graphicx}
\usepackage{cleveref}

\usepackage{tabularx}
\usepackage{colortbl}
\usepackage{caption}
\usepackage[caption=false]{subfig} 
\usepackage{appendix}

\newenvironment{myitemize}{%
\begin{itemize}[leftmargin=1em, itemsep=.1em, parsep=.1em, topsep=.1em,
    partopsep=.1em]}
{\end{itemize}}

\newenvironment{myenumerate}{%
\begin{enumerate}[leftmargin=1em, itemsep=.1em, parsep=.1em, topsep=.1em,
    partopsep=.1em]}
{\end{enumerate}}

\newtheorem{ex}{Example}

\newtheorem{example}[ex]{Example} 

\definecolor{lightblue}{rgb}{0.3,0.3,1}
\definecolor{lightred}{rgb}{1,0.3,0.3}

\begin{document}

\title{GPU Acceleration of SQL Analytics on Compressed Data}

\author{
    Zezhou Huang, Krystian Sakowski, Hans Lehnert, Wei Cui, Carlo Curino, Matteo Interlandi,\\ Marius Dumitru, Rathijit Sen
}
\affiliation{
    \institution{Microsoft}
    \country{}
}
\email{
    {zacharyhuang, krsakows, weicu, mariusd, 
    firstname.lastname}@microsoft.com
}

\begin{abstract}
GPUs are uniquely suited to accelerate (SQL) analytics workloads thanks to their massive compute parallelism and High Bandwidth Memory (HBM)---when datasets fit in the GPU HBM, performance is unparalleled. Unfortunately, GPU HBMs remain typically small when compared with lower-bandwidth CPU main memory. Besides brute-force scaling across many GPUs, current solutions to accelerate queries on large datasets include leveraging data partitioning and loading smaller data batches in GPU HBM, and hybrid execution with a connected device (e.g., CPUs). Unfortunately, these approaches are exposed to the limitations of lower main memory and host-to-device interconnect bandwidths, introduce additional I/O overheads, or incur higher costs. This is a substantial problem when trying to scale adoption of GPUs on larger datasets. Data compression can alleviate this bottleneck, but to avoid paying for costly decompression/decoding, an ideal solution must include computation primitives to operate directly on data in compressed form.
  
  This is the focus of our paper: a set of new methods for running queries directly on light-weight compressed data using schemes such as Run-Length Encoding (RLE), index encoding, bit-width reductions, and dictionary encoding. Our novelty includes operating on multiple RLE columns without decompression, handling heterogeneous column encodings, and leveraging PyTorch tensor operations for portability across devices. Experimental evaluations show speedups of an order of magnitude compared to state-of-the-art commercial CPU-only analytics systems, for real-world queries on a production dataset that would not fit into GPU memory uncompressed. This work paves the road for GPU adoption in a much broader set of use cases, and it is complementary to most other scale-out or fallback mechanisms.

  \vspace{-20pt}
\end{abstract}

\maketitle

\pagestyle{\vldbpagestyle}
\vspace{4ex}

\section{Introduction}

The massive compute parallelism and multi-TB/sec device memory bandwidths of modern GPUs make them attractive for accelerating SQL analytics queries~\cite{crystal_shanbhag_2020,book:gpudb:2021}. 
However, one challenge is that the high-bandwidth memory (HBM) capacity on GPUs is much less, usually by integer factors or an order of magnitude, compared to lower-bandwidth main memory capacity in modern high-end CPU servers. This restricts the dataset sizes that can be processed on GPUs without requiring data movement at run time over slow CPU-GPU interconnects\footnote{\label{fn1}
PCIe4 has a bandwidth of 31.5 GB/sec; 63 GB/sec for PCIe5. However, NVLink-C2C can achieve up to 450 GB/sec. 
This is in contrast to GPU's HBM bandwidths: e.g., A100s have a memory bandwidth of 2TB/sec, 3.35 TB/sec for H100s, and 5.3 TB/sec for MI300Xs. In comparison, a typical CPU server can have up to 0.5 TB/sec.}, or doing CPU-GPU hybrid processing~\cite{rosenfeld2022query}, or using multiple GPUs~\cite{Yogatama25-lancelot,Lutz20-pump-up-the-volume,maltenberger22-multigpu-sort,voltron-data,Rui20-multigpujoin} which increases costs. 

One approach to alleviate this bottleneck is to run queries on compressed data, which is the focus of this paper. We focus on light-weight encodings for compression, in particular, run-length encoding (RLE) for consecutively repeating values, index encoding for sparse data, dictionary encoding, and bit-width reductions. Our methods not only allow GPU query processing on datasets that would not fit uncompressed in the GPU HBM, but also achieve high performance by leveraging redundancy eliminated by compression.

Being able to operate directly on RLE data provides both space and time advantages. The representation uses constant space, \textit{regardless of the length of the run}. In contrast, an uncompressed representation uses space that is directly proportional to the number of values, that is, the length of the run. This reduces GPU HBM capacity requirements. Operating on RLE data can also be faster. For example, a SUM operation over an RLE data column would need to multiple the run values with the run lengths, then add the results over all runs, whereas in a plain/uncompressed representation, every value in the column needs to be scanned. Additionally, the space savings with RLE reduces I/O overheads for data movement between the GPUs and other devices, e.g., CPU-GPU data transfers over the (relatively) low-bandwidth PCIe bus.   

SQL query processing directly on RLE data, however, is challenging when operations involve multiple data columns. This is because of misalignment in runs between columns that can occur due to different data distributions as well as effect of predicate filters on different columns. Continuing with the above example, SUM (A+B) on two RLE columns A and B is not straightforward to implement since runs in A and B can start and end at different positions, making a direct step of adding the run values between the two columns impossible. While this can be handled with an iterative approach that loops through the different runs and keeps track of their start and end positions, such an approach is inefficient on parallel computational backends such as GPUs due to severe under-utilization of resources. Additional complexities arise with heterogeneous encoding schemes across columns, e.g., if A uses RLE, but B uses an index representation that is more suitable for non-repetitive sparse data.  

While prior works~\cite{hippogriffdb,tile-integer-compression,fastlanes-gpu} have recognized and exploited the space-savings advantages of RLE data, they usually decompress the data in the GPU memory hierarchy, shared memory, and registers before using them as inputs to relational operator implementations.
Some works study the potential for query processing over partially compressed data, but focus mostly on specific operations such as table scan~\cite{fang2010database} or join~\cite{yuan2013yin} without providing a comprehensive framework for all relational operators.
In contrast, we develop \textit{novel parallel algorithms and a framework for SQL relational operators to: (1) directly operate on encoded data as much as possible without decompressing them, and (2) be efficiently implemented on parallel devices such as GPUs}. Our methods provide query run time speedups of an order of magnitude compared to state-of-the-art CPU-only relational engines for representative production queries on a production dataset that is larger than the HBM capacity of a single A100 GPU in uncompressed form. To make our implementations easily portable to multiple accelerators, we use the same approach as TQP~\cite{surakav_he_2022, gandhi2022tensor} and implement SQL queries using PyTorch tensor programs, but extend it to show how PyTorch functions can be used to implement relational operators for compressed data.

\noindent In summary, we extend the state of the art in the following ways.
\begin{myitemize}
\item We show how SQL operations can be performed on data encoded for compression, particularly RLE data, in GPUs \textit{without expanding as much as possible}, in contrast to traditional approaches.
\item We demonstrate for the first time how this can be done in TQP with PyTorch tensor operations.
\item We show query run time speedups of an order of magnitude compared to state-of-the-art CPU-only database and analytical systems on a real-world dataset that is too large to fit into the HBM of a high-end GPU without compression.
\end{myitemize}
As far as we know this is the first work showcasing a comprehensive framework allowing to execute complex queries (e.g., including joins, group by, aggregation, projection and filter operators) end to end on lightweight-compressed data in GPU.

The rest of the paper is organized as follows. Section~\ref{sec:background} briefly describes TQP and the PyTorch primitives that we leverage in this work. Section~\ref{sec:encodings} presents the encoding schemes for compression that we use. Section~\ref{sec:core-ops} introduces our novel and parallel implementations of core transformations on compressed data. Sections~\ref{sec:logical}--\ref{sec:join} show how we use them in novel ways to efficiently implement relational operators on GPUs. Section~\ref{sec:experiments} presents experimental results and speedups with our techniques. Section~\ref{sec:related} discusses related work and Section~\ref{sec:conclude} concludes the paper.
\section{Background}
\label{sec:background}

We first describe TQP, our query execution framework, then introduce the PyTorch primitives used in our compressed operators.

\subsection{TQP}
\label{sec:background-tqp}

The Tensor Query Processor (TQP)~\cite{surakav_he_2022, tensor_tea_vldb_2022} converts SQL queries into tensor programs, then runs them using Tensor Computation Runtimes (TCRs), such as PyTorch~\cite{pytorch_paszke_2019}, on hardware backends such as GPUs~\cite{surakav_he_2022, gandhi2022tensor, gpudb-characterization-opt} and APUs~\cite{tqp-xbox}. This makes TQP easily portable to different devices while at the same time benefiting from the optimizations provided by the TCRs. In this work, we continue to use TQP's approach to get the same portability and performance advantages, but extend it to show how we can convert queries to tensor programs that can operate on lightweight-compressed data.

TQP supports input tabular data from files in Parquet~\cite{parquet} and CSV formats, and from in-memory formats such as NumPy~\cite{numpy} and Pandas DataFrames~\cite{pandas}. TQP converts each input column into PyTorch tensors. The conversion is straightforward for numeric and date columns. TQP does not currently support decimals, and instead represents them using floating-point numbers. ASCII string columns are value-encoded and dictionary-encoded. At the end of the conversion process, every column is represented using one or more numeric PyTorch tensors. The conversion step is done offline, before running queries. 

Given an input SQL query, TQP uses a query optimizer (currently, Apache Spark's~\cite{spark} Catalyst optimizer) to obtain a physical query plan. It then converts it into an equivalent tensor program~\cite{surakav_he_2022}. It loads tensors corresponding to the input columns needed by the query into the device memories---for GPUs, this is the High Bandwidth Memory (HBM) which is available as global memory on the GPUs. TQP loads and operates on entire columns, spanning all rows of the table, rather than splitting them up into smaller chunks. This is beneficial for good performance as it avoids overheads of repeated sets of kernel launches and improves GPU resource utilization, but requires the full column tensor (in addition to temporary results) to fit in the available HBM capacity. In this work, we continue with TQP's approach of loading and operating on full columns, but additionally allow compact representation for encoded data, thereby reducing the memory requirements, as we discuss next.

\subsection{PyTorch Primitives}
\label{sec:background-pytorch}

Tensors are the fundamental data structures in PyTorch. While tensors can be high-dimensional (for ML applications), in this work we use them as one-dimensional arrays that directly correspond to database columns. Our algorithms rely on several PyTorch primitives that have been highly optimized for ML tasks but proves effective for database query processing.

\noindent\textbf{Tensor Indexing []:} The bracket operator supports three modes: (1) single indexing (e.g., $data[0]$ selects the element at position 0), (2) multi-indexing with index tensors (e.g., $data[[0, 2]]$ selects elements at positions 0 and 2), and (3) filtering with boolean masks (e.g., $data[mask]$ returns elements where mask is True). For example, given $data=[10, 20, 30, 40]$: $data[[0, 2]]$ returns $[10, 30]$; $data[[True, False, True, False]]$ also returns $[10, 30]$.

\noindent\textbf{bucketize(input, boundaries, right):} Binary searches for bucket indices. With $right=True$, element $j$ goes to bucket $i$ if $boundaries[i-1]$ $ < input[j] \leq boundaries[i]$; with $right=False$, if $boundaries[i-1] \leq input[j] < boundaries[i]$. For example, $bucketize([1, 5, 3], [2, 4, 5])$ returns $[0, 2, 1]$ with $right=True$ (1$\leq$2, 4<5$\leq$5, 2<3$\leq$4); returns $[0, 3, 1]$ with $right=False$ (1<2, 5$\geq$5, 2$\leq$3<4).

\noindent\textbf{arange(end):} Generates a tensor with values from $0$ to $end-1$. For example, $arange(5)$ returns $[0, 1, 2, 3, 4]$.

\noindent\textbf{repeat\_interleave(input, repeats):} Repeats each element of $input$ consecutively by the corresponding count in $repeats$. For example, $repeat\_interleave([10, 20], [2, 1])$ returns $[10, 10, 20]$.

\noindent\textbf{unique(input):} Returns unique elements and inverse indices mapping original elements to their positions. For example, $unique([3, 1, 3, 2])$ returns unique $[1, 2, 3]$ and inverse indices $[2, 0, 2, 1]$.

\noindent\textbf{scatter(src, index, reduce):} Accumulates values from $src$ at positions specified by $index$ using the $reduce$ operation. Multiple values targeting the same index are combined according to the $reduce$. For example, $scatter([10, 20, 30], [0, 1, 0], reduce=sum)$ results in $[40, 20]$ (10+30 at index 0, 20 at index 1).

\section{Encodings for Compression and their Tensor representations}
\label{sec:encodings}

In order to enable query processing on compressed data, we expand the encodings supported in TQP.  
We add support for two additional light-weight encoding techniques---Run-Length Encoding (RLE) and indexing. RLE is useful for compactly representing sequences of consecutively repeating values. We use indexing in novel ways, including to separate outliers and enable compression through bit-width reduction, and for efficient representation of sparse data. 
Currently we do not support operations on data in heavy-weight compression formats, e.g., Snappy, zstd, LZ4, gzip, etc.\vspace{-2pt}

\subsection{Basic Encodings}

We will discuss multiple columns, each with multiple fields for their tensor representations. For notation clarity: we use subscripts to distinguish different columns (e.g., $c_1$, $c_2$), and superscripts for field access within each column's encoding (e.g., $c_1^{start}$, $c_{2}^{pos}$).

First, we list the types of basic encodings for columns that we support and their tensor representations. These are applied on top of any dictionary encodings of values, e.g., for string columns.
\begin{myitemize}
\item \textbf{Plain}: This is the existing tensor representation used by TQP as we discussed in Section~\ref{sec:background-tqp}. For every column, there is a 1:1 mapping from a row position in the column to the corresponding position in the tensor representing the values in the column.
\item \textbf{RLE}: We use three tensors to represent a set of RLE runs: (1) values ($c^{val}$), (2) start positions ($c^{start}$), and (3) end positions ($c^{end}$). Tensors $c^{val}$, $c^{start}$, and $c^{end}$ are of equal length, and the $i^{th}$ entry represents a run with value $c^{val}[i]$ spanning row numbers $c^{start}[i] \rightarrow c^{end}[i]$. The positions are zero-based and unique. Tensors $c^{val}$, $c^{start}$, $c^{end}$ are sorted by $c^{start}$ (equivalently, by $c^{end}$). Note that, we could have used a tensor for run lengths ($c^{len}$) instead of for end positions ($c^{end}$). We can choose either lengths or end positions based on convenience, and calculate the other at run time using the equation: $c^{len}=c^{end}-c^{start}+1$. Runs are non-overlapping by position.
\item \textbf{Index}: We use two tensors to represent index encoding: (1) values ($c^{val}$), and (2) positions ($c^{pos}$). Tensors $c^{val}$ and $c^{pos}$ are of equal length, sorted by $c^{pos}$, and the $i^{th}$ entry represents value $c^{val}[i]$ at row number $c^{pos}[i]$. The positions are zero-based and unique.
\end{myitemize}
While Plain encoding has an implicit 1:1 correspondence between row numbers and positions in the tensor representation, RLE and Index encoding  explicitly track row numbers. This also allows for efficient representation when there are gaps in the underlying data, e.g., when some portions of the column are deselected after application of filter predicates. Thus, for RLE, for two consecutive entries at tensor positions $i$ and $i+1$, we have $c^{start}_{i+1}\ge c^{end}_i + 1$. Similarly, for Index encoding, we have $c^{pos}_{i+1}\ge c^{pos}_i + 1$. In contrast, Plain encoding does not allow for gaps in the tensor representation. 

\subsection{Composite Encodings}

In addition to basic encodings, we also introduce the following novel composite encodings that combine Index with Plain and RLE to enable further compression.
\begin{myitemize}
\item \textbf{Plain + Index}: PyTorch requires a uniform data type for all elements within a single tensor. However, outliers can force the entire tensor to use a larger data type than necessary. To enable bit-width reduction with narrower data types, we can combine Plain with Index encoding. We store most values in one PyTorch tensor with a narrower type and represent the outliers separately using Index encoding. At positions corresponding to outliers, the Plain tensor contains uninitialized PyTorch tensor values (values happened to be in that memory location). These uninitialized values are never used during query execution. 
\item \textbf{RLE + Index}: RLE works best for continuous value segments. However, some columns may contain both continuous (pure) and non-continuous (impure) segments. While impure segments can be represented by a series of unit-length RLE runs, this can be inefficient since each run is represented by three elements (value, start, end). Instead, we can handle impure segments with Index encoding. The Index positions and the RLE intervals are disjoint in this composite encoding.
\end{myitemize}
Note that our bit-width reduction differs from traditional CPU compression techniques. While CPU approaches break columns into small groups to optimize bit width and combine with techniques like DELTA and FOR~\cite{c_store_abadi_2006, abadi_design_implementation_column_dbs_2013,zukowski2006super}, the small groups  suit poorly for GPU processing. Instead, we leverage PyTorch's preference for large tensors by using index-based outlier separation. After removing outliers, we apply centering to reduce bit width. Centering is similar to FOR, but instead of local reference values (typically minimum in each group) we use a global reference value of the mid-range for the entire column. This allows us to represent most values with narrower bit widths while handling outliers separately, achieving good compression without sacrificing GPU execution efficiency.

\begin{figure}
    \centering  \includegraphics[width=0.8\columnwidth]{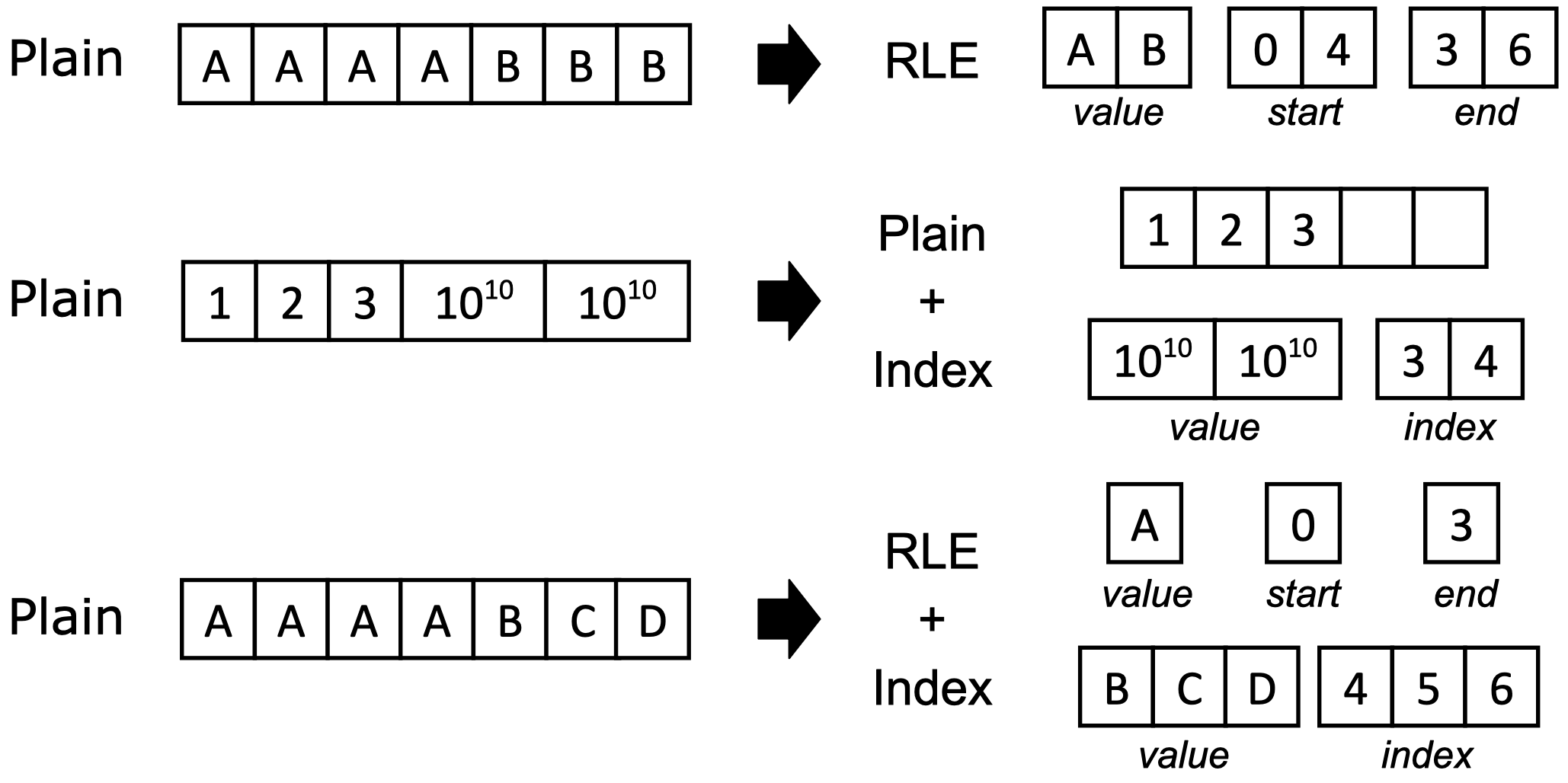}
    \vspace{-10pt}
    \caption{GPU-Optimized Tensor Data Representations.\vspace{-10pt}}
    \label{fig:representation}
    \vspace{-2ex}
\end{figure}

{
\vspace{-10pt}
\flushleft
\begin{example}
Consider the various optimized Tensor representations for Data Compression in \Cref{fig:representation}.
\end{example}
}
\textbf{Plain $\rightarrow$ RLE example}:
The sequence $[A,A,A,A,B,B,B]$ is compressed using run-length encoding. A repeats from positions 0 to 3, and B repeats from positions 4 to 6. We store this as $v=[A,B], s=[0,4], e=[3,6]$. This captures exactly where each value run begins and ends.

\textbf{Plain $\rightarrow$ Plain + Index example}:
When handling a column with values $[1,2,3,10^{10},10^{10}]$, the outliers at indices 3 and 4 would force the entire tensor to use a larger data type. We separate the data: regular values in Plain tensor $[1,2,3]$ and outliers in a separate value tensor $[10^{10},10^{10}]$ with their positions in index tensor $[3,4]$. In a dataset with 1 billion elements, if a few outliers require int64 but most values fit in int8, Plain representation would force all values into int64, using 8 GiB of memory (8 bytes × 1 billion). With Plain + Index, we store most values in int8 (1 GiB) plus a small overhead for outliers and their indices, saving  $\sim$7 GiB of memory while preserving all data values.

\textbf{Plain $\rightarrow$ RLE + Index example}:
The sequence $[A,A,A,A,B,C,D]$ has a consecutive run of A (positions 0-3) but B and C do not form runs. We use RLE for the A run ($v=[A], s=[0], e=[3]$) and Index encoding for the scattered values ($v=[B,C,D], p=[4,5,6]$).

\subsection{Data and Mask column 
 representations}

Our encoding schemes are versatile and can be used to represent: (1) data columns containing actual values from database tables and intermediate results, (2) mask columns that encode boolean predicates for selections and filters, and (3) internal data structures required by query operators. This unified representation approach allows our system to maintain a consistent execution model while adapting to different data characteristics and query requirements.

While encoding approaches for data and mask columns are similar, there are a few differences in their representations as follows.
\begin{myitemize}
\item The domain of values for masks is restricted to \{$T, F$\}. 
\item Since there are no outliers in values, the composite Plain + Index encoding is not applicable for masks.
\item In position-explicit encodings (RLE, Index) for masks, only positions corresponding to $T$ values are maintained in the tensor representations. Plain encoding tracks both $T$ and $F$ values.
\item In position-explicit encodings for masks, the value tensor ($v$) is not needed, since all tracked values are implicitly equal to $T$.
\end{myitemize}
Depending on the operator type, the operands can be data columns, mask columns, scalar expressions, or literals. Result columns can be data or masks. We will use the terms DataColumn and MaskColumn where needed to distinguish between these two types.

\section{Primitives}
\label{sec:core-ops}

We develop a set of novel parallel algorithms to implement fundamental operations on encoded data types. These operations include conversion between encodings, detecting containment and overlap between position-explicit encodings, positional range intersection and union, compaction, etc. Table~\ref{tab:fundamental-ops} lists the set of non-trivial operations. These operations can be used as building blocks to implement more complex operations.
In Sections~\ref{sec:logical}--\ref{sec:join} we describe how we implement relational operators using these fundamental operations.

\vspace{-2ex}
\begin{table}[ht]
    \centering
    \caption{Fundamental Operations for Encoded data.\vspace{-10pt}}
    \label{tab:fundamental-ops} 
    \resizebox{\columnwidth}{!}{
    \begin{tabular}{|@{ }c|c@{ }|}\hline
         \textbf{Primitive}&  \textbf{Description}\\\hline\hline
         \texttt{range\_intersect}& Intersection of RLE runs\\\hline
         \texttt{idx\_in\_rle}& Intersection of Index and RLE data\\\hline
         \texttt{idx\_in\_idx}& Intersection of Index columns\\\hline
         \texttt{rle\_contain\_idx}& Index positions contained within an RLE run\\\hline
         \texttt{range\_union}& Union of RLE runs\\\hline
         \texttt{merge\_sorted\_idx}& Merge sorted Index tensors\\\hline         
         \texttt{compact\_rle}& Remove gaps between runs of RLE data\\\hline
         \texttt{compact\_rle+index}& Remove gaps in RLE + Index data\\\hline
         \texttt{complement\_rle}& Complement of RLE intervals\\\hline
         \texttt{complement\_index}& Complement of Index positions\\\hline         
         \texttt{rle\_to\_index}& Convert RLE data to Index\\\hline
         \texttt{rle\_to\_plain}& Convert RLE data to Plain\\\hline
         \texttt{plain\_to\_rle}& Convert Plain data to RLE\\\hline
         \texttt{plain\_to\_rle+index}& Convert Plain data to Composite (RLE+Index)\\\hline
         \texttt{plain\_to\_plain+index}& Convert Plain data to Composite (Plain+Index)\\\hline         
    \end{tabular}
    }
    \vspace{-10pt}
\end{table}

Due to space limitation, in this section we include pseudo-code only for a few primitives, notably the intersection primitives that are heavily used for implementing relational operators.
Functions in \textbf{bold font} are provided by PyTorch and have been optimized by PyTorch developers for different backends including GPUs. 
Note that there are no explicit for loops or conditional checks in the codes, which helps to maximize utilization of the GPU parallelism.

\subsection{Range Intersection}

This operation finds the intersection of two sorted lists of ranges. We use this operation frequently in computations involving multiple RLE columns. At the end of the intersect, all columns will have the same number of runs and same start and end positions (same $s$ and $e$ tensors for all columns) corresponding to position ranges that are common to all input columns. 
If a range is split, the value is duplicated for the new ranges. Our goal is to minimize the number of range splits (i.e., maximize run lengths) in the result.

Our approach is inspired by bioinformatics techniques for chromosome range intersections~\cite{layer2013binary}, as detailed in \Cref{alg:range_intersect}. The "range\_intersect" algorithm efficiently computes the intersection of two RLE inputs, $c_1$ and $c_2$, by first bucketizing the start ($c_1^{start}$) and end ($c_1^{end}$) points of the first input relative to the second (Lines 1-2). It then counts intersections (Line 3) and generates index tensors $idx_1$ and $idx_2$ for each intersecting range (Lines 4-6), using the helper function \texttt{range\_arange} (Algorithm~\ref{alg:range_arange}) for $idx_2$. Unlike PyTorch's \textbf{arange}, which operates on single start/length values, \texttt{range\_arange} accepts tensors of starts and lengths to generate multiple concatenated sequences. Finally, the algorithm determines the start and end points of the intersections (Line 7), producing the resulting tensors $s$ and $e$. To optimize performance, the input with the fewer ranges should always be used as $c_1$.
\begin{algorithm}
    \caption{range\_intersect}
    \label{alg:range_intersect}
    \raggedright
    \textbf{Input:} RLE columns $c_1$, $c_2$\\
    \textbf{Output:} Intersection start and end positions $s$, $e$
    \begin{algorithmic}[1]
        \small
        \State $bin_s$ $\gets$ \textbf{bucketize}($c_1^{start}$, $c_2^{end}$, \textit{right=False})
        \State $bin_e$ $\gets$ \textbf{bucketize}($c_1^{end}$, $c_2^{start}$, \textit{right=True})
        \State \textit{cnt} $\gets$ $bin_e$ - $bin_s$
        \State \textit{arange} $\gets$ \textbf{arange}(\textbf{len}(\textit{cnt}))
        \State $idx_1$ $\gets$ \textbf{repeat\_interleave}(\textit{arange}, \textit{cnt})
        \State $idx_2$ $\gets$ \texttt{range\_arange}($bin_s$, \textit{cnt})
        \State $s, e$ $\gets$ \textbf{max}($c_1^{start}$[$idx_1$], $c_2^{start}$[$idx_2$]), \textbf{min}($c_1^{end}$[$idx_1$], $c_2^{end}$[$idx_2$])
        \State \Return $s$, $e$
    \end{algorithmic}
\end{algorithm}
\vspace{-12pt}
\begin{algorithm}
    \caption{range\_arange (helper function)}
    \label{alg:range_arange}
    \raggedright
    \textbf{Input:}  \textit{start}, \textit{length}\\
    \textbf{Output:} Concatenated sequence \textit{result}
    \begin{algorithmic}[1]
        \small
        \State \textit{t} $\gets$ \textbf{cumsum}(\textit{length})
        \State \textit{t} $\gets$ \textbf{cat}(([0], \textit{t}[:-1]))\hspace{10pt}// prepends 0, includes all but last element
        \State \textit{total\_size} $\gets$ \textbf{sum}(\textit{length})
        \State \textit{result} $\gets$ \textbf{repeat\_interleave}(\textit{start}, \textit{length})\\
        \hspace{4em}+\textbf{arange}(\textit{total\_length})\\
        \hspace{4em}-\textbf{repeat\_interleave}(\textit{t}, \textit{length})
        \State \Return \textit{result}
    \end{algorithmic}
\end{algorithm}
\vspace{-3pt}

\begin{figure}
    \centering
    \includegraphics[width=\linewidth]{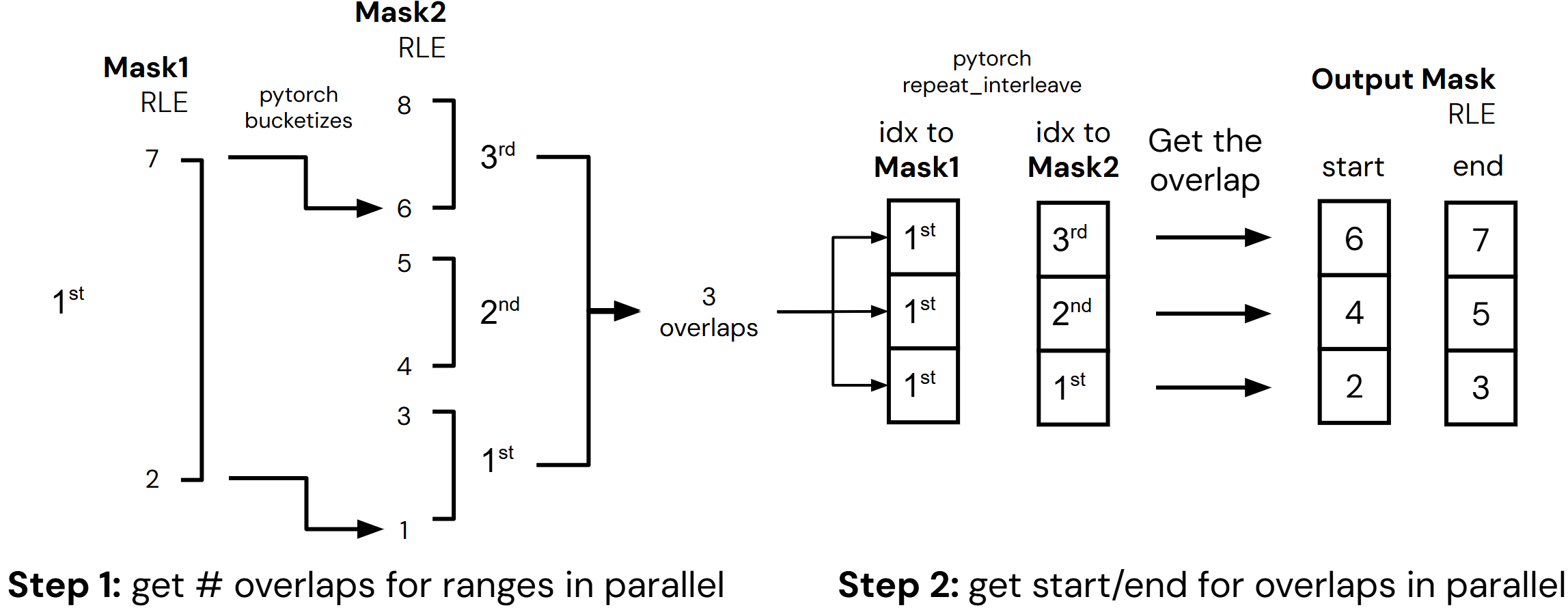}
    \vspace{-20pt}
    \caption{Illustration of range\_intersect algorithm for AND between RLE masks: bucketization identifies overlaps, then parallel PyTorch operations compute precise intersections.\vspace{-10pt}}
    \label{fig:rle_and}
    \vspace{-2ex}
\end{figure}
\vspace{-3pt}
\begin{example}
Consider the range intersection between the positional ranges of two RLE masks, $c_1$ and $c_2$, as shown in \Cref{fig:rle_and}.
Mask $c_1$ has a single range represented as $c_1^{start}=[2]$, $c_1^{end}=[7]$.
Mask $c_2$ has three ranges represented as $c_2^{start}=[1, 4, 6]$, $c_2^{end}=[3, 5, 8]$.
Recall that masks only represent ranges corresponding to a value of $T$.
\end{example}
\vspace{-3pt}
To compute their intersection using range\_intersect (\Cref{alg:range_intersect}):
\begin{myitemize}
    \item \textbf{Step 1: Bucketize start positions.} We bucketize the start position ($c_1^{start}$) of Mask $c_1$ relative to the end positions ($c_2^{end}$) of Mask $c_2$. $bin_s = \textbf{bucketize}([2], [3, 5, 8]) = [0]$. This means the start point 2 comes before the first end point 3.
    
    \item \textbf{Step 2: Bucketize end positions.} We bucketize the end positions ($c_1^{end}$) of Mask $c_1$ relative to the start positions ($c_2^{start}$) of Mask $c_2$. $bin_e = \textbf{bucketize}([7], [1, 4, 6], right=True) = [3]$. This means the end point $7$ comes after all start positions in $c_2$.
    
    \item \textbf{Step 3: Count overlaps.} We count the overlaps by subtracting $bin_s$ from $bin_e$. $cnt = bin_e - bin_s = [3] - [0] = [3]$. This indicates that the range in $c_1$ overlaps with all 3 ranges in $c_2$.
    
    \item \textbf{Step 4: Create index tensors.} We create index tensors that identify which ranges from each mask participate in each overlap. $idx_1 = \textbf{repeat\_interleave}([0], [3]) = [0, 0, 0]$ (indices into $c_1$) and $idx_2 = \textbf{range\_arange}([0], [3]) = [0, 1, 2]$ (indices into $c_2$).
    
    \item \textbf{Step 5: Compute intersection points.} We compute the actual intersection points by taking the maximum of start points and minimum of end points. $s = \textbf{max}(c_1^{start}[idx_1], c_2^{start}[idx_2]) = \textbf{max}([2, 2, 2], [1, 4, 6]) = [2, 4, 6]$ and $e = \textbf{min}(c_1^{end}[idx_1], \\c_2^{end}[idx_2]) = \textbf{min}([7, 7, 7], [3, 5, 8]) = [3, 5, 7]$.
\end{myitemize}

\noindent The output is an RLE mask with $s=[2, 4, 6]$ and $e=[3,5,7]$, representing the segments where both $c_1$ and $c_2$ have True values.

\subsection{Index Intersection}

These operations handle intersections for Index encoded data ($c^{pos}$). A common task is finding which positions ($c_{idx}^{pos}$) from an Index list fall within any RLE ranges ($c_{rle}^{start}, c_{rle}^{end}$). We provide two algorithms for this: \texttt{idx\_in\_rle} (Algorithm~\ref{alg:idx_in_rle}) and \texttt{rle\_contain\_idx} (Algorithm~\ref{alg:rle_contain_idx}). The main computational work involves the \texttt{bucketize} operation performed by each. \texttt{idx\_in\_rle} uses \texttt{bucketize}($c_{idx}^{pos}$, $c_{rle}^{start}$), while \texttt{rle\_contain\_idx} uses \texttt{bucketize}($c_{rle}^{start}$, $c_{idx}^{pos}$) and \texttt{bucketize}($c_{rle}^{end}$, $c_{idx}^{pos}$). The choice between them for optimal performance depends on the relative input sizes: \texttt{idx\_in\_rle} is generally preferred when $|c_{idx}^{pos}| \ll |c_{rle}^{start}|$. We also provide \texttt{idx\_in\_idx} (Algorithm~\ref{alg:idx_in_idx}) for intersecting two Index lists.
\vspace{-3pt}
\begin{example}[\texttt{idx\_in\_rle}]
Let us consider \texttt{idx\_in\_rle} with an Index column $c_{idx}$ and an RLE column $c_{rle}$.
Suppose $c_{idx}^{pos} = [2, 4, 7]$ and $c_{rle}^{start} = [0, 6]$, $c_{rle}^{end} = [2, 7]$.
We want to find which elements in $c_{idx}^{pos}$ fall within the RLE ranges [0-2] or [6-7] defined by $c_{rle}$.
\end{example}
\vspace{-3pt}
\begin{myitemize}
    \item \textbf{Step 1: Bucketize positions.} \texttt{bucketize}($c_{idx}^{pos}$, $c_{rle}^{start}$, \textit{right=True}) gives $[1, 1, 2]$. Subtracting 1 yields $bin = [0, 0, 1]$. This tells us which RLE range each position $p$ \textit{might} belong to (position 2 maps to range 0, 4 maps to range 0, 7 maps to range 1).
    \item \textbf{Step 2: Verify containment.} We check two conditions for each position $p$ in $c_{idx}^{pos}$: (1) \textit{bin} $\ge$ 0 ensures the position is not before the first start $s$ in $c_{rle}^{start}$, and (2) $p \le c_{rle}^{end}$[\textit{bin}] ensures the position is not past the end $e$ of its assigned RLE range.
        For position 2: \textit{bin}=0. $0 \ge 0$ ($T$) and $2 \le c_{rle}^{end}[0]=2$ ($T$). 2 is included.
        For position 4: \textit{bin}=0. $0 \ge 0$ ($T$) and $4 \le c_{rle}^{end}[0]=2$ ($F$). 4 is excluded.
        For position 7: \textit{bin}=1. $1 \ge 0$ ($T$) and $7 \le c_{rle}^{end}[1]=7$ ($T$). 7 is included.
    \item \textbf{Step 3: Apply mask.} The resulting mask is $[T, F, T]$. Applying this to $c_{idx}^{pos}$ gives the final result $p_{out}$ = [2, 7].\vspace{-10pt}
\end{myitemize}

\begin{minipage}[t]{0.55\columnwidth}
\begin{algorithm}[H]
    \caption{idx\_in\_rle}
    \label{alg:idx_in_rle}
    \raggedright
    \textbf{Input:} Index column $c_{idx}$, RLE $c_{rle}$\\
    \textbf{Output:} Result positions $p_{out}$
    \begin{algorithmic}[1]
        \small
        \State \textit{bin} $\gets$ \textbf{bucketize}($c_{idx}^{pos}$, $c_{rle}^{start}$, \textit{right=True})
        \State \textit{bin} $\gets$ \textit{bin} - 1
        \State \textit{mask} $\gets$ (\textit{bin} $\geq$ 0) $\land$ ($c_{idx}^{pos}$ $\leq$ $c_{rle}^{end}$[\textit{bin}])
        \State \Return $c_{idx}^{pos}$[\textit{mask}]
    \end{algorithmic}
\end{algorithm}
\end{minipage}\hfill
\begin{minipage}[t]{0.43\columnwidth}
\begin{algorithm}[H]
    \caption{idx\_in\_idx}
    \label{alg:idx_in_idx}
    \raggedright
    \textbf{Input:} Index columns $c_1$, $c_2$\\
    \textbf{Output:} Result positions $p_{out}$
    \begin{algorithmic}[1]
        \small
        \State \textit{bin} $\gets$ \textbf{bucketize}($c_1^{pos}$, $c_2^{pos}$, \textit{right=True}) - 1
        \State \textit{mask} $\gets$ (\textit{bin} $\geq$ 0) $\land$ ($c_1^{pos} = c_2^{pos}[bin]$)
        \State \Return $c_1^{pos}$[\textit{mask}]
    \end{algorithmic}
\end{algorithm}
\end{minipage}
\vspace{-5pt}
\begin{algorithm}
    \caption{rle\_contain\_idx}
    \label{alg:rle_contain_idx}
    \raggedright
    \textbf{Input:} Index column $c_{idx}$, RLE column $c_{rle}$\\
    \textbf{Output:} Result positions $p_{out}$
    \begin{algorithmic}[1]
        \small
        \State $bin_s$ $\gets$ \textbf{bucketize}($c_{rle}^{start}$, $c_{idx}^{pos}$)
        \State $bin_e$ $\gets$ \textbf{bucketize}($c_{rle}^{end}$, $c_{idx}^{pos}$, \textit{right=True}) - 1
        \State \textit{mask} $\gets$ ($bin_s \leq bin_e$)
        \State $bin_s, bin_e\gets$ $bin_s$[\textit{mask}], $bin_e$[\textit{mask}]
        \State \Return $c_{idx}^{pos}$[\textbf{rle\_to\_index}($bin_s$, $bin_e$)]
    \end{algorithmic}
\end{algorithm}
\vspace{-5pt}

\begin{example}[\texttt{rle\_contain\_idx}]
Let us illustrate \texttt{rle\_contain\_idx} using the same input data as the \texttt{idx\_in\_rle} example for direct comparison.
Suppose $c_{idx}^{pos} = [2, 4, 7]$ and the RLE column $c_{rle}$ has ranges defined by $c_{rle}^{start} = [0, 6]$ and $c_{rle}^{end} = [2, 7]$. The ranges are [0-2] and [6-7].
We want to find which positions in $c_{idx}^{pos}$ are contained within any of these RLE ranges.
\end{example}
\vspace{-5pt}
\begin{myitemize}
    \item \textbf{Step 1: Bucketize RLE starts.} We bucketize each RLE start position ($c_{rle}^{start}$) against the sorted index positions ($c_{idx}^{pos}$).
    $bin_s = \textbf{bucketize}(c_{rle}^{start}, c_{idx}^{pos}) = \textbf{bucketize}([0, 6], [2, 4, 7]) = [0, 2]$.
    (0 is before index 0 value '2'; 6 is before index 2 value '7').
    \item \textbf{Step 2: Bucketize RLE ends.} We bucketize each RLE end position ($c_{rle}^{end}$) against $c_{idx}^{pos}$, using $right=True$ and subtracting 1.
    $bin_e = \textbf{bucketize}(c_{rle}^{end}, c_{idx}^{pos}$, right=True) - 1 = \textbf{bucketize}([2, 7], [2, 4, 7], right=True) - 1 = [1, 3] - 1 = [0, 2].
    (2 is $\le$ index 0 value 2, giving index 1; 7 is $\le$ index 2 value 7, giving index 3. Then subtract 1).
    \item \textbf{Step 3: Create mask.} We create a mask where $bin_s \le bin_e$.
    $mask = (bin_s \le bin_e) = ([0, 2] \le [0, 2]) = [True, True]$.
    \item \textbf{Step 4: Apply mask.} We apply $mask$ to $bin_s$ and $bin_e$. They remain unchanged: $bin_s = [0, 2]$, $bin_e = [0, 2]$.
    \item \textbf{Step 5: Convert RLE indices to position indices.} We use \texttt{rle\_to\_index} to generate the flat list of indices into $c_{idx}^{pos}$ that correspond to the ranges defined by the masked $bin_s$ and $bin_e$.
    For range 1 ($s=0, e=0$): indices [0].
    For range 2 ($s=2, e=2$): indices [2].
    Combined indices: $[0, 2]$. Let this result be $idx_{flat}$.
    \item \textbf{Step 6: Gather positions.} We select the positions from $c_{idx}^{pos}$ using $idx_{flat}$:
    $p_{out} = c_{idx}^{pos}[idx_{flat}] = c_{idx}^{pos}[[0, 2]] = [2, 7]$.
\end{myitemize}
The final output is $p_{out} = [2, 7]$, which are exactly the positions from $c_{idx}^{pos}$ that fall within the RLE ranges [0-2] or [6-7]. This matches the output of the \texttt{idx\_in\_rle} example for the same input.

Finally, the \texttt{idx\_in\_idx} algorithm (Algorithm~\ref{alg:idx_in_idx}) similarly to \texttt{idx\_in\_rle} uses \texttt{bucketize} to find potential matches between two sorted position lists ($c_1^{pos}, c_2^{pos}$) and then verifying exact equality.\vspace{-5pt}

\subsection{Primitive Microbenchmarks}

\begin{figure}[ht]
    \centering
    \subfloat[\texttt{range\_intersect}\vspace{-10pt}]{\includegraphics[width=0.45\columnwidth]{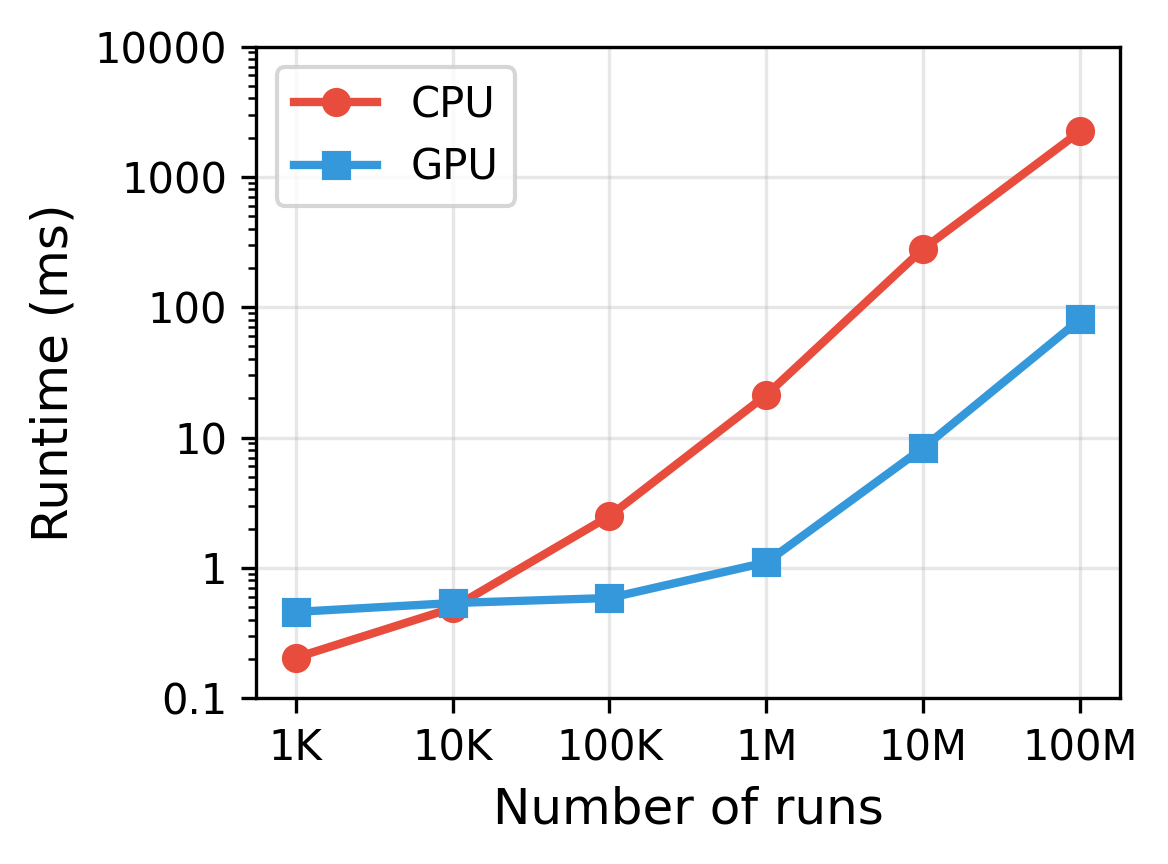}
    \label{fig:prim_range_intersect}\vspace{-15pt}}
    \subfloat[\texttt{idx\_in\_rle}\vspace{-10pt}]{\includegraphics[width=0.45\columnwidth]{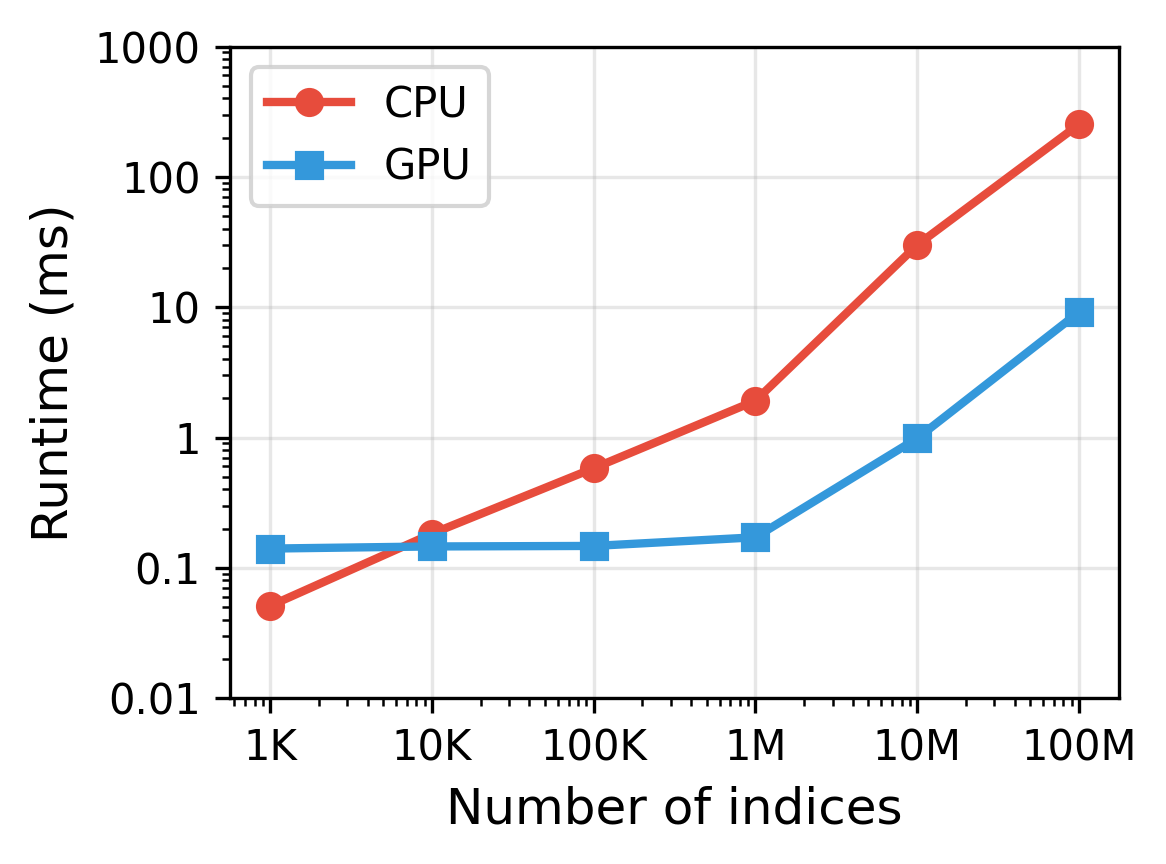}\label{fig:prim_idx_in_rle}\vspace{-15pt}}\\
    \subfloat[\texttt{idx\_in\_idx}\vspace{-10pt}]{\includegraphics[width=0.45\columnwidth]{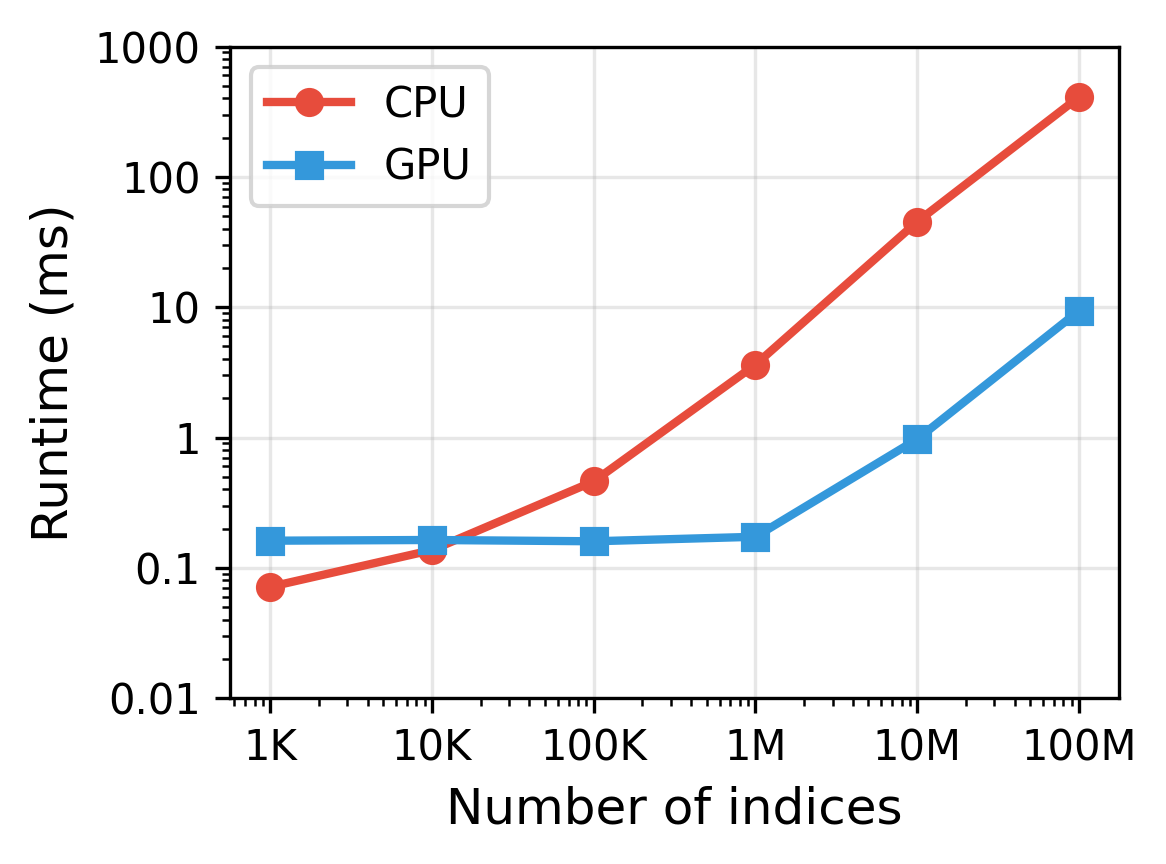}\label{fig:prim_idx_in_idx}\vspace{-15pt}}
    \subfloat[\texttt{rle\_contain\_idx}\vspace{-10pt}]{\includegraphics[width=0.45\columnwidth]{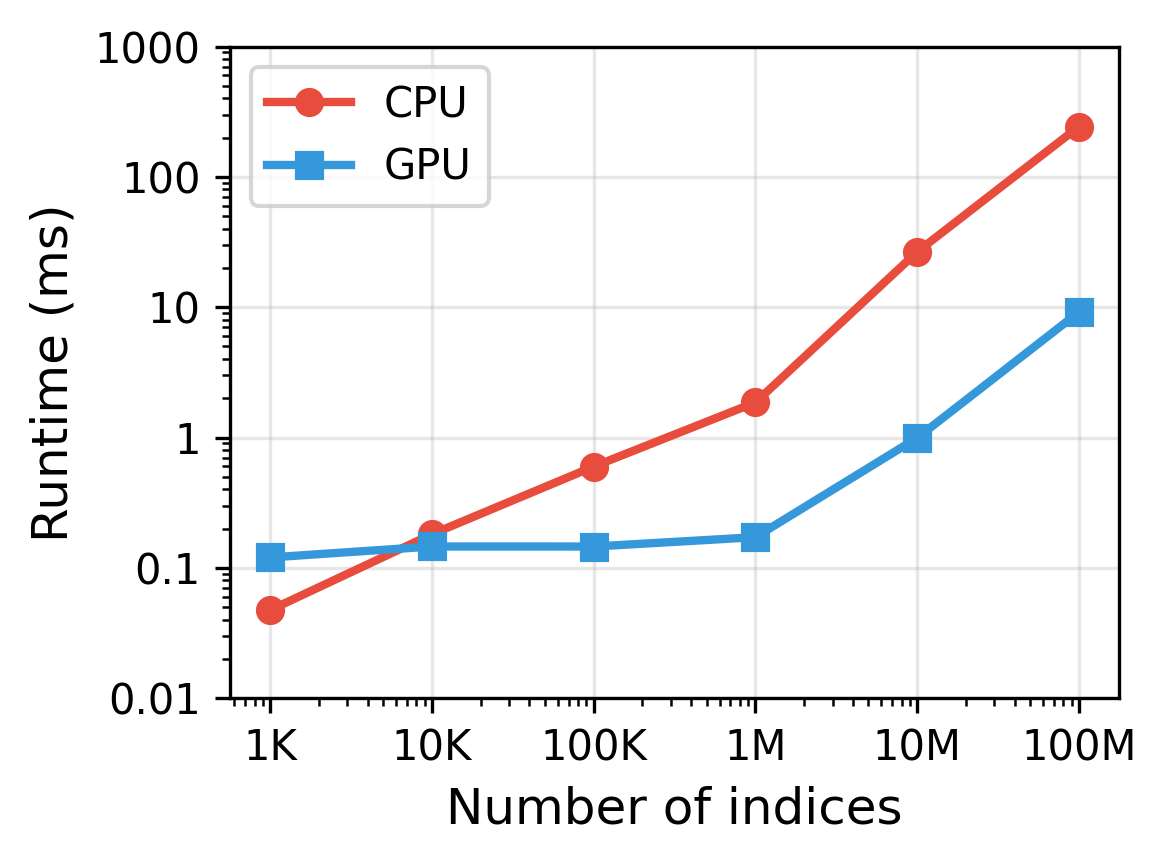}\label{fig:prim_rle_contain_idx}\vspace{-15pt}}
    \caption{CPU vs GPU performance for RLE primitives. GPU achieves $21$--$46\times$ speedup at scale (100M elements) with crossover points around 10K-100K elements.\vspace{-10pt}}
    \label{fig:primitives_comparison}
\end{figure}

We ran microbenchmarks comparing CPU and GPU implementations of our 4 fundamental primitives across input sizes from 1K to 100M elements on an Azure NC24ads A100 v4 VM~\cite{Azure-NCA100}. Both implementations use PyTorch, with CPU execution via \texttt{torch.device(`cpu')} on an AMD EPYC 7V13 (24 cores, 2.445 GHz) and GPU execution on an NVIDIA A100. \Cref{fig:primitives_comparison} shows that our GPU primitives achieve $21$--$46\times$ speedups over CPU implementations at large scales ($\geq 1$M elements). While CPU outperforms GPU by $1.5$--$2.3\times$ at small scales (1K elements) due to kernel launch overhead, GPU becomes advantageous at crossover points around 10K--100K elements, validating our design choice to optimize for large tensor query execution.
\vspace{-4ex}

\section{Logical Operations}
\label{sec:logical}

Logical operators (AND, OR, NOT) operate on \textit{MaskColumn}s as input operands and produce a \textit{MaskColumn} as result.

\begin{table*}
\vspace{-3ex}
\footnotesize
\centering
\caption{Implementation Design of Physical Operator for AND between \textit{Column}s (\S~\ref{sec:logical_and}), Alignment (\S~\ref{sec:arith-comp-select}) and Apply Join Index (\S~\ref{sec:apply_join_index}). \textcolor{lightred}{Unsorted RLE and Unsorted Index} are exclusively for Apply Join Index. For explicit-positional encoding (RLE or Index), the table's left section applies when $m_1$ and $m_2$ are sorted, with only the upper triangular matrix shown due to symmetry. Note: Outputs from \texttt{rle\_to\_index} (Index) and \texttt{rle\_to\_plain} (Plain) serve as input for subsequent lookups in this table.\vspace{-10pt}}
\label{tab:and_design}
\begin{tabular}{@{}l@{}@{}c@{}@{}c@{}@{}c@{}| c c@{}}
\toprule
& \textbf{\textcolor{black}{$m_1$: RLE}} & \textbf{\textcolor{black}{$m_1$: Plain}} & \textbf{\textcolor{black}{$m_1$: Index}} & \textbf{\textcolor{lightred}{$m_1$: Unsorted RLE}} & \textbf{\textcolor{lightred}{$m_1$: Unsorted Index}} \\
\midrule
        \textbf{\textcolor{lightblue}{$m_2$: RLE}} &  
        \begin{tabular}{c} $\left\{ 
        \begin{array}{c} 
        \text{\texttt{range\_intersect}(\textcolor{black}{$m_1$}, \textcolor{lightblue}{$m_2$})}
        \\ 
        \text{\texttt{range\_intersect}(\textcolor{lightblue}{$m_2$}, \textcolor{black}{$m_1$})}
        \end{array} 
        \right.$ \end{tabular} &
        \begin{tabular}{c} $\left\{ 
        \begin{array}{c} 
        \text{\texttt{rle\_to\_index}(\textcolor{lightblue}{$m_2$})}
        \\ 
        \text{\texttt{rle\_to\_plain}(\textcolor{lightblue}{$m_2$})}
        \end{array} 
        \right.$ \end{tabular}
        &
        \begin{tabular}{c} $\left\{ 
        \begin{array}{c} 
        \text{\texttt{idx\_in\_rle}(\textcolor{black}{$m_1$}, \textcolor{lightblue}{$m_2$})}
        \\ 
        \text{\texttt{rle\_contain\_idx}(\textcolor{black}{$m_1$}, \textcolor{lightblue}{$m_2$})}
        \end{array} 
        \right.$ \end{tabular}
        & \texttt{range\_intersect}(\textcolor{lightred}{$m_1$}, \textcolor{lightblue}{$m_2$}) 
        & \texttt{idx\_in\_rle}(\textcolor{lightred}{$m_1$}, \textcolor{lightblue}{$m_2$})
        \\
        \textbf{\textcolor{lightblue}{$m_2$: Plain}} &
        & \textcolor{black}{$m_1$} \& \textcolor{lightblue}{$m_2$}
        & \textcolor{black}{$m_1$}[\textcolor{lightblue}{$m_2$}[\textcolor{black}{$m_1$}]]
        & \texttt{rle\_to\_index}(\textcolor{lightred}{$m_1$})
        & \textcolor{lightred}{$m_1$}[\textcolor{lightblue}{$m_2$}[\textcolor{lightred}{$m_1$}]]
        \\
        \textbf{\textcolor{lightblue}{$m_2$: Index}} & & & 
        \begin{tabular}{c} $\left\{ 
        \begin{array}{c} 
        \text{\texttt{idx\_in\_idx}(\textcolor{black}{$m_1$}, \textcolor{lightblue}{$m_2$})}
        \\ 
        \text{\texttt{idx\_in\_idx}(\textcolor{lightblue}{$m_2$}, \textcolor{black}{$m_1$})}
        \end{array} 
        \right.$ \end{tabular}
         &
         \texttt{rle\_contain\_idx}(\textcolor{lightblue}{$m_2$}, \textcolor{lightred}{$m_1$})
         & \texttt{idx\_in\_idx}(\textcolor{lightred}{$m_1$}, \textcolor{lightblue}{$m_2$}) \\
        \bottomrule
\end{tabular}
\end{table*}
\begin{table}
\vspace{-2ex}
    \centering
    \footnotesize
    \caption{Output \textit{MaskColumn} encoding for AN. The table is symmetric and we only show the upper triangular matrix.\vspace{-10pt}}
    \label{tab:and_output}
    \resizebox{0.8\columnwidth}{!}{
    \begin{tabular}{l c c c}
        \toprule
         & \textbf{$m_1$: RLE} & \textbf{$m_1$: Plain} & \textbf{$m_1$: Index} \\
        \midrule
        \textbf{\textcolor{lightblue}{$m_2$: RLE}} & RLE& Plain/Index & Index \\
        \textbf{\textcolor{lightblue}{$m_2$: Plain}} & & Plain& Index \\
        \textbf{\textcolor{lightblue}{$m_2$: Index}} & & & Index \\
        \bottomrule
    \end{tabular}
    }\vspace{-8pt}
\end{table}

\subsection{AND}
\label{sec:logical_and}

We summarize the design of this operator in \Cref{tab:and_design}.
\begin{myitemize}
    \item \textbf{Plain mask AND Plain mask}: Use PyTorch's logical AND operator (\&). For example, $[T, F, T, F]$ AND $[T, T, F, F]$ yields $[T, F, F, F]$.
    
    \item \textbf{RLE mask AND RLE mask}: We implement this using the \texttt{range\_intersect} operation (\Cref{alg:range_intersect}).

    \item \textbf{RLE mask AND Plain mask}: Direct AND is inefficient due to searching each `True' value in Plain mask. Instead, convert RLE mask to Index (\texttt{rle\_to\_index}) or Plain (\texttt{rle\_to\_plain}) then apply AND following \Cref{tab:and_design}. The choice depends on the selectivity threshold (total elements / selected elements). If this ratio exceeds the threshold, convert to an Index mask; otherwise, convert to a Plain mask. By default, we choose 20, which was determined through offline profiling on our GPU system. This threshold balances RLE compression benefits against conversion costs—providing good performance when RLE is beneficial while keeping conversion overhead acceptable.

    \item \textbf{RLE mask AND Index mask}: Check if Index values fall within RLE ranges using PyTorch's \texttt{bucketize}. Use \texttt{idx\_in\_rle} or \texttt{rle\_contain\_idx} depending on relative mask sizes.

    \item \textbf{Plain mask AND Index mask}: Select Index values where Plain positions are True using PyTorch's subscript operator ($[]$).

    \item \textbf{Index mask AND Index mask}: Check if values in one Index tensor appear in the other using \texttt{bucketize} on the larger tensor.
\end{myitemize}

The output encoding for the AND operation appears in \Cref{tab:and_output}. Intuitively, the characteristics of the input layouts dictate the resulting output layout. For instance, if both inputs use RLE, their masks are typically continuous, so the output is also continuous and best encoded as RLE. However, if one input is RLE (ranges) and the other is Index (points), the resulting mask is likely discontinuous, making the Index representation more appropriate.

\noindent \textbf{Alternative Design.} For RLE mask AND Plain mask, our current design only converts RLE to Index/Plain representations, because it prioritizes simplicity and predictable performance. An alternative design could convert Plain to RLE instead. For instance, when a Plain mask undergoes multiple AND operations, it may become highly selective and potentially benefit from RLE compression. However, such conversion could introduce significant overhead.

\begin{figure}[ht]
    \centering
    \includegraphics[width=0.8\linewidth]{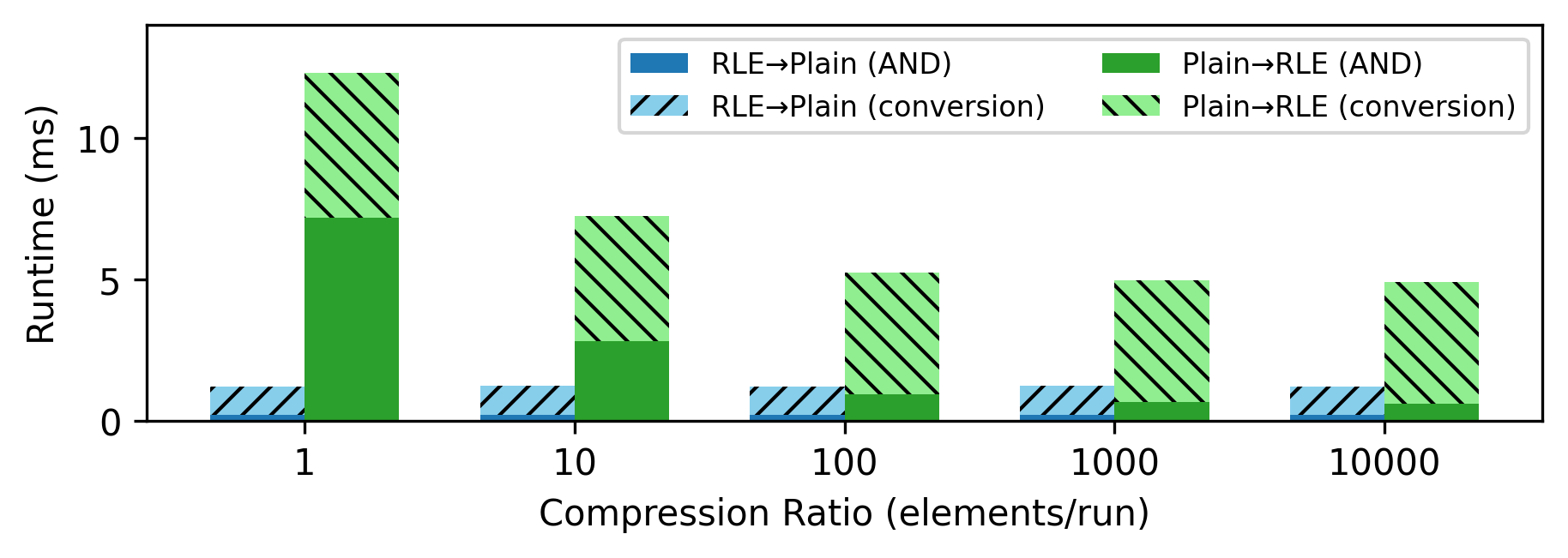}
    \vspace{-10pt}
    \caption{Performance comparison of alternative AND designs across Plain compression ratios. Alternative Design Plain $\rightarrow$ RLE has faster AND operation when Plain is highly compressible. However, the large overhead of Plain to RLE conversion negates the gain.}
    \label{fig:representation_choice}
    \vspace{-10pt}
\end{figure}

To illustrate, we conducted a microbenchmark for AND between RLE mask and Plain mask, comparing RLE$\rightarrow$Plain versus Plain$\rightarrow$RLE conversion strategies. The setup uses 100M elements with a fixed highly-compressed RLE mask and Plain masks with varying compression ratios (1 to 10K). \Cref{fig:representation_choice} shows RLE$\rightarrow$Plain is consistently $4.0\times$ to $10.2\times$ faster. Although the alternative design achieves faster AND operations with highly compressed data, conversion cost dominates: Plain$\rightarrow$RLE conversion incurs 4.29ms overhead versus only 1.02ms for RLE$\rightarrow$Plain conversion. This large conversion overhead negates the AND operation gains.

However, we acknowledge that, while for a single operator, converting Plain to RLE is too costly, the RLE representation may benefit downstream operations. To decide optimally though is challenging and potentially requires runtime statistics. We recognize there are interesting query optimization opportunities in this space, but leave these encoding decisions as future work.

\vspace{-1ex}

\subsection{OR}
\label{sec:logical_or}

We outline the design of this operator in \Cref{tab:or_design}, and show the encoding of the result in \Cref{tab:or_output}.\vspace{-5pt}

\begin{table}[ht]

\centering
\caption{Implementation design of Physical Operator for OR between \textit{MaskColumn}s $m_1$ and \textcolor{lightblue}{$m_2$}. Since OR is commutative, the table is symmetric and we only show the upper triangular matrix for better readability.\vspace{-10pt}}
\label{tab:or_design}
\resizebox{\columnwidth}{!}{
\begin{tabular}{@{}l@{}@{}c@{}@{}c@{}@{}c@{}}
\toprule
& \textbf{$m_1$: RLE} & \textbf{$m_1$: Plain} & \textbf{$m_1$: Index} \\
\midrule
        \textbf{\textcolor{lightblue}{$m_2$: RLE}} &   \text{\texttt{range\_union}(\textcolor{black}{$m_1$}, \textcolor{lightblue}{$m_2$})}  &
        \begin{tabular}{c} $\left\{ 
        \begin{array}{c} 
        \text{\texttt{rle\_to\_index}(\textcolor{lightblue}{$m_2$})}
        \\ 
        \text{\texttt{rle\_to\_plain}(\textcolor{lightblue}{$m_2$})}
        \end{array} 
        \right.$ \end{tabular}
        &
        \begin{tabular}{c} $\left\{ 
        \begin{array}{c} 
        \text{\texttt{idx\_in\_rle}(\textcolor{black}{$m_1$}, \textcolor{lightblue}{$m_2$})}
        \\ 
        \text{\texttt{rle\_contain\_idx}(\textcolor{black}{$m_1$}, \textcolor{lightblue}{$m_2$})}
        \end{array} 
        \right.$ \end{tabular}
        \\
       \textbf{\textcolor{lightblue}{$m_2$: Plain}} & & \textcolor{black}{$m_1$} | \textcolor{lightblue}{$m_2$} & \textcolor{lightblue}{$m_2$}[\textcolor{black}{$m_1$}] = $T$ \\
       \textbf{\textcolor{lightblue}{$m_2$: Index}} & & & 
       \begin{tabular}{c} $\left\{ 
        \begin{array}{c} 
        \text{\texttt{merge\_sorted\_idx}($m_1$, \textcolor{lightblue}{$m_2$}) }
        \\ 
        \text{\texttt{merge\_sorted\_idx}(\textcolor{lightblue}{$m_2$}, $m_1$) }
        \\
        \text{\texttt{concat\_sort}($m_1$, \textcolor{lightblue}{$m_2$}) }
        \end{array} 
        \right.$ \end{tabular}
        \\
        \bottomrule
\end{tabular}
}\vspace{-10pt}
\end{table}

\begin{table}[ht]
\vspace{-2ex}
    \centering
    \footnotesize
    \caption{Output \textit{MaskColumn} encoding for OR Operation. The table is symmetric and we only show the upper triangular matrix for better readability.\vspace{-10pt}}
    \label{tab:or_output}
    \resizebox{0.8\columnwidth}{!}{
    \begin{tabular}{l c c c}
        \toprule
         & \textbf{$m_1$: RLE} & \textbf{$m_1$: Plain} & \textbf{$m_1$: Index} \\
        \midrule
        \textbf{\textcolor{lightblue}{$m_2$: RLE}} & RLE & Plain & RLE + Index \\
        \textbf{\textcolor{lightblue}{$m_2$: Plain}} & & Plain& Plain \\
        \textbf{\textcolor{lightblue}{$m_2$: Index}} & & & Index \\
        \bottomrule
    \end{tabular}
    }\vspace{-10pt}
\end{table}

\begin{myitemize}
    \item \textbf{Plain mask OR Plain mask}: Use PyTorch's logical OR operator ($|$). For example, $[T, F, T, F]$ OR $[T, T, F, F]$ yields $[T, T, T, F]$.
    
    \item \textbf{RLE mask OR RLE mask}: Compute union of two sorted range lists by identifying start/end points of consecutive True values and determining non-overlapping segments.

    \item \textbf{RLE mask OR Plain mask}: Convert the RLE mask to Index (\texttt{rle\_to\_index}) or Plain (\texttt{rle\_to\_plain}) then apply OR operation. Similar to AND, the choice depends on the selectivity threshold with default value 20.

    \item \textbf{RLE Mask OR Index Mask}: Bucketize based on relative mask sizes using \texttt{idx\_in\_rle} or \texttt{rle\_contain\_idx}.

    \item \textbf{Plain Mask OR Index Mask}: Use PyTorch's subscript ([]) operator.

   \item \textbf{Index Mask OR Index Mask}: Merge sorted Index tensors using \texttt{merge\_sorted\_idx}. Bucketize the larger tensor for efficiency, track element origins with flags, then merge using conditional selection. For small tensors, concatenate and sort.
\end{myitemize}

\subsection{NOT}
\label{sec:logical_not}

The NOT operator takes a single \textit{MaskColumn} as input and produces an output \textit{MaskColumn}.
\begin{myitemize}
    \item \textbf{NOT Plain mask}: Use PyTorch's complement ($\sim$) operator for bitwise negation.

    \item \textbf{NOT RLE mask}: Use \texttt{complement\_rle} primitive to compute gaps between consecutive runs. Each gap starts at previous run end + 1 and ends at next run start - 1. Requires metadata to track total column rows.

    \item \textbf{NOT Index mask}: Use \texttt{complement\_index} primitive to compute gaps between indices, tracking column size. Output is in RLE format (not Index) because Index masks are sparse, making NOT results continuous and RLE-suited.
\end{myitemize}

\noindent Implementation details and examples are provided in \Cref{sec:not-operator-details}.

\subsection{Operating on Composite \textit{MaskColumn}s}
\label{sec:logical_op_composite_mask}

The previous sections discussed logical operations for non-composite \textit{MaskColumn}s. We now extend these operations to Composite masks (Plain + Index, or RLE + Index). Rather than implementing specialized operators for each composite configuration, we leverage a key insight: Composite \textit{MaskColumn}s can be conceptualized as the disjunction (OR operation) of their constituent mask tensors. This observation enables us to decompose logical operations on Composite masks into sequences of operations on their individual components, thereby reusing our existing non-composite operators. We apply Boolean algebra identities—specifically De Morgan's Laws~\cite{demorgan1847formal}, along with Associative and Distributive properties—to systematically transform these operations. Let $m^{rle}$ and $m^{idx}$ denote the RLE and Index components of Composite mask $m$, respectively.
\begin{myitemize}

\item \textbf{NOT}: $\neg (m_1^{rle} \vee m_1^{idx}) = (\neg m_1^{rle}) \wedge (\neg m_1^{idx})$

\item \textbf{OR}: $(\textcolor{lightblue}{m_1^{rle}} \vee \textcolor{lightblue}{m_1^{idx}}) \vee (\textcolor{lightred}{m_2^{rle}} \vee \textcolor{lightred}{m_2^{idx}}) = (\textcolor{lightblue}{m_1^{rle}}  \vee \textcolor{lightred}{m_2^{rle}}) \vee (\textcolor{lightblue}{m_1^{idx}} \vee \textcolor{lightred}{m_2^{idx}})$

\item \textbf{AND}:$(\textcolor{lightblue}{m_1^{rle}} \vee \textcolor{lightblue}{m_1^{idx}}) \wedge (\textcolor{lightred}{m_2^{rle}} \vee \textcolor{lightred}{m_2^{idx}})$\\
$= (\textcolor{lightblue}{m_1^{rle}} \wedge \textcolor{lightred}{m_2^{rle}}) \vee (\textcolor{lightblue}{m_1^{rle}} \wedge \textcolor{lightred}{m_2^{idx}}) 
\vee (\textcolor{lightblue}{m_1^{idx}} \wedge \textcolor{lightred}{m_2^{rle}}) \vee (\textcolor{lightblue}{m_1^{idx}} \wedge \textcolor{lightred}{m_2^{idx}})$

\end{myitemize}
Note that the sub-operations can all be performed in parallel using multiple CUDA streams . The encoding of the result \textit{MaskColumn} is RLE for NOT operator, and Composite for OR and AND operators.

\section{Arithmetic, Comparison, and Selection Operations}
\label{sec:arith-comp-select}

We focus on operators that perform point-wise/position-wise operations on values from two input columns, for positions that are common to both columns. This includes the logical AND operator (Section~\ref{sec:logical_and}), binary arithmetic operators, comparison operators, and selection operators with filter predicates. Joins need additional handling (Section~\ref{sec:join}). The logical OR operator (Section~\ref{sec:logical_or}) does not fit this category in the general case since the result covers a union of positions from the input columns which can be larger than the set of positions for any one column in the presence of gaps. 

The main challenge in performing these operations on compressed inputs is that, unlike Plain encoding, there is no alignment between positional representations of the inputs. Thus, point-wise operations are not directly possible. We need to first align the positional representations of the inputs, including the value tensors as needed, then perform the operations on the aligned segments. We call this \textit{Alignment}.

\begin{example}
Consider $c_1+c_2$ on input columns $c_1$ and $c_2$ both in RLE format. $c_1^{val}=[4,1,3], c_1^{start}=[0,10,20], c_1^{end} = [9,19,39]$; $c_2^{val}=[6,8], c_2^{start}=[0,15], c_2^{end}=[14,39]$. Thus, $c_1$ has 3 runs (lengths 10, 10, 20) and $c_2$ has 2 runs (lengths 15, 25). Similar examples apply for $-, *, /, <, >, =$ operators.
\end{example}
Due to run misalignment, we cannot directly do $c_1^{val}+c_2^{val}$. Misalignment can happen even with the same number of runs if run lengths differ.
We first align the runs of the input columns, similar to the AND operator (Section~\ref{sec:logical_and}) with an additional step of reconstructing the value tensors. After alignment, we get two RLE columns $r_1$ and $r_2$ with $r_1^{start} = r_2^{start} = [0,10,15,20]$, $r_1^{end} = r_2^{end} = [9,14,19,39]$, $r_1^{val}=[4,1,1,3]$, $r_2^{val}=[6,6,8,8]$. With identical position tensors, we can do point-wise addition: $r_1^{val}+r_2^{val}=[10,7,9,11]$.
Operations with scalar operands (e.g., $c*2, 1-c, c\geq 3$) are simple—no alignment needed, just operate on value tensors for RLE and Index encodings.

For selection operations (e.g., \textit{SELECT C FROM D WHERE P}), we compute \textit{MaskColumn} $m$ from predicate \textit{P}, align $m$ and \textit{C}, then apply $m$ to the tensor representations of $C$ if needed. For RLE and Index encodings, alignment performs selection (no final application needed). For Plain encodings, final mask application is required.
\section{Group-By-Aggregation Operations}
\label{sec:groupby}

For aggregation queries, the process is decomposed into two phases:
\begin{enumerate}[leftmargin=*]
    \item \textbf{Grouping}: Partitioning rows according to the group-by columns.
    \item \textbf{Aggregating}: Computing summary statistics (e.g., SUM, COUNT, AVG, MIN, MAX) for each identified group.
\end{enumerate}
Both phases are conceptually straightforward in PyTorch---function \texttt{torch.unique} can handle grouping by building inverse indices based on unique values in the group-by columns, and function \texttt{torch.scatter} can handle aggregation based on these inverse indices. However, the challenge lies in managing heterogeneous compression schemes across \textit{DataColumn}s. For example, consider a query that groups by two columns, one RLE-compressed and one Plain. The \texttt{torch.unique} function will not work because it requires aligned Plain \textit{DataColumn}s, but the RLE column values are not aligned with the Plain column. This heterogeneity prevents direct application of \texttt{torch.unique} and \texttt{torch.scatter} due to misalignment. We solve this by applying our Alignment technique (Section~\ref{sec:arith-comp-select}) to both phases.

\subsection{Grouping Phase}

The grouping phase constructs an inverse index that maps each row to its corresponding group; rows in the same group share the same data values across the group-by columns. It takes a set of aligned \textit{DataColumn}s for the group-by as input and returns an inverse index. The inverse index is a single tensor containing values from the range 0 to N-1, where N is the number of unique values across the group-by columns. The inverse index length is the same as the number of values in the group-by columns---for Plain data, it is the number of rows; for RLE and Index data, it is the number of RLE runs or index points.

\subsection{Aggregating Phase}

The aggregation operator takes a set of aligned \textit{DataColumn}s (for aggregation), the inverse index (from the grouping phase), and an aggregation function as input, and returns a set of tensors (one for each column to be aggregated) containing the aggregated results. The length of these tensors is the same as the number of unique values across the group-by columns.

To implement aggregation, we apply the \texttt{torch.scatter} function to the data columns, scattering based on the inverse indices. This approach efficiently computes aggregates like SUM, COUNT, AVG, STD, VAR, MIN, and MAX for each group. Note that if the data is RLE-compressed, each value must be repeated based on its corresponding RLE range size. We need to account for this when scattering the data. This depends on the aggregation function. A detailed walkthrough is provided in \Cref{sec:groupby-example}.
\begin{myitemize} 
    \item \textbf{MIN, MAX}: These functions are unaffected by compression and we only need to consider the value tensor: $min(v)$, $max(v)$. 
    \item \textbf{SUM, COUNT}: For RLE columns, instead of expanding the data and counting/summing individual values, we compute the run lengths: $l=e-s+1$. This is sufficient for COUNT. For SUM, we multiply the run lengths by the values ($v*l$). 
    \item \textbf{AVG, STD, VAR}: These functions can be computed as a post-processing step after SUM and COUNT. For example, to compute AVG, we divide SUM by COUNT. To compute STD and VAR, we first compute the sum of the squared values, then use SUM and COUNT to calculate STD and VAR. 
\end{myitemize}

\section{Join Operations}
\label{sec:join}

We reuse TQP's GPU-based hash join~\cite{surakav_he_2022}, used to join two Plain columns (from two tables) by building a hash table on one column (usually the smaller) and probing it with the other, but extend it to handle compressed columns (RLE and Index-encoded) as well. The join involves the following two steps.
\begin{myenumerate}
    \item \textbf{Get Join Index}: Two \textit{Join Index} tensors are computed---one for each input column---that reference the rows in each column which match according to a join predicate. These \textit{Join Index} tensors may be unsorted and contain duplicates (in case of one-to-many and many-to-many joins). 
    \item \textbf{Apply Join Index}: The \textit{Join Index} tensors are applied to the columns participating in the join, to create the join result.
\end{myenumerate}

\begin{figure}
    \centering
    \includegraphics[width=0.9\linewidth]{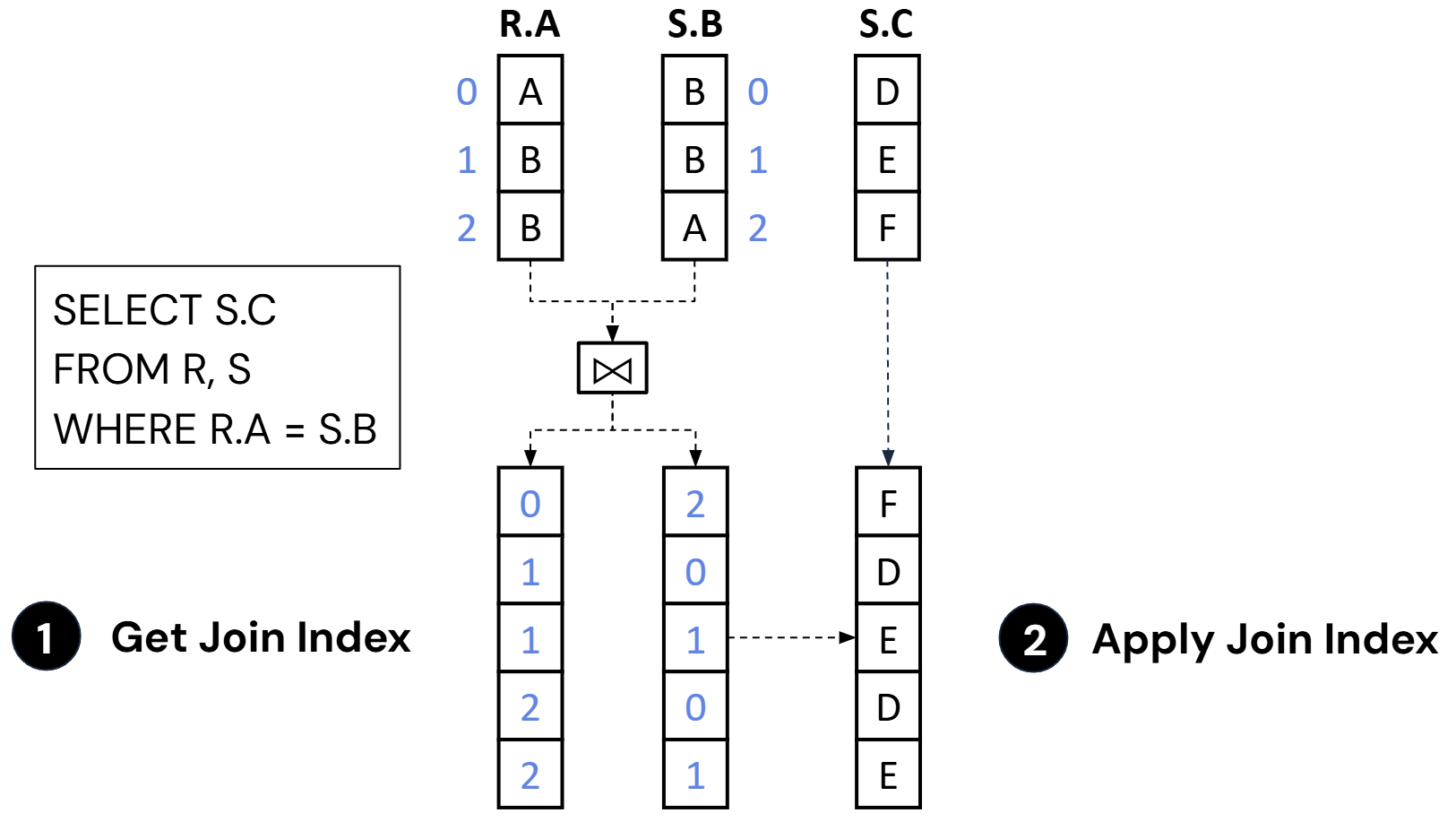}
    \vspace{-10pt}
    \caption{Illustration of join operation execution for the query ``SELECT S.C FROM R, S WHERE R.A = S.B''. 
    \vspace{-15pt}
    }
    \label{fig:join_operation}
\end{figure}
\begin{example}
Consider the join operation illustrated in \Cref{fig:join_operation} for the query ``SELECT S.C FROM R, S WHERE R.A = S.B''.
We have two tables:
Table R with column A containing values [A, B, B] at positions 0, 1, and 2.
Table S with column B containing values [B, B, A] and column C containing [D, E, F] at positions 0, 1, and 2.
\end{example} 
The join operation executes as follows:
\begin{myitemize}
    \item \textbf{Step 1: Get Join Index.} We identify all pairs of positions where R.A equals S.B. R.A[0] = 'A' matches S.B[2] = 'A', creating the pair (0, 2). R.A[1] = 'B' matches both S.B[0] = 'B' and S.B[1] = 'B', creating pairs (1, 0) and (1, 1). Similarly, R.A[2] = 'B' also matches both S.B[0] and S.B[1], creating pairs (2, 0) and (2, 1). The resulting Join Index tensors are: Left index (for R): [0, 1, 1, 2, 2] (positions in R.A) and Right index (for S): [2, 0, 1, 0, 1] (positions in S.B).
    
    \item \textbf{Step 2: Apply Join Index.} We use the right index tensor to retrieve the corresponding values from S.C. This gives us S.C[2] = 'F' from the first match, S.C[0] = 'D' and S.C[1] = 'E' from the second and third matches (with R.A[1]), and S.C[0] = 'D' and S.C[1] = 'E' again from the fourth and fifth matches (with R.A[2]).
\end{myitemize}
The final result of the join is the column [F, D, E, D, E], which contains all values from S.C that correspond to matching rows between R.A and S.B. Note that because joins can create one-to-many relationships, the same value may appear multiple times in the result, and the output size may be larger than either input table.

In the remainder of this section, we discuss how we extend the two steps to join columns encoded for compression, \textit{without first decompressing them}.

\subsection{Get Join Index}
\label{sec:join_index}
Given two \textit{DataColumn}s, our goal is to implement a hash join on the two columns, such that the output is two \textit{Join Index} tensors that reference the rows in each column which match according to a join predicate. Fortunately, the hashing and probing operations implementation for Plain columns can be reused directly for the value tensors of both RLE and Index \textit{DataColumn}s. The adaptations for each compressed layout are as follows.

\begin{myitemize}
    \item \textbf{RLE Data}: We perform a hash join on the RLE column's value tensor, treating each run like a single row in the hash table. The resulting index references runs rather than individual rows. To map these run indices back to actual row indices, we re-expand by combining each run's start/end range with the index tensor. 

    \item \textbf{Index Data}: We hash join on the column's value tensor, so the output indices initially reference the index entries themselves, not row positions. We then apply those indices to the original index tensor to recover the row positions for the final join.
\end{myitemize}
\noindent The output of the hash join are two \textit{Join Index} tensors, which are RLE or Index-encoded depending on the input \textit{DataColumn} encodings. (\Cref{tab:join_index}). 

Note that if either join column uses RLE, the matching index yields a one-to-many join and needs to be duplicated by the run length. If both join columns use RLE, then each pair of matching runs yields a many-to-many join, and the final run lengths are determined by the product of their lengths (by applying \Cref{alg:range_arange} to duplicate the runs). A detailed example of this join index generation process is provided in \Cref{sec:join-example}.
\vspace{-5pt}

\begin{table}[ht]
    \centering
    \caption{\textit{Join Index} encodings based on encodings of Input \textit{DataColumn}s. The \textit{Join Index} tensors are unsorted and may contain duplicates. In each entry $i,j$ of the table below, the top encoding is of the \textit{Join Index} for the join column indicated by $j$ and the \textcolor{lightblue}{bottom encoding} is of the \textit{Join Index} for the join column indicated by \textcolor{lightblue}{$i$}.\vspace{-10pt}}
    \label{tab:join_index}
    \resizebox{0.73\columnwidth}{!}{
    \begin{tabular}{l c c c}
        \toprule
         & \textcolor{black}{\textbf{RLE Data}} & \textcolor{black}{\textbf{Plain Data}} & \textcolor{black}{\textbf{Index Data}} \\
        \midrule
        \textcolor{lightblue}{\textbf{RLE Data}} & 
        \begin{tabular}{c} \textcolor{black}{RLE} \\ \hline \textcolor{lightblue}{RLE} \end{tabular} & 
        \begin{tabular}{c} \textcolor{black}{Index} \\ \hline \textcolor{lightblue}{RLE} \end{tabular} & 
        \begin{tabular}{c} \textcolor{black}{Index} \\ \hline \textcolor{lightblue}{RLE} \end{tabular} \\
        \midrule
        \textcolor{lightblue}{\textbf{Plain Data}} & 
        \begin{tabular}{c} \textcolor{black}{RLE} \\ \hline \textcolor{lightblue}{Index} \end{tabular} & 
        \begin{tabular}{c} \textcolor{black}{Index} \\ \hline \textcolor{lightblue}{Index} \end{tabular} & 
        \begin{tabular}{c} \textcolor{black}{Index} \\ \hline \textcolor{lightblue}{Index} \end{tabular} \\
        \midrule
        \textcolor{lightblue}{\textbf{Index Data}} & 
        \begin{tabular}{c} \textcolor{black}{RLE} \\ \hline \textcolor{lightblue}{Index} \end{tabular} & 
        \begin{tabular}{c} \textcolor{black}{Index} \\ \hline \textcolor{lightblue}{Index} \end{tabular} & 
        \begin{tabular}{c} \textcolor{black}{Index} \\ \hline \textcolor{lightblue}{Index} \end{tabular} \\
        \bottomrule
    \end{tabular}
    }\vspace{-16pt}
\end{table}

\subsection{Apply Join Index}
\label{sec:apply_join_index}

Once we have the \textit{Join Indices}, we apply them to all participating \textit{DataColumn}s to generate the joined result. This is straightforward for the join columns. For other columns, we build on the same approach we developed earlier for filtering/selecting data (see \Cref{sec:arith-comp-select}), but extend it to handle unsorted or duplicate entries.

\Cref{tab:and_design} outlines the extension needed for \textcolor{lightred}{Unsorted RLE} and \textcolor{lightred}{Unsorted Index} \textit{MaskColumn}s. For instance, suppose we want to apply an RLE \textit{Join Index} to an RLE \textit{DataColumn}. Previously, if both sides were sorted, we used \Cref{alg:range_intersect} to bucketize whichever side was smaller for performance. This was valid because \texttt{bucketize} requires sorted input. Now, if one side is unsorted, we must bucketize the sorted side so that each run or index from the unsorted side can be matched to the corresponding runs in the sorted data.

This step can be further optimized for one-to-many and one-to-one joins, since there are no duplicates in the \textit{Join Index} for the many-side of the join for the former case, and no duplicates in either \textit{Join Index} for the latter case.

\section{Experimental Setup and Results}
\label{sec:experiments}

We compare our tensor-based GPU query execution on compressed data against tensor-based GPU query execution on Plain data (with dictionary encoding, like TQP)\footnote{Execution plans on Plain data are an optimized version of the tensor programs generated by TQP. Plain performance are therefore equal or better than TQP. For more details on the query optimization rules applied, see Appendix~\ref{appendix:query-optimization}.}, and CPU-only query execution with SQL Server and Analysis Services~\cite{as}.
We run GPU experiments on an Azure NC24ads A100 v4 VM~\cite{Azure-NCA100} (pay-as-you-go @\$3.673/hour) having an NVIDIA A100 GPU with 80 GiB HBM, 24 vCPUs, and 220 GiB main memory.
For CPU-only analytics systems,
we use an Azure D64s v5 VM~\cite{Azure-Dv5} (pay-as-you-go @\$3.072/hour) with 64 vCPUs and 256 GiB main memory. 
We report warm query times, averaged over multiple runs, after the data is loaded into memory (GPU HBM for GPU and main memory for CPU experiments) and caches have been warmed up. Our assumption is that data already \textbf{resides} in GPU memory during query execution. This is a key benefit of compression: data becomes smaller and can fit in GPU memory, avoiding expensive PCIe transfers. We monitor GPU memory usage with the nvidia-smi profiling tool.

We select input table column encodings using simple heuristics.
\begin{myitemize}
    \item Columns under 1M rows use Plain encoding.
    \item Else, if the RLE compression ratio $> 20$, then use RLE.
    \item Else, if many single-element runs exist but longer runs still yield an RLE compression ratio $> 20$, then use RLE+Index.
    \item Else, if removing top/bottom $5\%$ of values allows a narrower type for the remaining data, then use Plain+Index.
    \item Else, use Plain encoding (possibly centered).
\end{myitemize}

\vspace{-2ex}
\begin{figure}[ht]
    \centering
    \includegraphics[width=0.99\columnwidth]{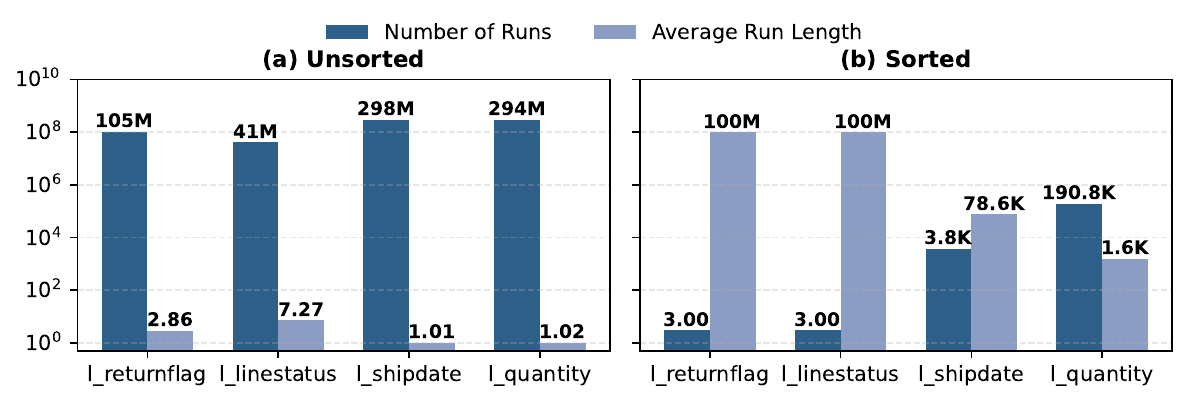}
\vspace{-2ex}
    \caption{Total \# of runs and average run lengths for fact table for TPC-H SF=50 Q1 before and after sort. Sorting greatly improves RLE compression, reducing runs from $>$105M to 3 while increasing average  length from 2.86 to $>$100M.}
    \label{fig:tpch_rle_stats}
\end{figure}

\begin{figure*}[ht]
    \centering
    \includegraphics[width=\textwidth]{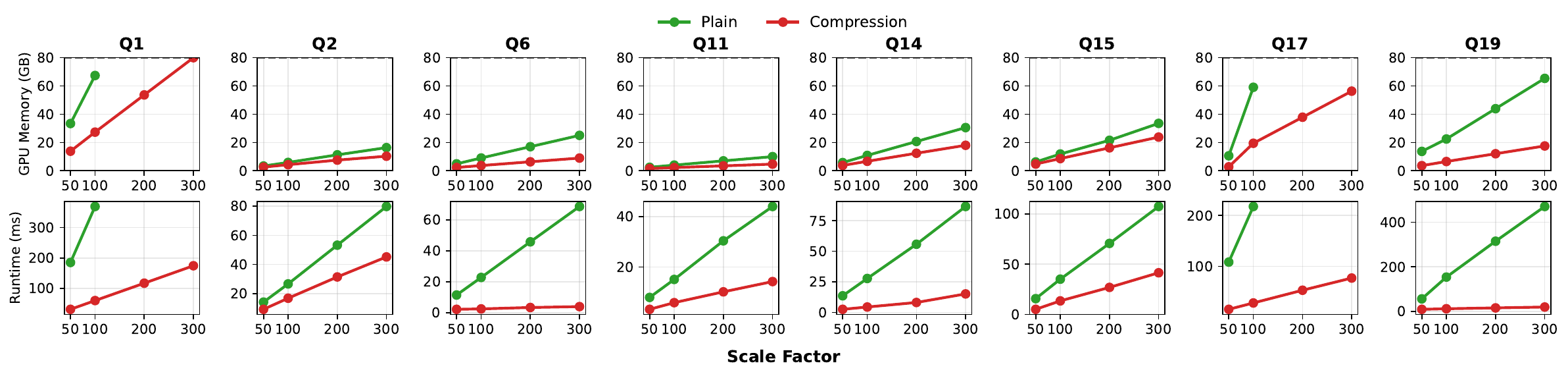}
    \vspace{-10pt}
    \caption{Peak GPU memory usage (top) and query run times (bottom) for TPC-H queries on Plain and Compressed input data across different scale factors. Compression achieves speedups up to $23.8\times$ while reducing memory usage by up to $3.7\times$.\vspace{-10pt}}
    \label{fig:tpch-peak-mem}
\end{figure*}

\begin{figure}[ht]
    \centering
    \includegraphics[width=0.8\columnwidth]{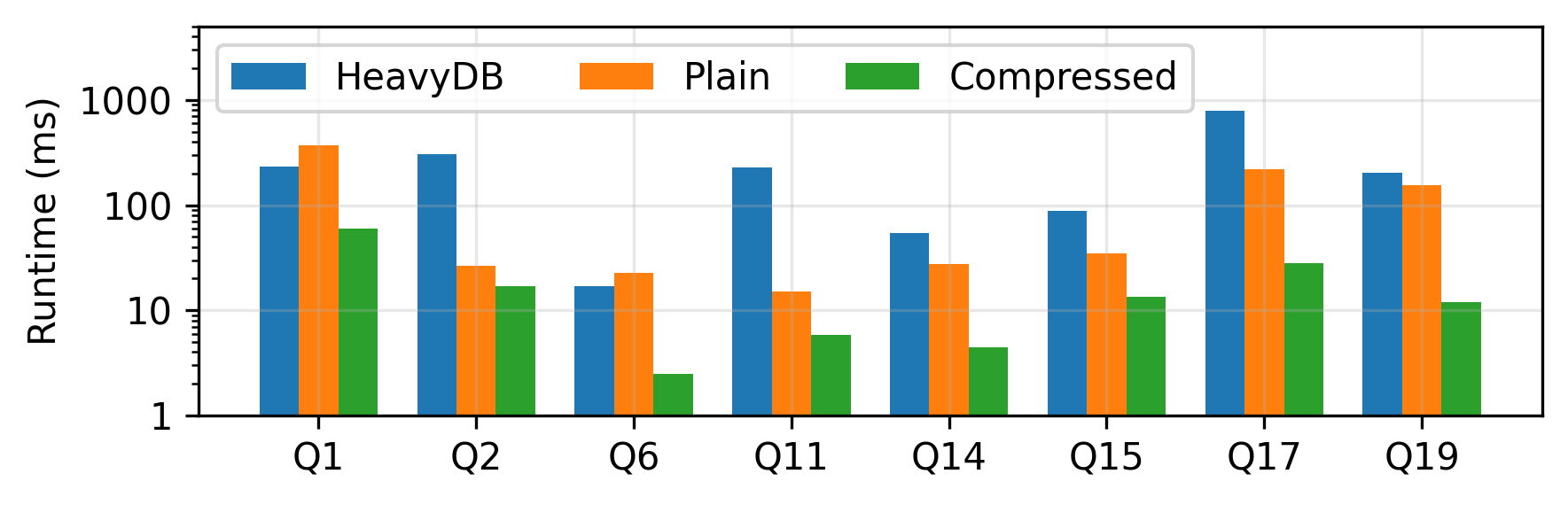}
    \vspace{-10pt}
    \caption{TPC-H query runtime between HeavyDB and compressed GPU execution at SF=100. Our compressed approach achieves $12.8\times$ geometric mean speedup over HeavyDB.\vspace{-5pt}}
    \label{fig:heavydb-comparison}
\end{figure}

\subsection{TPC-H Queries}
\label{sec:expts-tpch}

We evaluate our approach using a subset of TPC-H queries (Q1, Q2, Q6, Q11, Q14, Q15, Q17, Q19) at scale factors (SF) 50, 100, and 300. The TPC-H benchmark utilizes synthetic data which, compared to real-world datasets (discussed in Section~\ref{sec:expts-pbi}), exhibits limited sparsity and redundancy.

\subsubsection{Query-Specific Data Ordering}
\label{sec:expts-tpch-perquery-ordering}
To study the impact of data ordering on RLE compression effectiveness, we employ query-specific sort orders for the relevant tables based on the columns involved in each query's filters and joins. Each table is sorted using a global multi-column ordering (equivalent to SQL's ORDER BY clause). The specific column orderings used for each query are detailed in \Cref{appendix:tpch-ordering}. Sorting significantly improves RLE effectiveness by increasing average run lengths and reducing the total number of runs. For instance, as illustrated for the l\_returnflag column in Q1 at SF=50 (Figure~\ref{fig:tpch_rle_stats}), sorting reduces the number of runs from $>105M$ to just 3, while increasing the average run length from 2.86 to $>100M$. In contrast, general-purpose V-order~\cite{v-order} proved less effective for TPC-H even with skewness, due to high-cardinality columns (\Cref{sec:expts-tpch-skewed}).

Figure~\ref{fig:tpch-peak-mem} presents both peak GPU memory usage and query run times, comparing query execution performance on uncompressed (Plain) data versus compressed (RLE, Index) data formats across SF 50, 100, and 300.
Our evaluation reveals substantial benefits when processing compressed data directly on the GPU. Peak memory usage is significantly reduced; for example, at SF=300, Q19 requires 65.3 GiB with Plain data but only 17.5 GiB with compressed data (a $3.7\times$ reduction). Compression enables processing larger scale factors like SF=300 for queries such as Q1 and Q17, which exceed the 80 GiB GPU memory limit with Plain data.

Query run times are also dramatically reduced, with speedups often exceeding $10\times$. For instance, at SF=300, Q6 runs 17.3$\times$ faster (68.6 ms vs 3.96 ms) and Q19 runs 23.8$\times$ faster (470.1 ms vs 19.8 ms) on compressed data.
These results highlight the effectiveness of our techniques in improving both memory efficiency and performance for TPC-H queries on GPUs.

Using SQL Server query times for SF=50 as reference, the sum of times for the 8 queries running on GPU with compressed data was 12.8x lower for SF=50 and 2.6x lower for SF=300 (6x larger SF)! To compare with existing GPU database systems, we show HeavyDB~\cite{heavydb} performance in Figure~\ref{fig:heavydb-comparison}. HeavyDB outperforms uncompressed execution on Q1 and Q6, but underperforms on the rest. Overall, HeavyDB is $12.8\times$ slower than our compressed execution (geometric mean across all queries).

\subsubsection{Compression Quality Ablation Study}
\label{sec:expts-compression-ablation}
To study the impact of RLE compression quality on query performance, we experimented with Q17 and Q19 at SF=100. These queries demonstrate significant speedups with compression and their performance depends on the l\_partkey column in LINEITEM, allowing systematic variation of compression effectiveness. We created datasets with varying compression ratios by controlling run formation. Starting with 600M LINEITEM rows naturally grouped into 20M partkeys ($\sim30$ rows each), we systematically break runs by further dividing each row group into 2--16 smaller runs. This creates datasets with compression ratios from 30x (20M runs) to 1.87x (320M runs).

\begin{figure}[ht]
    \centering
    \includegraphics[width=0.9\columnwidth]{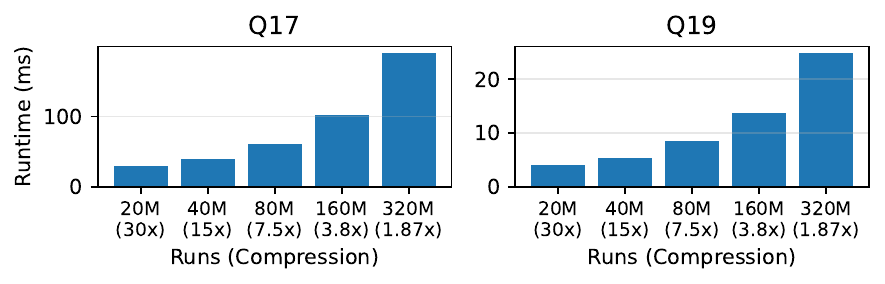}
    \vspace{-10pt}
    \caption{Query runtime degradation as compression ratio decreases for Q17 and Q19 at SF=100. Performance degrades $6\times$--$6.6\times$ as compression drops from 30$\times$ to 1.87$\times$.\vspace{-10pt}}
    \label{fig:run-breaking-runtime}
\end{figure}

Figure~\ref{fig:run-breaking-runtime} shows that query performance degrades as the number of runs increases and compression effectiveness decreases. With optimal compression (30x, 20M runs), Q17 executes in 28.87ms but degrades to 189.95ms with poor compression (1.87x, 320M runs)—a 6.6x slowdown. Similarly, Q19 runtime increases from 4.11ms to 24.83ms—a 6x slowdown. These results underscore the importance of good RLE compression for high query performance.

\subsection{Production queries}
\label{sec:expts-pbi}

To study the properties of production workloads, we evaluate our methods using representative production queries on a first-party production dataset, selected by the product team as a challenging case where CPU baselines perform well. It has a star schema with a large fact table having 2.94 Billion rows and 13 small dimension tables. It also has 4 `bridge' tables that enable joins between the fact table and 4 of the dimension tables.Unlike in the TPC-H study (Section~\ref{sec:expts-tpch}), we use the same input files, and thus the same row ordering for a given table, for all queries. We consider 2 ordering strategies: (1) On disk, each table has already been stored as V-ordered Parquet files~\cite{v-order}, which is enabled by default in Microsoft Fabric. This is not query-specific. (2) We also experiment with a query-specific sorting strategy based on cardinality ordering, sorting all query columns by cardinality (smallest first). This corresponds to real-world scenarios where BI customers optimize specific queries using simple rules.

We use three queries for this study. Q1 uses 10 columns while Q2 and Q3 use 12 columns each. Together, the queries use 15 columns (detailed column usage is provided in \Cref{tab:pbi_query_cols}) which is a subset of the total columns in the fact table. Relational operations in the queries include predicate filters on dimension tables, joins and semi-joins, and group-by aggregation (SUM). Q1 includes 7 semi-joins and 2 primary-key foreign-key (PK-FK) joins, Q2 and Q3 each includes 10 semi-joins and 1 PK-FK join. Q2 and Q3 have a similar template but differ in a filter predicate.

\subsubsection{Compression Effectiveness}
\begin{figure}[ht]
    \centering
    \includegraphics[width=0.8\columnwidth]{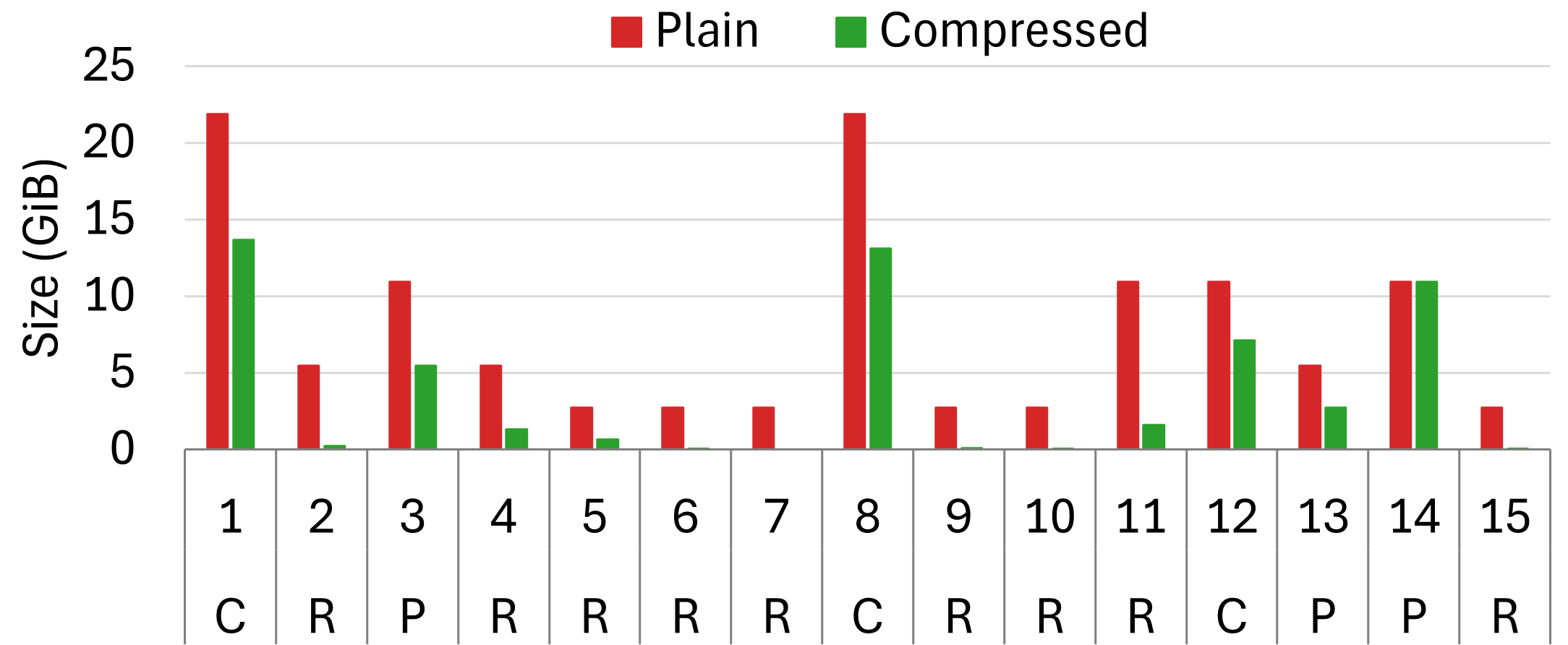}
    \vspace{-10pt}
    \caption{Input column sizes for the 15 columns in the fact table for Plain and Compressed representations. The top row in the x-axis labels shows the column numbers while the bottom shows the encoding for each column in compressed representation---R: RLE, C: Composite (Plain + Index), P: Plain with bit-width reduction.\vspace{-10pt}}
    \label{fig:pbi_col_sizes}
\end{figure}

To study compression benefits on real-world data, Figure~\ref{fig:pbi_col_sizes} shows the sizes for the in-memory representation of the 15 columns of the fact table. The series `Plain' refers to the default Plain encoding (with dictionary encoding) but with no other compression schemes applied.

\begin{figure}[ht]
    \centering
    \includegraphics[width=0.8\columnwidth]{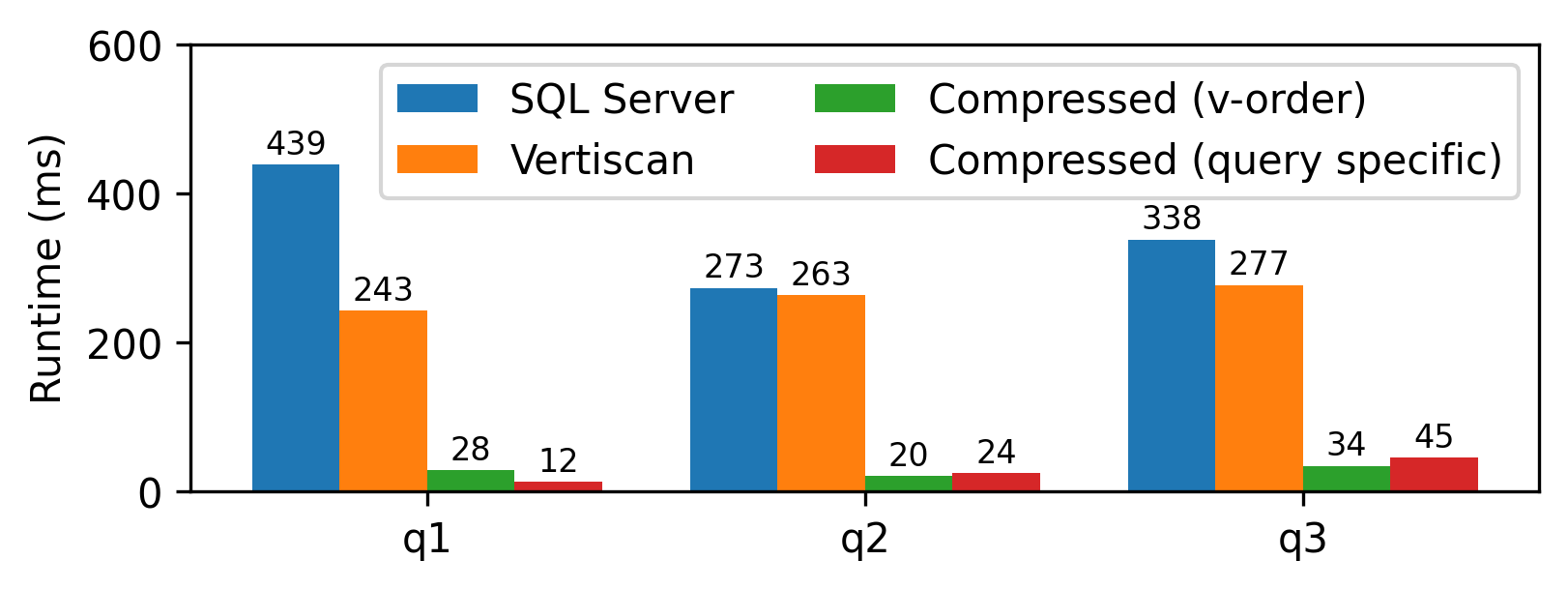}
    \vspace{-10pt}
    \caption{Query run times with Microsoft SQL Server on CPUs, Analysis Services on CPUs, and with our techniques for compressed input data on an A100 GPU.\vspace{-10pt}}
    \label{fig:pbi_query_times_comparison}
\end{figure}

The total size for the 15 columns is 120.36 GiB with Plain and does not fit into the 80 GiB GPU HBM. Considering only the subset of columns needed by the specific queries, Q1 needs 87.53 GiB and Q2 and Q3 need 82.06 GiB, which is still larger than the GPU HBM capacity.
The bottom row of the x-axis shows the encoding scheme for Compressed---`C': Composite (Plain + Index), `R': RLE, `P': Plain with bit-width reduction based on the range of values. The total size for 15 columns with Compressed is 56.84 GiB. For Q1 columns it is 43.69 GiB and for Q2 and Q3 it is 30.86 GiB, which can fit in the GPU HBM and also leave space for intermediate results. For reference, the on-disk V-ordered file size for the projected fact table for the 15 columns is 29.87 GiB.

This production dataset has substantial opportunities for RLE compression. 7 of the 15 fact table columns were compressed using RLE, achieving significant compression ratios. For example, column 7 has a single run of 2.94B rows, while column 11 achieves 6.9$\times$ compression (1.59 GiB vs 10.94 GiB for Plain representation). Detailed RLE statistics and compression analysis are provided in \Cref{appendix:pbi-compression}.
We also measured data transfer times from CPU to GPU using \texttt{tensor.to('cuda')} on our A100 80GB PCIe system (PCIe 4.0 x16 configuration). For compressed data, Q1 columns (45.85 GB) transfer in 2.16 seconds, while Q2 columns (26.18 GB) transfer in 1.29 seconds. These transfer times achieve 21.2 GB/s and 20.3 GB/s bandwidth respectively, representing $\sim64\%$ of the PCIe 4.0 x16 theoretical maximum ($\sim31.5 GB/s$ ).

\subsubsection{Query Performance Analysis}
To study performance gains on production queries, Figure~\ref{fig:pbi_query_times_comparison} shows query run times on state-of-the-art CPU-based commercial RDBMS, Microsoft SQL Server, and analytical data engine, Analysis Services. Our techniques lead to substantial speedups---15.75 (Q1), 13.66 (Q2), 9.82 (Q3), 12.76 (total) over SQL Server and 8.72 (Q1), 13.16 (Q2), 8.05 (Q3), 9.52 (total) over Analysis Services.
The cardinality-based sorting strategy achieved run times of 12.26 ms (Q1), 24.27 ms (Q2), and 45.19 ms (Q3), comparable to V-ordering. V-order performed better for Q2 and Q3 despite being query-agnostic, while cardinality-based sorting creates highly skewed column size distributions. This demonstrates V-order's effectiveness without query-specific knowledge, while cardinality-based sorting provides a simple competitive alternative.
In \Cref{sec:scalability-analysis}, we present a scalability analysis showing that compressed execution enables processing datasets up to 222\% (6.52B rows) of the original size within GPU memory, while Plain execution fails at 50\% capacity. Compressed queries also scale better: Q2 and Q3 GPU times on Plain data at just 20\% dataset size already exceed CPU times for the full 100\% dataset.

\subsection{Public BI Datasets}
\label{sec:expts-public-bi}

To further validate our approach on diverse real-world workloads, we evaluate our techniques using publicly available BI datasets~\cite{vogelsgesang2018get}. These are collected from Tableau Public, a platform hosting over 60K user-generated BI visualizations spanning government, healthcare, transportation, e-commerce, and other domains. These datasets reflect realistic production scenarios with diverse characteristics.

Of the 34 datasets, 20 ($59\%$) contain columns meeting the compression ratio threshold ($>20$) for RLE compression. Among these 20 datasets with compression opportunities, 14 have adequate size ($\geq 1$M rows) for meaningful GPU evaluation. We evaluate the first 3 qualified queries per dataset that utilize compressible columns, resulting in 38 total queries. Unlike our TPC-H experiments which used query-specific ordering, all datasets are sorted using V-order~\cite{v-order} across all columns.

\begin{figure}[ht]
    \centering
    \includegraphics[width=0.99\columnwidth]{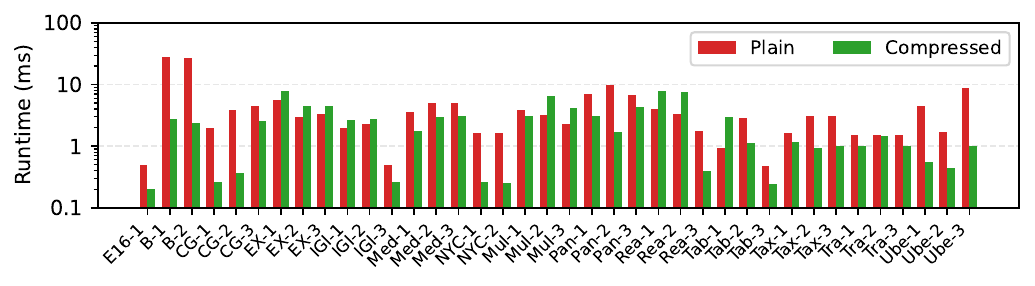}
    \vspace{-10pt}
    \caption{Query runtime for public BI datasets, comparing Plain vs Compressed data execution on GPU. Compression benefits $73.7\%$ of queries with $2.02\times$ geometric mean speedup and maximum speedup of $11.27\times$.\vspace{-10pt}}
    \label{fig:public-bi-runtime}
\end{figure}

\Cref{fig:public-bi-runtime} shows compression provides significant benefits in 28 of 38 queries ($73.7\%$) with maximum speedup of $11.27\times$ and geometric mean speedup of $2.02\times$. There are also 10 queries that show slowdowns up to $3.13\times$. This occurs when queries have few RLE columns, and they are directly operated (e.g., AND operations) with plain columns; this introduces additional overhead from RLE-to-plain decompression. Overall, real-world BI queries demonstrate high compressibility and our execution framework shows effective acceleration for the majority of workloads.

\vspace{-1ex}
\section {Related Work}
\label{sec:related}

\noindent\textbf{GPU acceleration for SQL analytics}: There has been considerable research into using GPUs to accelerate SQL analytical queries~\cite{crystal_shanbhag_2020,gpudb-characterization-opt,surakav_he_2022, yuan2013yin,rosenfeld2022query,Yogatama22-mordred,sioulas19-partitioned-radix-join,Yogatama25-lancelot,yuan25-vortex,chrysogelos19-hetexchange,wu25-gpu-joins-groupby,Hu22-tcudb,Rui17-fastequijoin,hong25-themis,deng24-prefetching,Tomas15-groupby,heavydb,voltron-data,book:gpudb:2021,blazingsql_2022,Lutz20-pump-up-the-volume,maltenberger22-multigpu-sort,Rui20-multigpujoin}. 
The majority of these works, either consider input data to be already in plain format(e.g.,~\cite{blazingsql_2022,heavydb,voltron-data,surakav_he_2022}), or decompress it before executing queries (e.g., ~\cite{crystal_shanbhag_2020}). 
Several of this works also only consider using GPUs to accelerate a subset of query operators (e.g.,~\cite{sioulas19-partitioned-radix-join,wu25-gpu-joins-groupby,Lutz20-pump-up-the-volume}). Compared to this previous research, we (1) support end-to-end GPU acceleration of queries; and (2) we are the first one providing a framework allowing to run queries in GPU and directly on compressed data.

\noindent\textbf{Query Execution over Compressed Data}: The domain of query execution over compressed data has seen considerable research. Numerous studies have optimized dictionary encoding~\cite{codecdb_jiang_2021,lee2014joins,damme2020morphstore}, among which C-store~\cite{c_store_abadi_2006, abadi_design_implementation_column_dbs_2013} stands out as especially relevant. C-store introduces optimization strategies, such as rewriting summations as products over RLE and performing selections directly on the dictionary. Despite these advancements, two significant limitations exist: 
(1) Constricted design space: Previous works do not consider more complex scenarios where two RLE representations might select masks from each other, or where RLE representations are employed as group-by columns. Furthermore, we introduce a novel index representation, whose interaction with RLE has not previously been considered. The challenge in these studies is the amplified complexity associated with operator implementation~\cite{c_store_abadi_2006,zukowski2006super}. For our case, this complexity is mitigated because we utilize PyTorch-based operators.
Our work presents a comprehensive exploration of these scenarios.
(2) CPU-centric designs: The majority of prior work focuses on CPU operations. In contrast, our designs and operators are fine-tuned specifically for GPUs, which fundamentally differ from CPUs. A GPU's architecture emphasizes parallelism, necessitating each thread to manage simple tasks. This distinction raises a plethora of intriguing design questions.

\noindent\textbf{Compression on GPU}: 
Several works apply GPUs for query execution ~\cite{crystal_shanbhag_2020,gpudb-characterization-opt, surakav_he_2022, tensor_tea_vldb_2022,gandhi2022tensor,tqp-xbox} but are limited to only basic compression of dictionary encoding.
With the challenges posed by limited GPU memory and PCIe bandwidth, many researchers have turned their attention to harnessing GPUs for compression ~\cite{fang2010database,hippogriffdb,rosenfeld2022query,sitaridi2016gpu,tile-integer-compression,fastlanes-gpu}. However, this compression typically facilitates only data transfer, and once the data is onboard the GPU, it often undergoes decompression, albeit potentially in a GPU-optimized manner. Existing works largely concentrate on refining the compression and decompression stages without addressing the direct end-to-end execution of queries in the encoded space.

\noindent\textbf{RLE compression in data formats}: RLE compression has wide support in both storage and in-memory data formats~\cite{parquet, arrow-ree, kuschewski2023btrblocks, fastlanes, vortex}. Reordering table rows to improve RLE compression has been leveraged in Microsoft Fabric~\cite{v-order}, DuckDB~\cite{duckdb-reorder-compression}, and others. We present new methods for SQL query processing to exploit the compression better for higher performance and smaller memory footprints.
\vspace{-1pt}

\vspace{-1pt}
\section{Conclusion}
We presented new methods for leveraging light-weight compression schemes to execute SQL analytics queries on GPUs directly on compressed data, with substantial reductions in GPU memory requirements and query run times. Our framework includes primitives for operating on encoded data, and implementations of relational operators including selection, group-by, aggregate, and join, that avoid decompression as far as possible. Our techniques result in significant acceleration of representative production queries on a real-world dataset compared to state-of-the-art commercial CPU analytics systems, and GPU query execution on uncompressed data.
\label{sec:conclude}

\clearpage

\balance

\appendix
\section{Operator Implementation Details and Examples}
\label{sec:implementation-details}

This section provides detailed algorithmic implementations and concrete examples for key operations in our compressed tensor query execution system.

\subsection{NOT Operator Implementation}
\label{sec:not-operator-details}

The NOT operator requires specialized algorithms for different mask representations. For RLE masks, we compute the gaps between runs to create the inverted result (\Cref{alg:not_rle}). For Index masks, we convert sparse indices to continuous ranges representing all non-selected positions (\Cref{alg:not_index}). These algorithms are illustrated with concrete examples below.

\begin{algorithm}[H]
    \caption{not\_rle}
    \label{alg:not_rle}
    \raggedright
    \textbf{Input:}  $start$, $end$, \textit{total\_size}\\
    \textbf{Output:}  $start'$, $end'$
    \begin{algorithmic}[1]
        \small
        \State $start'$ $\gets$ \textbf{cat}(([-1], $end$)) + 1
        \State $end'$ $\gets$ \textbf{cat}(($start$,[\textit{total\_size}])) - 1
        \State \textit{mask} $\gets$ $start'$ $\leq$ $end'$ 
        \State \Return  $start'$[\textit{mask}], $end'$[\textit{mask}]
    \end{algorithmic}
\end{algorithm}

\begin{algorithm}[H]
    \caption{not\_index}
    \label{alg:not_index}
    \raggedright
    \textbf{Input:} \textit{val}, \textit{total\_size}\\
    \textbf{Output:}  $start$, $end$
    \begin{algorithmic}[1]
        \small
        \State $start$ $\gets$ \textbf{cat}(([-1], \textit{val})) + 1
        \State $end$ $\gets$ \textbf{cat}((\textit{val}, [\textit{total\_size}]))  - 1
        \State \textit{mask} $\gets$ ($start$ $\leq$ $end$) $\land$ ($start$  $<$ \textit{total\_size}) $\land$ ($end$  $\geq 0$)
        \State \Return $start$[\textit{mask}], $end$[\textit{mask}]
    \end{algorithmic}
\end{algorithm}

\begin{figure}[h]
    \centering
    \includegraphics[width=0.9\linewidth]{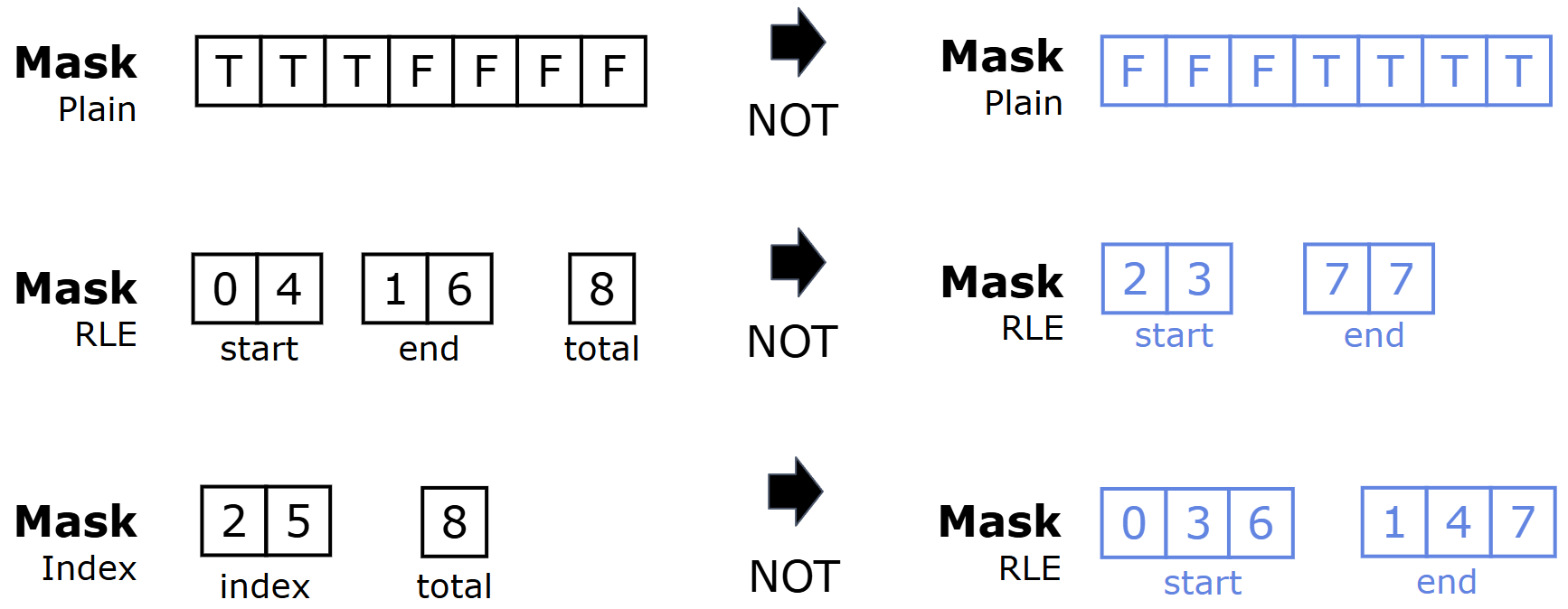}
    \vspace{-5pt}
    \caption{Illustration of NOT logical operator.
    \vspace{-10pt}}
    \label{fig:not_logical}
\end{figure}

\begin{example}
Consider applying the NOT operator to the three different mask representations shown in \Cref{fig:not_logical}:
\end{example}
\begin{myitemize}
    \item \textbf{NOT on Plain mask.} For a Plain mask $[T,T,T,F,F,F,F]$, we simply apply PyTorch's bitwise negation operator ($\sim$) to flip each boolean value, resulting in $[F,F,F,T,T,T,T]$.
    
    \item \textbf{NOT on RLE mask.} For an RLE mask with $s=[0,4]$, $e=[1,6]$, and $total\_size=8$, we apply the \texttt{complement\_rle} primitive. First, we compute the potential start points, $s'$, for inverted ranges by concatenating $[-1]$ with the original end positions and adding 1. $s' = [0,2,7]$. Then, we compute potential end points, $e'$, by concatenating the original start values with $[total\_size]$ and subtracting 1. $e' = [-1,3,7]$. We keep only the valid ranges where $s' \leq e'$, which gives us $s'=[2,7]$ and $e'=[3,7]$. This is an RLE mask covering the gaps between the original runs.
    
    \item \textbf{Step 3: NOT on Index mask.} For an Index mask with $p=[2,5]$ and $total\_size=8$, we apply the \texttt{complement\_index} primitive. We compute potential start points by concatenating $[-1]$ with the original indices and adding 1. $s = [0,3,6]$. We compute potential end points by concatenating the original indices with $[total\_size]$ and subtracting 1. $e = [1,4,7]$. After applying a validity mask to ensure $s \leq e$, $s < total\_size$, and $e \geq 0$, all position ranges remain valid. This is an RLE Mask with runs covering all positions not in the original Index mask.
\end{myitemize}

\subsection{GroupBy Aggregation Example}
\label{sec:groupby-example}

We provide a detailed walkthrough of how groupby aggregation works on RLE-compressed data, illustrating the three-step process of grouping, alignment, and aggregation.

\begin{figure}[h]
    \centering
    \includegraphics[width=1\linewidth]{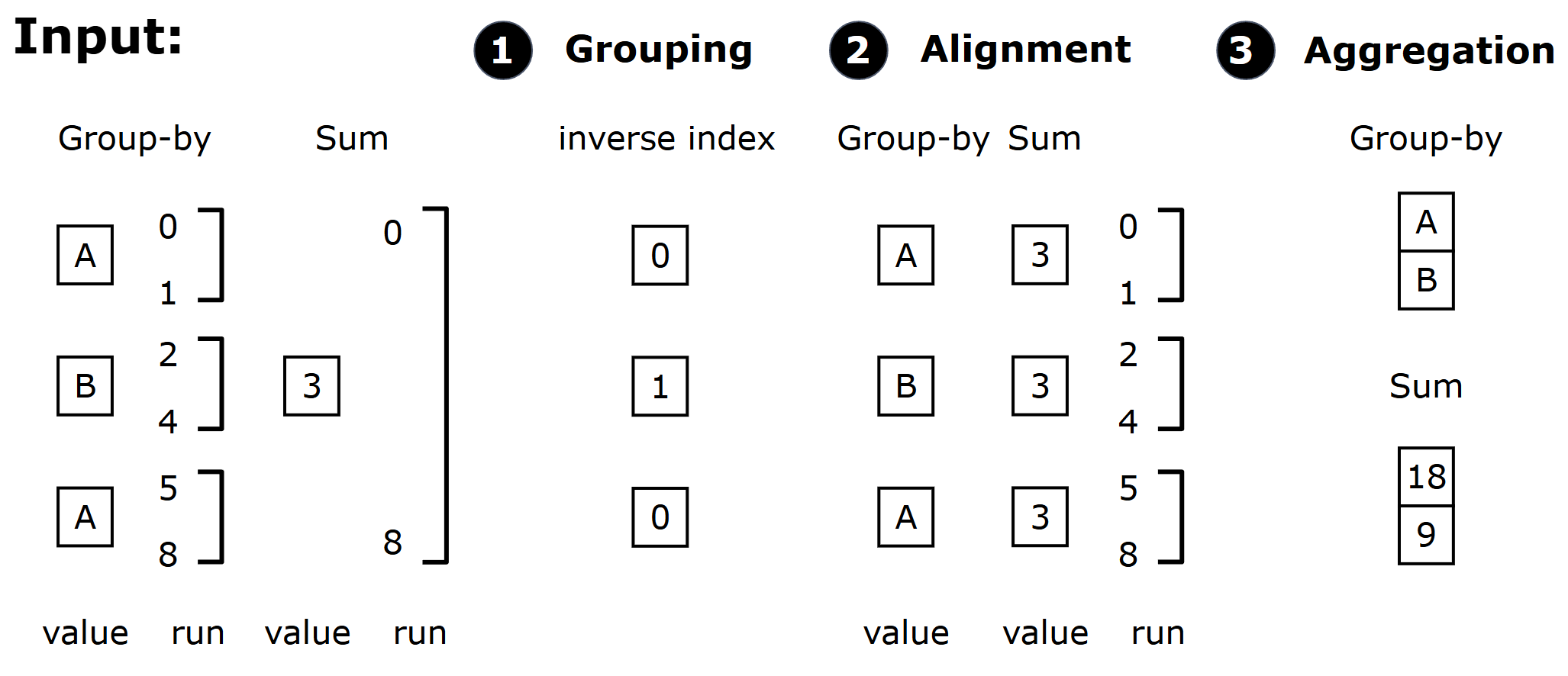}
    \vspace{-20pt}
    \caption{RLE-compressed data Group-by Aggregation: Grouping, Alignment and Aggregation steps illustrated.
    \vspace{-10pt}}
    \label{fig:aggregation}
\end{figure}

\begin{example}
Consider a SQL query "SELECT SUM(B) GROUP BY A" on RLE-compressed data as shown in \Cref{fig:aggregation}. The input consists of a group-by column A with values $[A, B, A]$ spanning runs 0--1, 2--4, 5--8 and a sum column B with value $3$, start $[0]$ and end $[8]$.
\end{example}
\begin{myitemize}
    \item \textbf{Step 1: Grouping.} We apply the grouping operation on the group-by column to identify unique groups. Using torch.unique on the values $[A, B, A]$, we identify two unique groups (A and B) and generate an inverse index, $x=[0, 1, 0]$ that maps each run to its corresponding group: the first run (group-by value A) maps to group 0, the second run (group-by value B) maps to group 1, and the third run (group-by value A) again maps to group 0. This creates a temporary RLE column $g$ with $v=x$ (where $x$ is the inverse index just computed) and with $s$ and $e$ the same as that of the group-by column ($A$).
    \item \textbf{Step 2: Alignment.} Next, we align $g$ and the aggregation column ($B$). After alignment, as shown in the figure, both columns share the same runs $[0-1]$, $[2-4]$, and $[5-8]$. The aggregation column B is represented with aligned values $[3, 3, 3]$ for these runs.
    \item \textbf{Step 3: Aggregation.} We do this in two steps---(a) perform a parallel aggregation for each segment in the aligned aggregation column by computing $v \times l$ using its values $[3, 3, 3]$ and run lengths $[2, 3, 4]$, resulting in $[3 \times 2, 3 \times 3, 3 \times 4] = [6, 9, 12]$, (b) perform a parallel aggregation for each group-by key by applying \texttt{torch.scatter} on the tensor $[6, 9, 12]$ using the inverse index $[0, 1, 0]$. The scatter results in values $6+12=18$ (for group A, index 0) and $9$ (for group B, index 1). We retrieve the group-by keys $[A, B]$ corresponding to the inverse indices $[0, 1]$. 
\end{myitemize}

\subsection{Join Index Generation Example}
\label{sec:join-example}

We illustrate how join indices are generated between columns with different encodings, showing the three-step process of value matching, expansion based on run lengths, and final index generation.

\begin{figure}[h]
    \centering
    \includegraphics[width=0.67\linewidth]{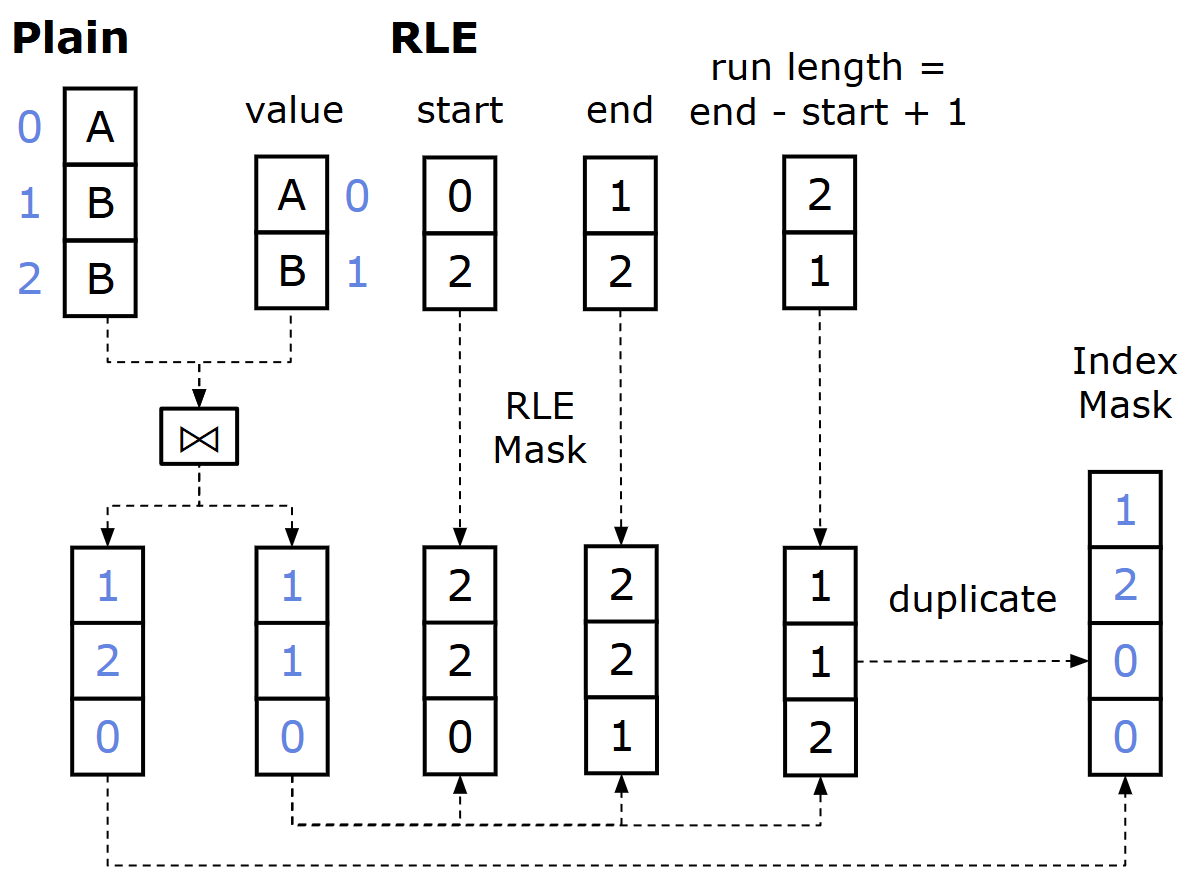}
    \vspace{-10pt}
    \caption{ Illustration of computing Join Indices between Plain and RLE columns. 
    \vspace{-5pt}
    }
    \label{fig:get_join_index}
\end{figure}

\begin{example}
Consider generating Join Indices between a Plain DataColumn and an RLE DataColumn as shown in \Cref{fig:get_join_index}.
The Plain DataColumn contains values [A, B, B] at positions 0, 1, and 2.
The RLE DataColumn contains values $v=[A, B]$ with start positions $s=[0, 2]$, end positions $e=[1, 2]$, and run lengths $l=[2, 1]$ ($l=e-s+1$).
\end{example}
\begin{myitemize}
    \item \textbf{Step 1: Perform hash join on values.} We execute a hash join between the value arrays from both columns. This identifies that Plain[0] = 'A' matches RLE.value[0] = 'A', and Plain[1] = Plain[2] = 'B' matches RLE.value[1] = 'B'. The initial \textit{Join Indices} are: [1, 2, 0] for the Plain column and [1, 1, 0] for the RLE column.
    
    \item \textbf{Step 2: Expand \textit{Join Index} for Plain according to run lengths.} For each matched RLE run, we need to duplicate the corresponding Plain indices based on the run length. The RLE run with index 0 (value 'A') has a run length of 2, so the corresponding Plain index 0 is duplicated twice. The RLE run with index 1 (value 'B') has a run length of 1, so the corresponding Plain indices 1 and 2 are not duplicated.
    
    \item \textbf{Step 3: Generate final Join Indices.} The expanded \textit{Join Index} for Plain becomes [1, 2, 0, 0] and the \textit{Join Index} for RLE remains as RLE with $v=[1,1,0], s=[2,2,0], e=[2,2,1]$.
\end{myitemize}

\section{TPC-H Experiment Details}
\label{appendix:tpch-details}

\subsection{Query-Specific Column Orderings}
\label{appendix:tpch-ordering}

\Cref{tab:tpch-ordering} details the query-specific column orderings used for TPC-H queries to maximize RLE compression effectiveness. Each table is sorted using a global multi-column ordering (equivalent to SQL's ORDER BY clause) based on the columns involved in each query's filters and joins.
\begin{table}[b]
    \footnotesize
    \centering
    \caption{Query-specific column orderings for TPC-H.}
    \label{tab:tpch-ordering}   
    \vspace{-10px}
    \begin{tabular}{|@{ }c@{ }|c@{ }|c|}\hline
         \textbf{Query}&  \textbf{Ordering Columns}& \textbf{Table}\\\hline\hline
         Q1& l\_returnflag, l\_linestatus, l\_shipdate, l\_quantity& LINEITEM\\\hline
         \multirow{2}{*}{Q2}& ps\_suppkey, ps\_partkey& PARTSUPP\\\cline{2-3}
         &p\_size, p\_type, p\_mfgr, p\_partkey& PART\\\hline
         Q6& l\_quantity, l\_discount, l\_shipdate& LINEITEM\\\hline  
         \multirow{2}{*}{Q11}& s\_nationkey& SUPPLIER\\\cline{2-3}
         & ps\_suppkey& PARTSUPP\\\hline
         \multirow{2}{*}{Q14}& l\_shipdate& LINEITEM\\\cline{2-3}
         & p\_type& PART\\\hline                
         Q15& l\_shipdate& LINEITEM\\\hline
         \multirow{2}{*}{Q17}& l\_partkey& LINEITEM\\\cline{2-3}
         & p\_brand, p\_container, p\_partkey& PART\\\hline
         \multirow{2}{*}{Q19}& l\_partkey& LINEITEM\\\cline{2-3}
         & p\_brand, p\_container, p\_size, p\_partkey& PART\\\hline
    \end{tabular}
\end{table}

\subsection{TPC-H with Skewed Data}
\label{sec:expts-tpch-skewed}

Previous query-specific ordering corresponds to BI scenarios where known queries can be optimized. We also experiment with general ordering using V-order~\cite{v-order} on TPC-H with skew. V-order applies sorting on tables, dictionary encoding, and other techniques to improve compression while supporting fast reads. We experimented with TPC-H using V-order that is not query-specific, on TPC-H SF=50 with skewness parameter z=1. However, only Q1 showed performance speedup, while the rest did not improve. We find that skew alone is insufficient for effective RLE compression. Only Q1 shows speedup because the group-by columns (l\_returnflag and l\_linestatus) have low cardinality and are naturally RLE-friendly. For other queries, most columns are not suitable for RLE compression, resulting in higher query times due to decompression overhead when operating with Plain columns.

\begin{figure}[ht]
    \centering
    \includegraphics[width=0.8\columnwidth]{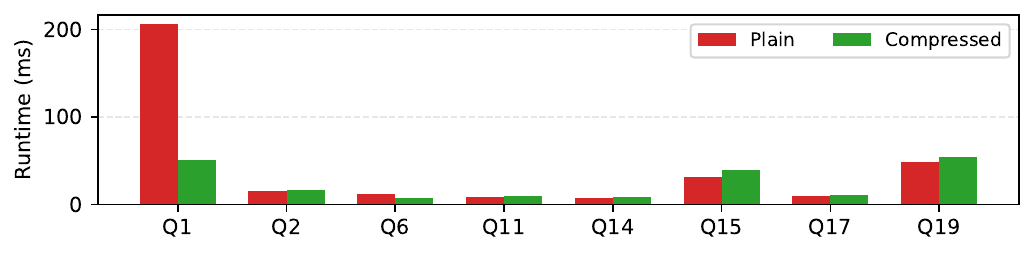}
    \caption{TPC-H query run times with skewed data (z=1) at SF=50, comparing Plain vs Compressed data execution on GPU. Only Q1 shows improvement due to low-cardinality columns suitable for RLE compression.}
    \label{fig:tpch-skew-runtime}
\end{figure}

\section{Production Workload Experiment Details}
\label{appendix:pbi-details}

This section provides comprehensive details of our production dataset experiments, including dataset characteristics, compression visualizations, and scalability analysis.

\subsection{Fact Table Column Usage}
\label{appendix:pbi-columns}

Table~\ref{tab:pbi_query_cols} shows the columns accessed by each production query evaluated in \Cref{sec:expts-pbi}. The queries operate on a star schema with a fact table containing 2.94 billion rows. 
\begin{table}[H]
    \centering
    \caption{Columns of the fact table used by the production queries. Column numbers are anonymized due to proprietary constraints.\vspace{-10pt}}
    \label{tab:pbi_query_cols}
    \resizebox{0.85\columnwidth}{!}{
    \begin{tabular}{|c|c|c|}\hline
         \textbf{Query}& \textbf{Columns of fact table used}& \textbf{Total used} \\\hline
         Q1 & 1, 2, 3, 4, 5, 10, 11, 12, 13, 14& 10\\\hline
         Q2 & 2, 3, 5, 6, 7, 8, 9, 10, 11, 12, 13, 15& 12\\\hline
         Q3 & 2, 3, 5, 6, 7, 8, 9, 10, 11, 12, 13, 15& 12\\\hline
    \end{tabular}
    }
\end{table}

Q1 accesses 10 columns and performs 7 semi-joins and 2 PK-FK joins. Q2 and Q3 each access 12 columns and perform 10 semi-joins and 1 PK-FK join. Together, the three queries use 15 distinct columns from the fact table. The total size for these 15 columns is 120.36 GiB with Plain encoding (exceeding the 80 GiB GPU HBM capacity) and 56.84 GiB with compressed encodings.

\subsection{RLE Compression Characteristics}
\label{appendix:pbi-compression}

\begin{figure}[ht]
    \centering
    \includegraphics[width=0.99\columnwidth]{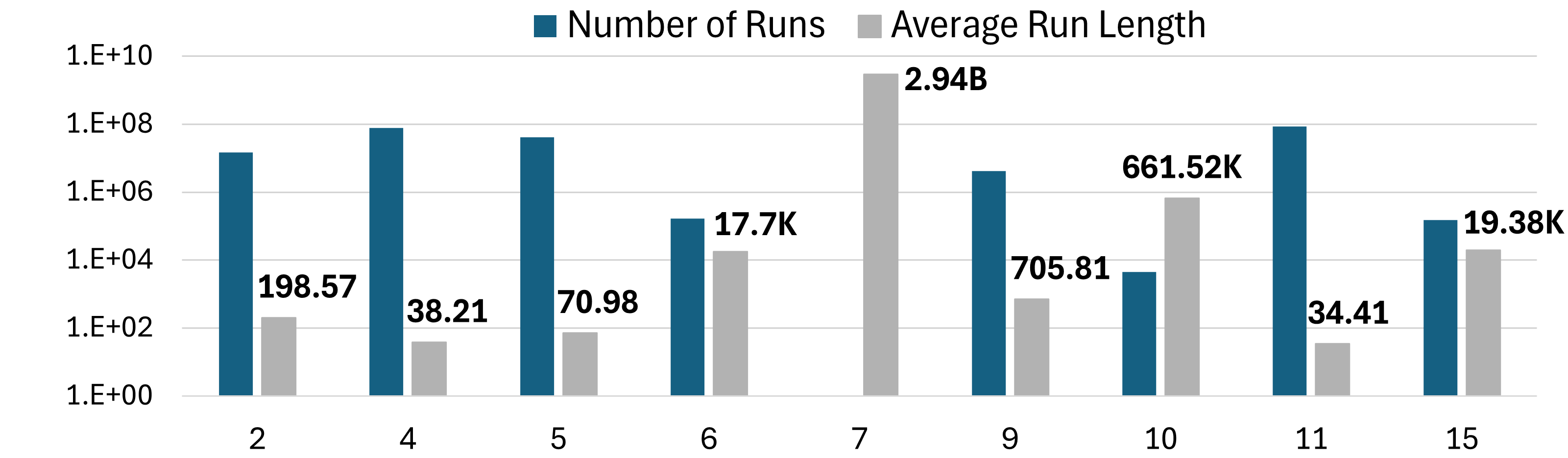}
    \vspace{-10pt}
    \caption{Total number of runs and average run lengths (with data labels) for RLE-compressed fact table columns.\vspace{-5pt}}
    \label{fig:pbi_rle_stats}
\end{figure}

The production dataset demonstrates substantial opportunities for RLE compression across multiple columns. Figure~\ref{fig:pbi_rle_stats} shows the detailed compression statistics for the 7 RLE-compressed columns.

Column 7 exhibits the highest compression with a single run of 2.94B rows (same value for all rows), representing perfect compression. Average run lengths for other columns range from several tens to hundreds of thousands, indicating varying degrees of data locality and repetition. 

For example, column 11 has 85.35M runs with an average run length of 34.41. Each run requires 20 bytes for representation---4 bytes for value ($v$), and 8 bytes each for start ($s$) and end ($e$) positions. This totals 1.59 GiB memory for the column, which is significantly less than the 10.94 GiB (= 4 × 2.94B) needed for the Plain representation, achieving a 6.9× compression ratio. Further memory savings are possible by using narrower data types for the start and end positions based on the actual data range.

\subsection{Scalability Analysis}
\label{sec:scalability-analysis}

To study performance scaling with data size, we study scalability characteristics by varying the input data size. For this we create smaller datasets with the first 5\% (0.15B), 20\% (0.6B), and 50\% (1.5B) rows of the fact table. 

\begin{figure*}[h]
\centering
{\includegraphics[width=0.3\textwidth]{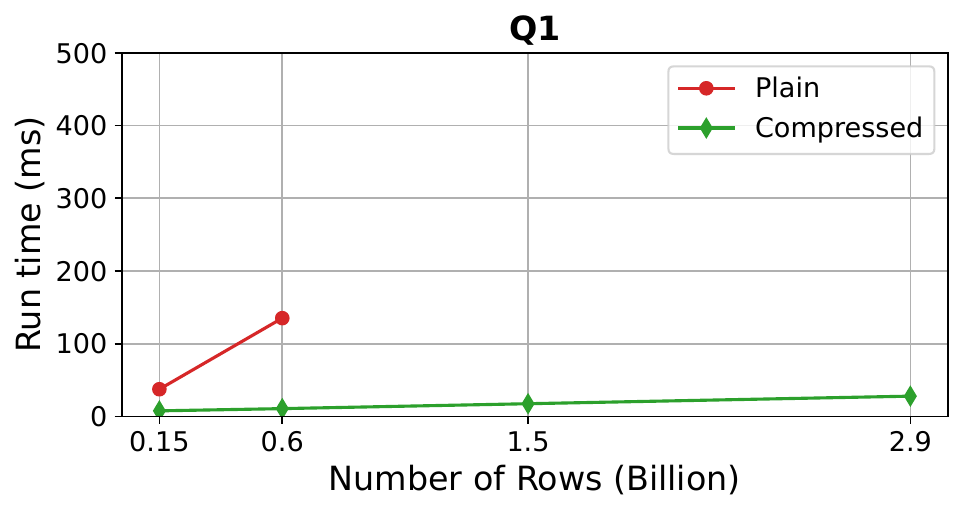}}
{\includegraphics[width=0.3\textwidth]{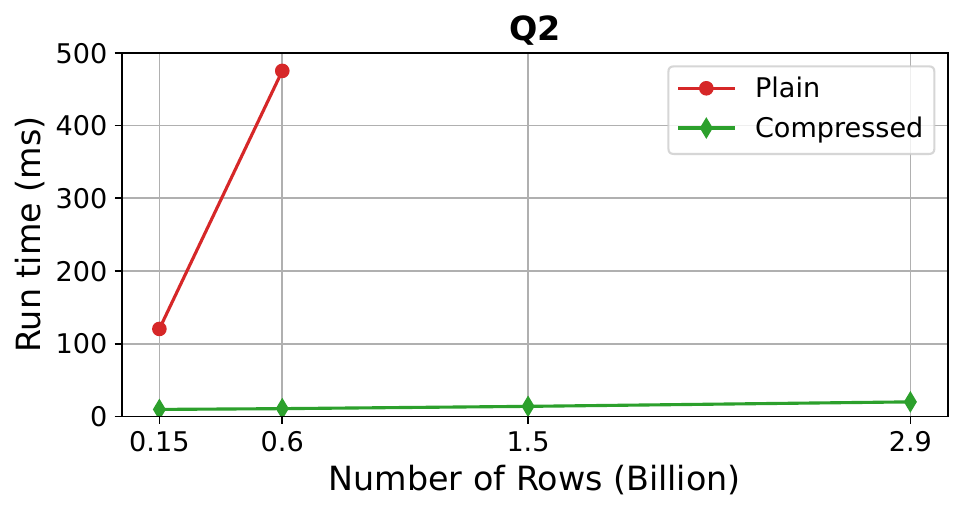}}
{\includegraphics[width=0.3\textwidth]{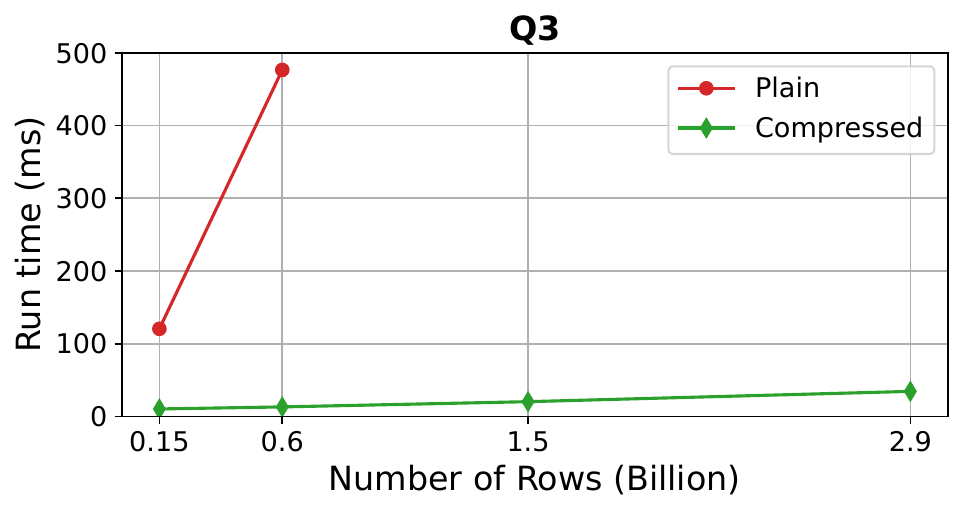}}
\vspace{-14pt}
\caption{Run times for production queries, using A100 GPU, on Plain and Compressed data for different table sizes.\vspace{-5pt}}
\label{fig:pbi-times}
\end{figure*}

Figure~\ref{fig:pbi-times} shows good performance scalability since GPU query times on compressed data increase much slower with data size compared to GPU query times on plain data. Queries cannot be run on Plain data for the 50\% (and 100\%) dataset due to running out of GPU HBM capacity, and Q2 and Q3 GPU times for just the 20\% dataset exceed that of SQL Server and Analysis Services on CPUs for the 100\% dataset! Thus, query execution on compressed data is necessary to unlock the GPU acceleration potential for this workload.
We could fit and process even larger datasets in the GPU HBM with our compression schemes. 

\begin{figure*}[h]
\centering
{\includegraphics[width=0.3\textwidth]{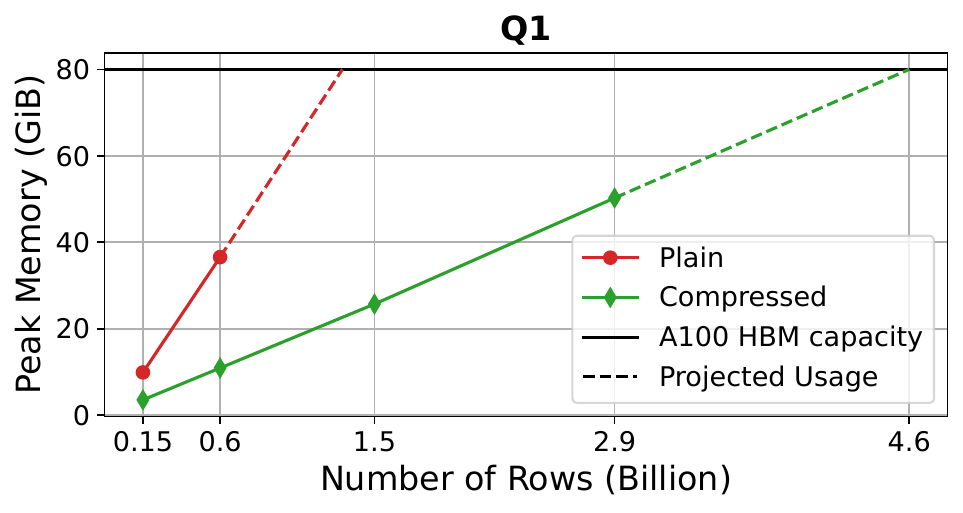}}
{\includegraphics[width=0.3\textwidth]{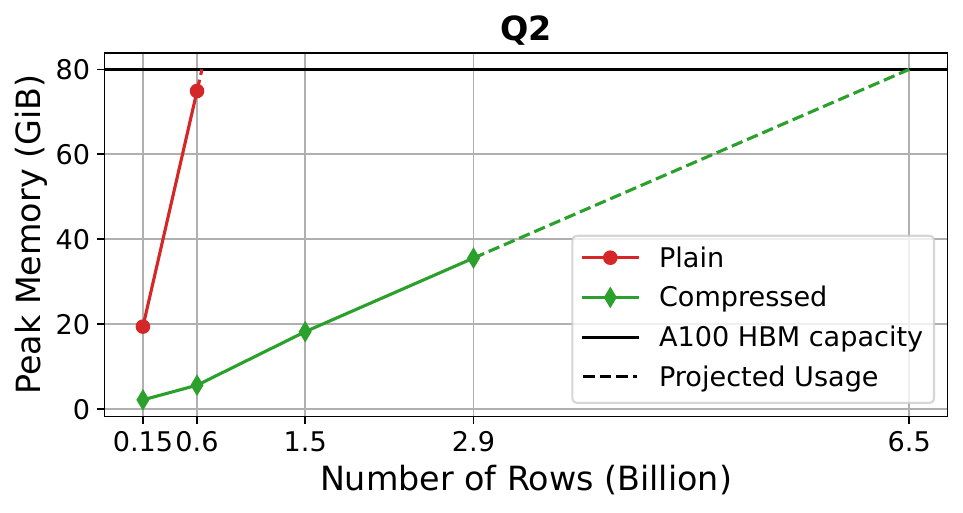}}
{\includegraphics[width=0.3\textwidth]{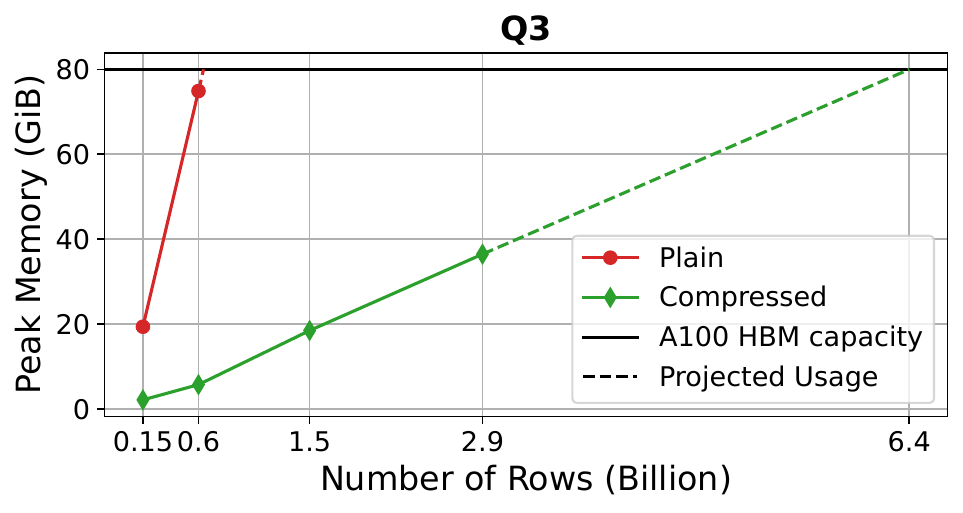}}
\vspace{-14pt}
\caption{Peak GPU memory used and projected for production queries on Plain and Compressed data for different table sizes.\vspace{-5pt}}
\label{fig:pbi-memory}
\end{figure*}

Figure~\ref{fig:pbi-memory} shows actuals and projections, using a linear model, of peak GPU memory consumption during query runs as we increase the fact table size. For Plain, we can only fit less than 50\% of fact table rows whereas with Compressed we could go up to 157\% (4.62B) for Q1, 222\% (6.52B) for Q2, and 216\% (6.36B) for Q3.

\section{Query Optimization Rules}
\label{appendix:query-optimization}

Our query execution plans are based on optimized versions of the tensor program query plans from TQP, with the following optimizations applied:

\begin{itemize}[leftmargin=*,topsep=0pt]
    \item \textbf{Apply predicates to RLE columns before Plain columns}: RLE columns can be filtered more efficiently due to their compressed representation, and they are potentially highly selective. When an RLE predicate is selective, it produces an Index mask that can be applied to Plain columns, avoiding the need to evaluate the Plain predicate on all rows.
    \item \textbf{Composite predicate evaluation on RLE columns}: When an RLE column has multiple predicates (e.g., \texttt{col > 10 AND col < 100}), we evaluate all predicates directly on the RLE value tensor to produce a single boolean mask, then apply it to the start and end positions once. This avoids the overhead of creating intermediate RLE masks and repeatedly applying the range intersection algorithm (\Cref{alg:range_intersect}). This is in contrast with Plain columns, where predicates are evaluated one-by-one to produce individual masks that are then combined with AND operations, as Plain mask AND operations are much less costly than RLE range intersections. 
    \item \textbf{Join ordering to prioritize RLE join columns}: If a table participates in multiple join and/or semi-join operations, we do the joins/semi-joins involving the RLE columns before the Plain columns. This helps to avoid/reduce fragmentation of RLE runs that could otherwise happen if the joins/semi-joins are performed with Plain columns first. In the spirit of predicate pushdown, prioritizing RLE joins allows early filtering of entire runs when join conditions are not satisfied, reducing the data volume for subsequent operations.
    \item \textbf{Avoiding redundant filter operations in groupby-aggregate}: If in a filter-groupby-aggregate query pattern, the groupby columns are RLE, then we apply the filter operations just on the groupby columns and not on the aggregate columns. This is because the RLE groupby columns will track the filtered ranges through the adjusted start and end positions, and will automatically only consider rows for aggregation within those ranges. Thus, explicitly applying the filter operations on the aggregate columns as well is functionally correct, but redundant. 
\end{itemize}

\noindent We find that these optimizations also benefit Plain processing, explaining why our Plain implementation outperforms baseline TQP: (1) \textbf{selection pushdown} - evaluating selections and (semi-)joins over potentially highly selective columns first produces masks that reduce work on other columns, and (2) \textbf{NULL optimization} - our query plans exclude NULL handling operators and assume non-NULL data, allowing \texttt{count(*)} to be computed once for all aggregations rather than per-column counts. We acknowledge that the lack of NULL support is a limitation. These optimizations are currently applied manually; incorporating them into a data encoding-aware query optimizer and supporting NULL handling in compressed representations are left as future work.


\begin{thebibliography}{54}


\ifx \showCODEN    \undefined \def \showCODEN     #1{\unskip}     \fi
\ifx \showDOI      \undefined \def \showDOI       #1{#1}\fi
\ifx \showISBNx    \undefined \def \showISBNx     #1{\unskip}     \fi
\ifx \showISBNxiii \undefined \def \showISBNxiii  #1{\unskip}     \fi
\ifx \showISSN     \undefined \def \showISSN      #1{\unskip}     \fi
\ifx \showLCCN     \undefined \def \showLCCN      #1{\unskip}     \fi
\ifx \shownote     \undefined \def \shownote      #1{#1}          \fi
\ifx \showarticletitle \undefined \def \showarticletitle #1{#1}   \fi
\ifx \showURL      \undefined \def \showURL       {\relax}        \fi
\providecommand\bibfield[2]{#2}
\providecommand\bibinfo[2]{#2}
\providecommand\natexlab[1]{#1}
\providecommand\showeprint[2][]{arXiv:#2}

\bibitem[arr(ssed)]%
        {arrow-ree}
 \bibinfo{year}{2025 (last accessed)}\natexlab{}.
\newblock \bibinfo{title}{Arrow Columnar Format: Run-End Encoded Layout}.
\newblock \bibinfo{howpublished}{[Online] Available from: \url{https://arrow.apache.org/docs/format/Columnar.html\#run-end-encoded-layout}}.
\newblock


\bibitem[vor(ssed)]%
        {vortex}
 \bibinfo{year}{2025 (last accessed)}\natexlab{}.
\newblock \bibinfo{title}{Vortex}.
\newblock \bibinfo{howpublished}{[Online] Available from: \url{https://github.com/spiraldb/vortex}}.
\newblock


\bibitem[Abadi et~al\mbox{.}(2013)]%
        {abadi_design_implementation_column_dbs_2013}
\bibfield{author}{\bibinfo{person}{Daniel Abadi}, \bibinfo{person}{Peter Boncz}, {and} \bibinfo{person}{Stavros Harizopoulos}.} \bibinfo{year}{2013}\natexlab{}.
\newblock \bibinfo{booktitle}{\emph{The Design and Implementation of Modern Column-Oriented Database Systems}}.
\newblock \bibinfo{publisher}{Now Publishers Inc.}, \bibinfo{address}{Hanover, MA, USA}.
\newblock
\showISBNx{1601987544}


\bibitem[Abadi et~al\mbox{.}(2006)]%
        {c_store_abadi_2006}
\bibfield{author}{\bibinfo{person}{Daniel Abadi}, \bibinfo{person}{Samuel Madden}, {and} \bibinfo{person}{Miguel Ferreira}.} \bibinfo{year}{2006}\natexlab{}.
\newblock \showarticletitle{Integrating Compression and Execution in Column-Oriented Database Systems}. In \bibinfo{booktitle}{\emph{Proceedings of the 2006 ACM SIGMOD International Conference on Management of Data}} (Chicago, IL, USA) \emph{(\bibinfo{series}{SIGMOD '06})}. \bibinfo{publisher}{Association for Computing Machinery}, \bibinfo{address}{New York, NY, USA}, \bibinfo{pages}{671–682}.
\newblock
\showISBNx{1595934340}
\urldef\tempurl%
\url{https://doi.org/10.1145/1142473.1142548}
\showDOI{\tempurl}


\bibitem[Afroozeh and Boncz(2023)]%
        {fastlanes}
\bibfield{author}{\bibinfo{person}{Azim Afroozeh} {and} \bibinfo{person}{Peter Boncz}.} \bibinfo{year}{2023}\natexlab{}.
\newblock \showarticletitle{The FastLanes Compression Layout: Decoding > 100 Billion Integers per Second with Scalar Code}.
\newblock \bibinfo{journal}{\emph{PVLDB}} \bibinfo{volume}{16}, \bibinfo{number}{9} (\bibinfo{date}{May} \bibinfo{year}{2023}), \bibinfo{pages}{2132--2144}.
\newblock
\showISSN{2150-8097}
\urldef\tempurl%
\url{https://doi.org/10.14778/3598581.3598587}
\showDOI{\tempurl}


\bibitem[Afroozeh et~al\mbox{.}(2024)]%
        {fastlanes-gpu}
\bibfield{author}{\bibinfo{person}{Azim Afroozeh}, \bibinfo{person}{Lotte Felius}, {and} \bibinfo{person}{Peter Boncz}.} \bibinfo{year}{2024}\natexlab{}.
\newblock \showarticletitle{Accelerating GPU Data Processing using FastLanes Compression}. In \bibinfo{booktitle}{\emph{Proceedings of the 20th International Workshop on Data Management on New Hardware}} (Santiago, AA, Chile) \emph{(\bibinfo{series}{DaMoN '24})}. \bibinfo{publisher}{Association for Computing Machinery}, \bibinfo{address}{New York, NY, USA}, Article \bibinfo{articleno}{8}, \bibinfo{numpages}{11}~pages.
\newblock
\showISBNx{9798400706677}
\urldef\tempurl%
\url{https://doi.org/10.1145/3662010.3663450}
\showDOI{\tempurl}


\bibitem[Asada et~al\mbox{.}(2022)]%
        {tensor_tea_vldb_2022}
\bibfield{author}{\bibinfo{person}{Yuki Asada}, \bibinfo{person}{Victor Fu}, \bibinfo{person}{Apurva Gandhi}, \bibinfo{person}{Advitya Gemawat}, \bibinfo{person}{Lihao Zhang}, \bibinfo{person}{Dong He}, \bibinfo{person}{Vivek Gupta}, \bibinfo{person}{Ehi Nosakhare}, \bibinfo{person}{Dalitso Banda}, \bibinfo{person}{Rathijit Sen}, {and} \bibinfo{person}{Matteo Interlandi}.} \bibinfo{year}{2022}\natexlab{}.
\newblock \showarticletitle{Share the tensor tea: how databases can leverage the machine learning ecosystem}.
\newblock \bibinfo{journal}{\emph{PVLDB}} (\bibinfo{year}{2022}), \bibinfo{pages}{3598--3601}.
\newblock


\bibitem[BlazingSQL(2021)]%
        {blazingsql_2022}
\bibfield{author}{\bibinfo{person}{BlazingSQL}.} \bibinfo{year}{2021}\natexlab{}.
\newblock \bibinfo{title}{BlazingSQL}.
\newblock \bibinfo{howpublished}{\url{https://github.com/BlazingDB/blazingsql}}.
\newblock


\bibitem[Cao et~al\mbox{.}(2023)]%
        {gpudb-characterization-opt}
\bibfield{author}{\bibinfo{person}{Jiashen Cao}, \bibinfo{person}{Rathijit Sen}, \bibinfo{person}{Matteo Interlandi}, \bibinfo{person}{Joy Arulraj}, {and} \bibinfo{person}{Hyesoon Kim}.} \bibinfo{year}{2023}\natexlab{}.
\newblock \showarticletitle{GPU Database Systems Characterization and Optimization}.
\newblock \bibinfo{journal}{\emph{PVLDB}} \bibinfo{volume}{17}, \bibinfo{number}{3} (\bibinfo{date}{Nov.} \bibinfo{year}{2023}), \bibinfo{pages}{441--454}.
\newblock
\showISSN{2150-8097}
\urldef\tempurl%
\url{https://doi.org/10.14778/3632093.3632107}
\showDOI{\tempurl}


\bibitem[Chrysogelos et~al\mbox{.}(2019)]%
        {chrysogelos19-hetexchange}
\bibfield{author}{\bibinfo{person}{Periklis Chrysogelos}, \bibinfo{person}{Manos Karpathiotakis}, \bibinfo{person}{Raja Appuswamy}, {and} \bibinfo{person}{Anastasia Ailamaki}.} \bibinfo{year}{2019}\natexlab{}.
\newblock \showarticletitle{HetExchange: encapsulating heterogeneous CPU-GPU parallelism in JIT compiled engines}.
\newblock \bibinfo{journal}{\emph{Proc. VLDB Endow.}} \bibinfo{volume}{12}, \bibinfo{number}{5} (\bibinfo{date}{Jan.} \bibinfo{year}{2019}), \bibinfo{pages}{544–556}.
\newblock
\showISSN{2150-8097}
\urldef\tempurl%
\url{https://doi.org/10.14778/3303753.3303760}
\showDOI{\tempurl}


\bibitem[Cui et~al\mbox{.}(2023)]%
        {tqp-xbox}
\bibfield{author}{\bibinfo{person}{Wei Cui}, \bibinfo{person}{Qianxi Zhang}, \bibinfo{person}{Spyros Blanas}, \bibinfo{person}{Jes\'{u}s Camacho-Rodr\'{\i}guez}, \bibinfo{person}{Brandon Haynes}, \bibinfo{person}{Yinan Li}, \bibinfo{person}{Ravi Ramamurthy}, \bibinfo{person}{Peng Cheng}, \bibinfo{person}{Rathijit Sen}, {and} \bibinfo{person}{Matteo Interlandi}.} \bibinfo{year}{2023}\natexlab{}.
\newblock \showarticletitle{Query Processing on Gaming Consoles}. In \bibinfo{booktitle}{\emph{Proceedings of the 19th International Workshop on Data Management on New Hardware}} (Seattle, WA, USA) \emph{(\bibinfo{series}{DaMoN '23})}. \bibinfo{publisher}{Association for Computing Machinery}, \bibinfo{address}{New York, NY, USA}, \bibinfo{pages}{86--88}.
\newblock
\showISBNx{9798400701917}
\urldef\tempurl%
\url{https://doi.org/10.1145/3592980.3595313}
\showDOI{\tempurl}


\bibitem[Damme et~al\mbox{.}(2020)]%
        {damme2020morphstore}
\bibfield{author}{\bibinfo{person}{Patrick Damme}, \bibinfo{person}{Annett Ungeth{"u}m}, \bibinfo{person}{Johannes Pietrzyk}, \bibinfo{person}{Alexander Krause}, \bibinfo{person}{Dirk Habich}, {and} \bibinfo{person}{Wolfgang Lehner}.} \bibinfo{year}{2020}\natexlab{}.
\newblock \showarticletitle{Morphstore: Analytical query engine with a holistic compression-enabled processing model}.
\newblock \bibinfo{journal}{\emph{arXiv preprint arXiv:2004.09350}} (\bibinfo{year}{2020}).
\newblock


\bibitem[Data(2025)]%
        {voltron-data}
\bibfield{author}{\bibinfo{person}{Voltron Data}.} \bibinfo{year}{(last accessed) 2025}\natexlab{}.
\newblock \bibinfo{title}{Theseus The Enterprise SQL Engine}.
\newblock \bibinfo{howpublished}{\url{https://voltrondata.com/}}.
\newblock


\bibitem[De~Morgan(1847)]%
        {demorgan1847formal}
\bibfield{author}{\bibinfo{person}{Augustus De~Morgan}.} \bibinfo{year}{1847}\natexlab{}.
\newblock \bibinfo{booktitle}{\emph{Formal Logic: Or, The Calculus of Inference, Necessary and Probable}}.
\newblock \bibinfo{publisher}{Taylor and Walton}, \bibinfo{address}{London}.
\newblock


\bibitem[Deng et~al\mbox{.}(2024)]%
        {deng24-prefetching}
\bibfield{author}{\bibinfo{person}{Yangshen Deng}, \bibinfo{person}{Shiwen Chen}, \bibinfo{person}{Zhaoyang Hong}, {and} \bibinfo{person}{Bo Tang}.} \bibinfo{year}{2024}\natexlab{}.
\newblock \showarticletitle{How Does Software Prefetching Work on GPU Query Processing?}. In \bibinfo{booktitle}{\emph{Proceedings of the 20th International Workshop on Data Management on New Hardware}} (Santiago, AA, Chile) \emph{(\bibinfo{series}{DaMoN '24})}. \bibinfo{publisher}{Association for Computing Machinery}, \bibinfo{address}{New York, NY, USA}, Article \bibinfo{articleno}{5}, \bibinfo{numpages}{9}~pages.
\newblock
\showISBNx{9798400706677}
\urldef\tempurl%
\url{https://doi.org/10.1145/3662010.3663445}
\showDOI{\tempurl}


\bibitem[Fang et~al\mbox{.}(2010)]%
        {fang2010database}
\bibfield{author}{\bibinfo{person}{Wenbin Fang}, \bibinfo{person}{Bingsheng He}, {and} \bibinfo{person}{Qiong Luo}.} \bibinfo{year}{2010}\natexlab{}.
\newblock \showarticletitle{Database compression on graphics processors}.
\newblock \bibinfo{journal}{\emph{Proceedings of the VLDB Endowment}} \bibinfo{volume}{3}, \bibinfo{number}{1-2} (\bibinfo{year}{2010}), \bibinfo{pages}{670--680}.
\newblock


\bibitem[Foundation(2021)]%
        {parquet}
\bibfield{author}{\bibinfo{person}{Apache~Software Foundation}.} \bibinfo{year}{2021}\natexlab{}.
\newblock \bibinfo{title}{Apache Parquet}.
\newblock
\newblock
\urldef\tempurl%
\url{https://parquet.apache.org/}
\showURL{%
\tempurl}
\newblock
\shownote{Accessed: 2025-03-25}.


\bibitem[Gandhi et~al\mbox{.}(2022)]%
        {gandhi2022tensor}
\bibfield{author}{\bibinfo{person}{Apurva Gandhi}, \bibinfo{person}{Yuki Asada}, \bibinfo{person}{Victor Fu}, \bibinfo{person}{Advitya Gemawat}, \bibinfo{person}{Lihao Zhang}, \bibinfo{person}{Rathijit Sen}, \bibinfo{person}{Carlo Curino}, \bibinfo{person}{Jes{\'u}s Camacho-Rodr{\'\i}guez}, {and} \bibinfo{person}{Matteo Interlandi}.} \bibinfo{year}{2022}\natexlab{}.
\newblock \showarticletitle{The Tensor Data Platform: Towards an AI-centric Database System}. In \bibinfo{booktitle}{\emph{CIDR}}.
\newblock


\bibitem[He et~al\mbox{.}(2022)]%
        {surakav_he_2022}
\bibfield{author}{\bibinfo{person}{Dong He}, \bibinfo{person}{Supun~C Nakandala}, \bibinfo{person}{Dalitso Banda}, \bibinfo{person}{Rathijit Sen}, \bibinfo{person}{Karla Saur}, \bibinfo{person}{Kwanghyun Park}, \bibinfo{person}{Carlo Curino}, \bibinfo{person}{Jes\'{u}s Camacho-Rodr\'{\i}guez}, \bibinfo{person}{Konstantinos Karanasos}, {and} \bibinfo{person}{Matteo Interlandi}.} \bibinfo{year}{2022}\natexlab{}.
\newblock \showarticletitle{Query Processing on Tensor Computation Runtimes}.
\newblock \bibinfo{journal}{\emph{PVLDB}} (\bibinfo{year}{2022}), \bibinfo{pages}{2811--2825}.
\newblock


\bibitem[HeavyDB(ssed)]%
        {heavydb}
\bibfield{author}{\bibinfo{person}{HeavyDB}.} \bibinfo{year}{2025 (last accessed)}\natexlab{}.
\newblock \bibinfo{title}{HeavyDB}.
\newblock \bibinfo{howpublished}{\url{https://github.com/heavyai/heavydb}}.
\newblock


\bibitem[Hong et~al\mbox{.}(2025)]%
        {hong25-themis}
\bibfield{author}{\bibinfo{person}{Kijae Hong}, \bibinfo{person}{Kyoungmin Kim}, \bibinfo{person}{Young-Koo Lee}, \bibinfo{person}{Yang-Sae Moon}, \bibinfo{person}{Sourav~S Bhowmick}, {and} \bibinfo{person}{Wook-Shin Han}.} \bibinfo{year}{2025}\natexlab{}.
\newblock \showarticletitle{Themis: A GPU-Accelerated Relational Query Execution Engine}.
\newblock \bibinfo{journal}{\emph{Proc. VLDB Endow.}} \bibinfo{volume}{18}, \bibinfo{number}{2} (\bibinfo{date}{Feb.} \bibinfo{year}{2025}), \bibinfo{pages}{426–438}.
\newblock
\showISSN{2150-8097}
\urldef\tempurl%
\url{https://doi.org/10.14778/3705829.3705856}
\showDOI{\tempurl}


\bibitem[Hu et~al\mbox{.}(2022)]%
        {Hu22-tcudb}
\bibfield{author}{\bibinfo{person}{Yu-Ching Hu}, \bibinfo{person}{Yuliang Li}, {and} \bibinfo{person}{Hung-Wei Tseng}.} \bibinfo{year}{2022}\natexlab{}.
\newblock \showarticletitle{TCUDB: Accelerating Database with Tensor Processors}. In \bibinfo{booktitle}{\emph{Proceedings of the 2022 International Conference on Management of Data}} (Philadelphia, PA, USA) \emph{(\bibinfo{series}{SIGMOD '22})}. \bibinfo{publisher}{Association for Computing Machinery}, \bibinfo{address}{New York, NY, USA}, \bibinfo{pages}{1360--1374}.
\newblock
\showISBNx{9781450392495}
\urldef\tempurl%
\url{https://doi.org/10.1145/3514221.3517869}
\showDOI{\tempurl}


\bibitem[Jiang et~al\mbox{.}(2021)]%
        {codecdb_jiang_2021}
\bibfield{author}{\bibinfo{person}{Hao Jiang}, \bibinfo{person}{Chunwei Liu}, \bibinfo{person}{John Paparrizos}, \bibinfo{person}{Andrew~A. Chien}, \bibinfo{person}{Jihong Ma}, {and} \bibinfo{person}{Aaron~J. Elmore}.} \bibinfo{year}{2021}\natexlab{}.
\newblock \showarticletitle{Good to the Last Bit: Data-Driven Encoding with CodecDB}. In \bibinfo{booktitle}{\emph{Proceedings of the 2021 International Conference on Management of Data}} (Virtual Event, China) \emph{(\bibinfo{series}{SIGMOD '21})}. \bibinfo{publisher}{Association for Computing Machinery}, \bibinfo{address}{New York, NY, USA}, \bibinfo{pages}{843–856}.
\newblock
\showISBNx{9781450383431}
\urldef\tempurl%
\url{https://doi.org/10.1145/3448016.3457283}
\showDOI{\tempurl}


\bibitem[Karnagel et~al\mbox{.}(2015)]%
        {Tomas15-groupby}
\bibfield{author}{\bibinfo{person}{Tomas Karnagel}, \bibinfo{person}{Ren{\'e} M{\"u}ller}, {and} \bibinfo{person}{Guy~M. Lohman}.} \bibinfo{year}{2015}\natexlab{}.
\newblock \showarticletitle{Optimizing GPU-accelerated Group-By and Aggregation}. In \bibinfo{booktitle}{\emph{ADMS@VLDB}}.
\newblock
\urldef\tempurl%
\url{https://api.semanticscholar.org/CorpusID:5017248}
\showURL{%
\tempurl}


\bibitem[Kuschewski et~al\mbox{.}(2023)]%
        {kuschewski2023btrblocks}
\bibfield{author}{\bibinfo{person}{Maximilian Kuschewski}, \bibinfo{person}{David Sauerwein}, \bibinfo{person}{Adnan Alhomssi}, {and} \bibinfo{person}{Viktor Leis}.} \bibinfo{year}{2023}\natexlab{}.
\newblock \showarticletitle{BtrBlocks: Efficient Columnar Compression for Data Lakes}. In \bibinfo{booktitle}{\emph{Proceedings of the 2023 ACM SIGMOD International Conference on Management of Data (SIGMOD~'23)}}. \bibinfo{pages}{2205--2217}.
\newblock


\bibitem[Layer et~al\mbox{.}(2013)]%
        {layer2013binary}
\bibfield{author}{\bibinfo{person}{Ryan~M Layer}, \bibinfo{person}{Kevin Skadron}, \bibinfo{person}{Gabriel Robins}, \bibinfo{person}{Ira~M Hall}, {and} \bibinfo{person}{Aaron~R Quinlan}.} \bibinfo{year}{2013}\natexlab{}.
\newblock \showarticletitle{Binary Interval Search: a scalable algorithm for counting interval intersections}.
\newblock \bibinfo{journal}{\emph{Bioinformatics}} \bibinfo{volume}{29}, \bibinfo{number}{1} (\bibinfo{year}{2013}), \bibinfo{pages}{1--7}.
\newblock


\bibitem[Lee et~al\mbox{.}(2014)]%
        {lee2014joins}
\bibfield{author}{\bibinfo{person}{Jae-Gil Lee}, \bibinfo{person}{Guy Lohman}, \bibinfo{person}{Konstantinos Morfonios}, \bibinfo{person}{Keshava Murthy}, \bibinfo{person}{Ippokratis Pandis}, \bibinfo{person}{Lin Qiao}, \bibinfo{person}{Vijayshankar Raman}, \bibinfo{person}{Vincent~Kulandai Samy}, \bibinfo{person}{Richard Sidle}, \bibinfo{person}{Knut Stolze}, {et~al\mbox{.}}} \bibinfo{year}{2014}\natexlab{}.
\newblock \showarticletitle{Joins on encoded and partitioned data}.
\newblock \bibinfo{journal}{\emph{Proceedings of the VLDB Endowment}} (\bibinfo{year}{2014}).
\newblock


\bibitem[Li et~al\mbox{.}(2016)]%
        {hippogriffdb}
\bibfield{author}{\bibinfo{person}{Jing Li}, \bibinfo{person}{Hung-Wei Tseng}, \bibinfo{person}{Chunbin Lin}, \bibinfo{person}{Yannis Papakonstantinou}, {and} \bibinfo{person}{Steven Swanson}.} \bibinfo{year}{2016}\natexlab{}.
\newblock \showarticletitle{HippogriffDB: balancing I/O and GPU bandwidth in big data analytics}.
\newblock \bibinfo{journal}{\emph{PVLDB}} \bibinfo{volume}{9}, \bibinfo{number}{14} (\bibinfo{date}{Oct.} \bibinfo{year}{2016}), \bibinfo{pages}{1647–1658}.
\newblock
\showISSN{2150--8097}
\urldef\tempurl%
\url{https://doi.org/10.14778/3007328.3007331}
\showDOI{\tempurl}


\bibitem[Lutz et~al\mbox{.}(2020)]%
        {Lutz20-pump-up-the-volume}
\bibfield{author}{\bibinfo{person}{Clemens Lutz}, \bibinfo{person}{Sebastian Bre\ss{}}, \bibinfo{person}{Steffen Zeuch}, \bibinfo{person}{Tilmann Rabl}, {and} \bibinfo{person}{Volker Markl}.} \bibinfo{year}{2020}\natexlab{}.
\newblock \showarticletitle{Pump Up the Volume: Processing Large Data on GPUs with Fast Interconnects}. In \bibinfo{booktitle}{\emph{Proceedings of the 2020 ACM SIGMOD International Conference on Management of Data}} (Portland, OR, USA) \emph{(\bibinfo{series}{SIGMOD '20})}. \bibinfo{publisher}{Association for Computing Machinery}, \bibinfo{address}{New York, NY, USA}, \bibinfo{pages}{1633--1649}.
\newblock
\showISBNx{9781450367356}
\urldef\tempurl%
\url{https://doi.org/10.1145/3318464.3389705}
\showDOI{\tempurl}


\bibitem[Maltenberger et~al\mbox{.}(2022)]%
        {maltenberger22-multigpu-sort}
\bibfield{author}{\bibinfo{person}{Tobias Maltenberger}, \bibinfo{person}{Ivan Ilic}, \bibinfo{person}{Ilin Tolovski}, {and} \bibinfo{person}{Tilmann Rabl}.} \bibinfo{year}{2022}\natexlab{}.
\newblock \showarticletitle{Evaluating Multi-GPU Sorting with Modern Interconnects}. In \bibinfo{booktitle}{\emph{Proceedings of the 2022 International Conference on Management of Data}} (Philadelphia, PA, USA) \emph{(\bibinfo{series}{SIGMOD '22})}. \bibinfo{publisher}{Association for Computing Machinery}, \bibinfo{address}{New York, NY, USA}, \bibinfo{pages}{1795–1809}.
\newblock
\showISBNx{9781450392495}
\urldef\tempurl%
\url{https://doi.org/10.1145/3514221.3517842}
\showDOI{\tempurl}


\bibitem[McKinney(2010)]%
        {pandas}
\bibfield{author}{\bibinfo{person}{Wes McKinney}.} \bibinfo{year}{2010}\natexlab{}.
\newblock \showarticletitle{Data structures for statistical computing in python}. In \bibinfo{booktitle}{\emph{Proceedings of the 9th Python in Science Conference}}, Vol.~\bibinfo{volume}{445}. \bibinfo{pages}{51--56}.
\newblock


\bibitem[Microsoft(2024a)]%
        {Azure-Dv5}
\bibfield{author}{\bibinfo{person}{Microsoft}.} \bibinfo{year}{2024}\natexlab{a}.
\newblock \bibinfo{title}{Dv5 sizes series}.
\newblock
\newblock
\urldef\tempurl%
\url{https://learn.microsoft.com/en-us/azure/virtual-machines/sizes/general-purpose/dv5-series}
\showURL{%
\tempurl}


\bibitem[Microsoft(2024b)]%
        {Azure-NCA100}
\bibfield{author}{\bibinfo{person}{Microsoft}.} \bibinfo{year}{2024}\natexlab{b}.
\newblock \bibinfo{title}{NC\_A100\_v4 sizes series}.
\newblock
\newblock
\urldef\tempurl%
\url{https://learn.microsoft.com/en-us/azure/virtual-machines/sizes/gpu-accelerated/nca100v4-series}
\showURL{%
\tempurl}


\bibitem[Microsoft(2024c)]%
        {v-order}
\bibfield{author}{\bibinfo{person}{Microsoft}.} \bibinfo{year}{2024}\natexlab{c}.
\newblock \bibinfo{title}{Understand V-Order for Microsoft Fabric Warehouse}.
\newblock \bibinfo{howpublished}{[Online] Available from: \url{https://learn.microsoft.com/en-us/fabric/data-warehouse/v-order}}.
\newblock


\bibitem[Microsoft(2025)]%
        {as}
\bibfield{author}{\bibinfo{person}{Microsoft}.} \bibinfo{year}{2025}\natexlab{}.
\newblock \bibinfo{title}{What is Analysis Services?}
\newblock \bibinfo{howpublished}{[Online] Available from: \url{https://learn.microsoft.com/en-us/analysis-services/analysis-services-overview?view=asallproducts-allversions}}.
\newblock


\bibitem[Paszke et~al\mbox{.}(2019)]%
        {pytorch_paszke_2019}
\bibfield{author}{\bibinfo{person}{Adam Paszke}, \bibinfo{person}{Sam Gross}, \bibinfo{person}{Francisco Massa}, \bibinfo{person}{Adam Lerer}, \bibinfo{person}{James Bradbury}, \bibinfo{person}{Gregory Chanan}, \bibinfo{person}{Trevor Killeen}, \bibinfo{person}{Zeming Lin}, \bibinfo{person}{Natalia Gimelshein}, \bibinfo{person}{Luca Antiga}, \bibinfo{person}{Alban Desmaison}, \bibinfo{person}{Andreas Kopf}, \bibinfo{person}{Edward Yang}, \bibinfo{person}{Zachary DeVito}, \bibinfo{person}{Martin Raison}, \bibinfo{person}{Alykhan Tejani}, \bibinfo{person}{Sasank Chilamkurthy}, \bibinfo{person}{Benoit Steiner}, \bibinfo{person}{Lu Fang}, \bibinfo{person}{Junjie Bai}, {and} \bibinfo{person}{Soumith Chintala}.} \bibinfo{year}{2019}\natexlab{}.
\newblock \showarticletitle{PyTorch: An Imperative Style, High-Performance Deep Learning Library}.
\newblock In \bibinfo{booktitle}{\emph{NeurIPS}}. \bibinfo{pages}{8024--8035}.
\newblock


\bibitem[Paul et~al\mbox{.}(2021)]%
        {book:gpudb:2021}
\bibfield{author}{\bibinfo{person}{Johns Paul}, \bibinfo{person}{Shengliang Lu}, {and} \bibinfo{person}{Bingsheng He}.} \bibinfo{year}{2021}\natexlab{}.
\newblock \bibinfo{booktitle}{\emph{Database Systems on GPUs}}.
\newblock \bibinfo{publisher}{Now Foundations and Trends}.
\newblock


\bibitem[Rosenfeld et~al\mbox{.}(2022)]%
        {rosenfeld2022query}
\bibfield{author}{\bibinfo{person}{Viktor Rosenfeld}, \bibinfo{person}{Sebastian Bre{\ss}}, {and} \bibinfo{person}{Volker Markl}.} \bibinfo{year}{2022}\natexlab{}.
\newblock \showarticletitle{Query processing on heterogeneous CPU/GPU systems}.
\newblock \bibinfo{journal}{\emph{ACM Computing Surveys (CSUR)}} \bibinfo{volume}{55}, \bibinfo{number}{1} (\bibinfo{year}{2022}), \bibinfo{pages}{1--38}.
\newblock


\bibitem[Rui et~al\mbox{.}(2020)]%
        {Rui20-multigpujoin}
\bibfield{author}{\bibinfo{person}{Ran Rui}, \bibinfo{person}{Hao Li}, {and} \bibinfo{person}{Yi-Cheng Tu}.} \bibinfo{year}{2020}\natexlab{}.
\newblock \showarticletitle{Efficient Join Algorithms for Large Database Tables in a Multi-GPU Environment}.
\newblock \bibinfo{journal}{\emph{Proc. VLDB Endow.}} \bibinfo{volume}{14}, \bibinfo{number}{4} (\bibinfo{date}{dec} \bibinfo{year}{2020}), \bibinfo{pages}{708--720}.
\newblock
\showISSN{2150-8097}
\urldef\tempurl%
\url{https://doi.org/10.14778/3436905.3436927}
\showDOI{\tempurl}


\bibitem[Rui and Tu(2017)]%
        {Rui17-fastequijoin}
\bibfield{author}{\bibinfo{person}{Ran Rui} {and} \bibinfo{person}{Yi-Cheng Tu}.} \bibinfo{year}{2017}\natexlab{}.
\newblock \showarticletitle{Fast Equi-Join Algorithms on GPUs: Design and Implementation}. In \bibinfo{booktitle}{\emph{Proceedings of the 29th International Conference on Scientific and Statistical Database Management}} (Chicago, IL, USA) \emph{(\bibinfo{series}{SSDBM '17})}. \bibinfo{publisher}{Association for Computing Machinery}, \bibinfo{address}{New York, NY, USA}, Article \bibinfo{articleno}{17}, \bibinfo{numpages}{12}~pages.
\newblock
\showISBNx{9781450352826}
\urldef\tempurl%
\url{https://doi.org/10.1145/3085504.3085521}
\showDOI{\tempurl}


\bibitem[Shanbhag et~al\mbox{.}(2020)]%
        {crystal_shanbhag_2020}
\bibfield{author}{\bibinfo{person}{Anil Shanbhag}, \bibinfo{person}{Samuel Madden}, {and} \bibinfo{person}{Xiangyao Yu}.} \bibinfo{year}{2020}\natexlab{}.
\newblock \showarticletitle{A {{Study}} of the {{Fundamental Performance Characteristics}} of {{GPUs}} and {{CPUs}} for {{Database Analytics}}}. In \bibinfo{booktitle}{\emph{SIGMOD}}. \bibinfo{pages}{1617--1632}.
\newblock


\bibitem[Shanbhag et~al\mbox{.}(2022)]%
        {tile-integer-compression}
\bibfield{author}{\bibinfo{person}{Anil Shanbhag}, \bibinfo{person}{Bobbi~W. Yogatama}, \bibinfo{person}{Xiangyao Yu}, {and} \bibinfo{person}{Samuel Madden}.} \bibinfo{year}{2022}\natexlab{}.
\newblock \showarticletitle{Tile-based Lightweight Integer Compression in GPU}. In \bibinfo{booktitle}{\emph{Proceedings of the 2022 International Conference on Management of Data}} (Philadelphia, PA, USA) \emph{(\bibinfo{series}{SIGMOD '22})}. \bibinfo{publisher}{Association for Computing Machinery}, \bibinfo{address}{New York, NY, USA}, \bibinfo{pages}{1390--1403}.
\newblock
\showISBNx{9781450392495}
\urldef\tempurl%
\url{https://doi.org/10.1145/3514221.3526132}
\showDOI{\tempurl}


\bibitem[Sioulas et~al\mbox{.}(2019)]%
        {sioulas19-partitioned-radix-join}
\bibfield{author}{\bibinfo{person}{Panagiotis Sioulas}, \bibinfo{person}{Periklis Chrysogelos}, \bibinfo{person}{Manos Karpathiotakis}, \bibinfo{person}{Raja Appuswamy}, {and} \bibinfo{person}{Anastasia Ailamaki}.} \bibinfo{year}{2019}\natexlab{}.
\newblock \showarticletitle{Hardware-Conscious Hash-Joins on GPUs}. In \bibinfo{booktitle}{\emph{2019 IEEE 35th International Conference on Data Engineering (ICDE)}}. \bibinfo{pages}{698--709}.
\newblock
\urldef\tempurl%
\url{https://doi.org/10.1109/ICDE.2019.00068}
\showDOI{\tempurl}


\bibitem[Sitaridi(2016)]%
        {sitaridi2016gpu}
\bibfield{author}{\bibinfo{person}{Evangelia Sitaridi}.} \bibinfo{year}{2016}\natexlab{}.
\newblock \bibinfo{booktitle}{\emph{GPU-acceleration of in-memory data analytics}}.
\newblock \bibinfo{publisher}{Columbia University}.
\newblock


\bibitem[Stam(2022)]%
        {duckdb-reorder-compression}
\bibfield{author}{\bibinfo{person}{Jim Stam}.} \bibinfo{year}{2022}\natexlab{}.
\newblock \emph{\bibinfo{title}{Low overhead self-optimizing storage for compression in DuckDB}}.
\newblock \bibinfo{thesistype}{Master's\ thesis}. \bibinfo{school}{Universiteit van Amsterdam--Vrije Universiteit Amsterdam}.
\newblock


\bibitem[Travis~E(2006)]%
        {numpy}
\bibfield{author}{\bibinfo{person}{Oliphant Travis~E}.} \bibinfo{year}{2006}\natexlab{}.
\newblock \bibinfo{title}{NumPy}.
\newblock \bibinfo{howpublished}{\url{http://www.numpy.org/}}.
\newblock


\bibitem[Vogelsgesang et~al\mbox{.}(2018)]%
        {vogelsgesang2018get}
\bibfield{author}{\bibinfo{person}{Adrian Vogelsgesang}, \bibinfo{person}{Michael Haubenschild}, \bibinfo{person}{Jan Finis}, \bibinfo{person}{Alfons Kemper}, \bibinfo{person}{Viktor Leis}, \bibinfo{person}{Tobias M{\"u}hlbauer}, \bibinfo{person}{Thomas Neumann}, {and} \bibinfo{person}{Manuel Then}.} \bibinfo{year}{2018}\natexlab{}.
\newblock \showarticletitle{Get real: How benchmarks fail to represent the real world}. In \bibinfo{booktitle}{\emph{Proceedings of the Workshop on Testing Database Systems}}. \bibinfo{pages}{1--6}.
\newblock


\bibitem[Wu et~al\mbox{.}(2025)]%
        {wu25-gpu-joins-groupby}
\bibfield{author}{\bibinfo{person}{Bowen Wu}, \bibinfo{person}{Dimitrios Koutsoukos}, {and} \bibinfo{person}{Gustavo Alonso}.} \bibinfo{year}{2025}\natexlab{}.
\newblock \showarticletitle{Efficiently Processing Joins and Grouped Aggregations on GPUs}.
\newblock \bibinfo{journal}{\emph{Proc. ACM Manag. Data}} \bibinfo{volume}{3}, \bibinfo{number}{1}, Article \bibinfo{articleno}{39} (\bibinfo{date}{Feb.} \bibinfo{year}{2025}), \bibinfo{numpages}{27}~pages.
\newblock
\urldef\tempurl%
\url{https://doi.org/10.1145/3709689}
\showDOI{\tempurl}


\bibitem[Yogatama et~al\mbox{.}(2025)]%
        {Yogatama25-lancelot}
\bibfield{author}{\bibinfo{person}{Bobbi Yogatama}, \bibinfo{person}{Weiwei Gong}, {and} \bibinfo{person}{Xiangyao Yu}.} \bibinfo{year}{2025}\natexlab{}.
\newblock \showarticletitle{Scaling your Hybrid CPU-GPU DBMS to Multiple GPUs}.
\newblock \bibinfo{journal}{\emph{Proc. VLDB Endow.}} \bibinfo{volume}{17}, \bibinfo{number}{13} (\bibinfo{date}{Feb.} \bibinfo{year}{2025}), \bibinfo{pages}{4709–4722}.
\newblock
\showISSN{2150-8097}
\urldef\tempurl%
\url{https://doi.org/10.14778/3704965.3704977}
\showDOI{\tempurl}


\bibitem[Yogatama et~al\mbox{.}(2022)]%
        {Yogatama22-mordred}
\bibfield{author}{\bibinfo{person}{Bobbi~W. Yogatama}, \bibinfo{person}{Weiwei Gong}, {and} \bibinfo{person}{Xiangyao Yu}.} \bibinfo{year}{2022}\natexlab{}.
\newblock \showarticletitle{Orchestrating data placement and query execution in heterogeneous CPU-GPU DBMS}.
\newblock \bibinfo{journal}{\emph{Proc. VLDB Endow.}} \bibinfo{volume}{15}, \bibinfo{number}{11} (\bibinfo{date}{July} \bibinfo{year}{2022}), \bibinfo{pages}{2491–2503}.
\newblock
\showISSN{2150-8097}
\urldef\tempurl%
\url{https://doi.org/10.14778/3551793.3551809}
\showDOI{\tempurl}


\bibitem[Yuan et~al\mbox{.}(2025)]%
        {yuan25-vortex}
\bibfield{author}{\bibinfo{person}{Yichao Yuan}, \bibinfo{person}{Advait Iyer}, \bibinfo{person}{Lin Ma}, {and} \bibinfo{person}{Nishil Talati}.} \bibinfo{year}{2025}\natexlab{}.
\newblock \bibinfo{title}{Vortex: Overcoming Memory Capacity Limitations in GPU-Accelerated Large-Scale Data Analytics}.
\newblock
\newblock
\showeprint[arxiv]{2502.09541}~[cs.DB]
\urldef\tempurl%
\url{https://arxiv.org/abs/2502.09541}
\showURL{%
\tempurl}


\bibitem[Yuan et~al\mbox{.}(2013)]%
        {yuan2013yin}
\bibfield{author}{\bibinfo{person}{Yuan Yuan}, \bibinfo{person}{Rubao Lee}, {and} \bibinfo{person}{Xiaodong Zhang}.} \bibinfo{year}{2013}\natexlab{}.
\newblock \showarticletitle{The Yin and Yang of Processing Data Warehousing Queries on GPU Devices}. In \bibinfo{booktitle}{\emph{Proceedings of the VLDB Endowment (PVLDB)}}, Vol.~\bibinfo{volume}{6}. \bibinfo{pages}{817--828}.
\newblock


\bibitem[Zaharia et~al\mbox{.}(2016)]%
        {spark}
\bibfield{author}{\bibinfo{person}{Matei Zaharia}, \bibinfo{person}{Reynold~S. Xin}, \bibinfo{person}{Patrick Wendell}, \bibinfo{person}{Tathagata Das}, \bibinfo{person}{Michael Armbrust}, \bibinfo{person}{Ankur Dave}, \bibinfo{person}{Xiangrui Meng}, \bibinfo{person}{Josh Rosen}, \bibinfo{person}{Shivaram Venkataraman}, \bibinfo{person}{Michael~J. Franklin}, \bibinfo{person}{Ali Ghodsi}, \bibinfo{person}{Joseph Gonzalez}, \bibinfo{person}{Scott Shenker}, {and} \bibinfo{person}{Ion Stoica}.} \bibinfo{year}{2016}\natexlab{}.
\newblock \showarticletitle{Apache Spark: a unified engine for big data processing}.
\newblock \bibinfo{journal}{\emph{Commun. ACM}} \bibinfo{volume}{59}, \bibinfo{number}{11} (\bibinfo{date}{Oct.} \bibinfo{year}{2016}), \bibinfo{pages}{56–65}.
\newblock
\showISSN{0001-0782}
\urldef\tempurl%
\url{https://doi.org/10.1145/2934664}
\showDOI{\tempurl}


\bibitem[Zukowski et~al\mbox{.}(2006)]%
        {zukowski2006super}
\bibfield{author}{\bibinfo{person}{Marcin Zukowski}, \bibinfo{person}{Sandor Heman}, \bibinfo{person}{Niels Nes}, {and} \bibinfo{person}{Peter Boncz}.} \bibinfo{year}{2006}\natexlab{}.
\newblock \showarticletitle{Super-scalar RAM-CPU cache compression}. In \bibinfo{booktitle}{\emph{22nd International Conference on Data Engineering (ICDE'06)}}. IEEE, \bibinfo{pages}{59--59}.
\newblock


\end{thebibliography}
\end{document}